\documentclass[%
superscriptaddress,
floats,
floatfix,
twocolumn,
nofootinbib,
 amsmath,amssymb,
 aps,
floatfix,
]{revtex4-1}

\usepackage{array}
\usepackage{dcolumn}% Align table columns on decimal point
\usepackage{bm}% bold math
\usepackage[colorlinks, pdfborder={0 0 0}, plainpages=false]{hyperref}
\usepackage{graphicx}
\usepackage{xspace}
\usepackage[usenames,dvipsnames]{color}
\usepackage{amssymb}
\usepackage[normalem]{ulem} %for \sout
\usepackage{lineno}
\usepackage{letltxmacro}
\usepackage{amsfonts}
\usepackage{amsmath}
\usepackage{amssymb}
\usepackage{bm}
\usepackage{dcolumn}
\usepackage{epsfig}
\usepackage{graphicx}
\usepackage{graphics}
\usepackage[latin1]{inputenc}
\usepackage{latexsym}
\usepackage{rotating}
\usepackage{hyperref}
\usepackage[usenames]{color}
\usepackage{float}
\usepackage{xspace} 
\usepackage{mathrsfs}
\usepackage{subfigure}
\usepackage{multirow}
\usepackage{booktabs}
\usepackage{soul}

\usepackage{colortbl}
\usepackage{etoolbox}
\usepackage[table]{xcolor}
\def\prd{{\it Phys. Rev.} D~}

\def\apjl{{\it Astrophys. J.} Lett~}
\def\apj{{\it Astrophys. J.}}
\def\Msun{M_\odot}
\def\PRD{{\it Phys. Rev.} D~}

\def\mnras{{\it MNRAS~}}

\def\lesssim{\mathrel{\hbox{\rlap{\hbox{\lower4pt\hbox{$\sim$}}}\hbox{$<$}}}}
\def\gtrsim{\mathrel{\hbox{\rlap{\hbox{\lower4pt\hbox{$\sim$}}}\hbox{$>$}}}}
\def\alt{\mathrel{\hbox{\rlap{\hbox{\lower4pt\hbox{$\sim$}}}\hbox{$<$}}}}
\def\agt{\mathrel{\hbox{\rlap{\hbox{\lower4pt\hbox{$\sim$}}}\hbox{$>$}}}}

\def\gta{\ifmmode {\mathbin{\lower 3pt\hbox   %> or of order
    {$\,\rlap{\raise 5pt\hbox{$\char'076$}}\mathchar"7218\,$}}}
    \else {${\mathbin{\lower 3pt\hbox
    {$\rlap{\raise 5pt\hbox{$\char'076$}}\mathchar"7218\,$}}}
    $}\fi}
\def\lta{\ifmmode {\,\mathbin{\lower 3pt\hbox   %< or of order
    {$\,\rlap{\raise 5pt\hbox{$\char'074$}}\mathchar"7218\,$}}}
    \else {${\mathbin{\lower 3pt\hbox
    {$\rlap{\raise 5pt\hbox{$\char'074$}}\mathchar"7218\,$}}}
    $}\fi}

\newcommand{\msun}{{\rm M}_{\odot}}
\newcommand{\beq}{\begin{equation}}
\newcommand{\eeq}{\end{equation}}
\newcommand{\bea}{\begin{eqnarray}}
\newcommand{\eea}{\end{eqnarray}}

\definecolor{darkperiwinkle}{RGB}{102, 102, 128}

\newcommand{\NCSA}{\affiliation{NCSA, University of Illinois at Urbana-Champaign, Urbana, Illinois 61801, USA}}

\newcommand{\ANCSA}{\affiliation{Department of Astronomy, University of Illinois at Urbana-Champaign, Urbana, Illinois 61801, USA}}
\newcommand{\PNCSA}{\affiliation{Department of Physics, University of Illinois at Urbana-Champaign, Urbana, Illinois 61801, USA}}
\newcommand{\CITA}{\affiliation{Canadian Institute for Theoretical Astrophysics, 60 St.~George Street, University of Toronto, Toronto, ON M5S 3H8, Canada}}
\newcommand{\SPIN}{\affiliation{Students Pushing Innovation (SPIN) undergraduate intern at NCSA, University of Illinois at Urbana-Champaign, Urbana, Illinois 61801, USA}} 

\newcommand{\JPL}{\affiliation{Jet Propulsion Laboratory, 4800 Oak Grove Drive, Pasadena, California 91109, USA}}
\newcommand{\CS}{\affiliation{Department of Computer Science, University of Illinois at Urbana-Champaign, Urbana, Illinois, 61801}}

\newcommand{\CAM}{\affiliation{Institute of Astronomy, University of Cambridge, Madingley Road, Cambridge CB3 0HA, UK}}
\newcommand{\ED}{\affiliation{School of Mathematics, University of Edinburgh and Biomathematics and Statistics Scotland, James Clerk Maxwell Building, Peter Guthrie Tait Road, Edinburgh EH9 3FD, UK}}

\newcommand{\DAMTP}{\affiliation{DAMTP, Centre for Mathematical Sciences, University of Cambridge, Wilberforce Road, Cambridge CB3 0WA, UK}}
\newcommand{\CENTRA}{\affiliation{IST-CENTRA, Departamento de F{\'i}sica, Avenida Rovisco Pais 1, 1049 Lisboa, Portugal}}
\newcommand{\COR}{\affiliation{Cornell Center for Astrophysics and Planetary Science, Cornell University, Ithaca, New York 14853, USA}}
\newcommand{\UNI}{\affiliation{The University of Illinois Laboratory High School, University of Illinois at Urbana-Champaign, Urbana, Illinois 61801, USA}}
\newcommand{\AEI}{\affiliation{Max-Planck-Institut f\"ur Gravitationsphysik, Albert-Einstein-Institut,
Am M\"uhlenberg 1, D-14476 Golm, Germany}}
\newcommand{\AR}{\affiliation{Department of Physics, University of Arizona, Tucson, AZ 85721, USA}}

\definecolor{light-gray}{gray}{0.9}

\begin{document}

%\preprint{APS/123-QED}

\title{\textbf{E}ccentric, \textbf{N}onspinning, \textbf{I}nspiral \textbf{G}aussian-process \textbf{M}erger \textbf{A}pproximant for the detection and characterization of eccentric binary black hole mergers}
\author{E. A. Huerta}
\email{elihu@illinois.edu}\NCSA\ANCSA
\author{C. J. Moore}\CENTRA\DAMTP
\author{Prayush Kumar}\CITA\COR
\author{Daniel George}\NCSA\ANCSA
\author{Alvin J. K. Chua}\JPL\CAM
\author{Roland Haas}\NCSA
\author{Erik Wessel}\NCSA\AR
\author{Daniel Johnson}\NCSA\PNCSA\CS\SPIN
\author{Derek Glennon}\NCSA\PNCSA\ANCSA\SPIN
\author{Adam Rebei}\NCSA\UNI
\author{A. Miguel Holgado}\NCSA\ANCSA
\author{Jonathan R. Gair}\ED
\author{Harald P. Pfeiffer}\CITA\AEI

\date{\today}% It is always \today, today,

\begin{abstract}
We present \texttt{ENIGMA}, a time domain, inspiral-merger-ringdown waveform model that describes non-spinning binary black holes systems that evolve on moderately eccentric orbits. The inspiral evolution is described using a consistent combination of post-Newtonian theory, self-force and black hole perturbation theory. Assuming eccentric binaries that circularize prior to coalescence, we smoothly match the eccentric inspiral with a stand-alone, quasi-circular merger, which is constructed using machine learning algorithms that are trained with quasi-circular numerical relativity waveforms. We show that \texttt{ENIGMA} reproduces with excellent accuracy the dynamics of quasi-circular compact binaries. We validate \texttt{ENIGMA} using a set of \texttt{Einstein Toolkit} eccentric numerical relativity waveforms, which describe eccentric binary black hole mergers with mass-ratios between \(1 \leq q \leq 5.5\), and eccentricities \(e_0 \lesssim 0.2\) ten orbits before merger. We use this model to explore in detail the physics that can be extracted with moderately eccentric, non-spinning  binary black hole mergers. In particular, we use \texttt{ENIGMA} to show that the gravitational wave transients GW150914, GW151226, GW170104, GW170814 and GW170608 can be effectively recovered with spinning, quasi-circular templates 
if the eccentricity of these events at a gravitational wave frequency of 10Hz satisfies \(e_0\leq \{0.175,\, 0.125,\,0.175,\,0.175,\, 0.125\}\), respectively.  We show that if these systems have eccentricities \(e_0~\sim 0.1\) at a gravitational wave frequency of 10Hz, they can be misclassified as quasi-circular binaries due to parameter space degeneracies between eccentricity and spin corrections. Using our catalog of eccentric numerical relativity simulations, we discuss the importance of including higher-order waveform multipoles in gravitational wave searches of eccentric binary black hole mergers. 
\end{abstract}

\pacs{Valid PACS appear here}% PACS, the Physics and Astronomy
\maketitle

%%%%%%%%%%%%%%%%%%%%%%%%%%%%%%%%%%%%%%%%%%%%%
%%%%%%%%%%%%%%%%%%%%%%%%%%%%%%%%%%%%%%%%%%%%%
\section{Introduction}
\label{intro}

The detection of gravitational waves (GWs) from binary black hole (BBH) mergers and the first binary neutron star (BNS) inspiral~\cite{DI:2016,secondBBH:2016,thirddetection,fourth:2017,GW170608,bnsdet:2017}, by the advanced Laser Interferometer Gravitational-wave Observatory (aLIGO)~\cite{DII:2016,LSC:2015} and the European advanced Virgo (aVirgo) detector~\cite{Virgo:2015}, has ushered in a revolution in astrophysics.

These groundbreaking discoveries have provided conclusive evidence that stellar mass BBHs form and coalesce within the age of the Universe~\cite{D9:2016}, and that their astrophysical properties~\cite{bbhswithligo:2016,D4:2016} are consistent with Einstein's theory of general relativity~\cite{gr}. Furthermore, the detection of two colliding NSs with GWs and broadband electromagnetic observations has confirmed that BNS mergers are the central engines of short gamma ray bursts (sGRBs)~\cite{bnsdet:2017,mma:2017arXiv}, and the cosmic factories where about half of all elements heavier than iron are produced \cite{2017arXiv171005836T}.

Ongoing improvements in the sensitivity of
the aLIGO and aVirgo detectors at lower frequencies will enable detailed studies on the
astrophysical content of GW signals. Since eccentricity modifies the amplitude and frequency evolution of GWs at lower frequencies, before it is radiated away due to GW emission~\citep{Peters:1964,peters,Huerta:2017a,ian:2017}, GW observations within the next few years with the aLIGO and aVirgo detectors will provide unique opportunities to search for and detect eccentric binary mergers.

Eccentricity is one of the cleanest signatures for the existence of compact binary systems formed in dense stellar environments~\citep{Clausen:2013,Samsing:2014,sam:2017ApJ...840L..14S,sam:171107452S,carl:171204937R}. Therefore, identifying and
carefully measuring the imprints of eccentricity in GW signals will enable new and detailed studies of 
astrophysical processes taking place in core-collapsed globular clusters and galactic nuclei, which would otherwise
remain inaccessible~\citep{Samsing:2014,gon:2017,hpoang:2017,Van:2016,Wen:2003,Osburn:2016,Maccarone:2007,Strader:2012,cho:2013ApJ,Anto:2015arXiv,CR:2015PRL,Carl:2016arXiv}.  

Current flagship matched-filtering GW searches are highly optimized for the detection of quasi-circular, spin-aligned compact binary sources~\cite{2016CQGra..33u5004U}, and burst-like GW signals~\cite{D6:2016}. GW sources that do not fall into these categories may be missed by these algorithms, as shown in~\cite{Huerta:2017a,Tiwari:2016,Huerta:2014,Huerta:2013a}. Given the proven detection capabilities of the aLIGO and aVirgo detectors, it is timely and relevant to develop tools to confirm or rule out the existence of these GW sources~\cite{Tiwari:2016}. This article focusses on the development of an inspiral-merger-ringdown (IMR) waveform model that is adequate for the detection and characterization of compact binary populations that form in dense stellar environments, and which are expected to enter the aLIGO frequency band with moderate eccentricities~\cite{Anto:2015arXiv,Samsing:2014,Anton:2014}.

The model we introduce in this article, the \texttt{ENIGMA} (\textbf{E}ccentric, \textbf{N}on-spinning, \textbf{I}nspiral-\textbf{G}aussian-process \textbf{M}erger \textbf{A}pproximant) waveform model, builds on recent work to accurately describe quasi-circular and moderately eccentric IMR BBH mergers~\cite{Huerta:2017a}. \texttt{ENIGMA} does not require the calibration of free parameters for its construction; instead, it combines in a novel way analytical and numerical relativity (NR) using machine learning algorithms~\cite{2014PhRvL.113y1101M,gpr:2016PhRvD}, which are a special class of algorithms that can \textit{learn} from examples to solve new problems without being explicitly re-programmed. In different words, the same algorithm can be used across science domains by just changing the dataset used to train it~\cite{DL-Book}. Specifically, we use Gaussian Process Regression (GPR)~\cite{gpebook} to interpolate numerical relativity simulations of the merger and ringdown across the waveform parameter space. GPR is well suited to this due to its flexibility, the fact that it makes minimal assumptions about the underlying data, and the ease with which it can be extended to higher dimensional interpolation problems. This last point will be particularly important when we come to extend the model to include BH spins in future work.

The main motivation to develop \texttt{ENIGMA} is to systematically quantify the importance of orbital eccentricity in the detection of GW sources with aLIGO. As we show in this article, \texttt{ENIGMA} can accurately reproduce the dynamics of quasi-circular binaries and the true features of eccentric NR simulations. These features are of paramount importance to clearly associate deviations from quasi-circularity to the physics of eccentric compact binary coalescence, and not to intrinsic waveform model inaccuracies. We use \texttt{ENIGMA} to estimate the minimum value of eccentricity which may be discernable with aLIGO observations of eccentric BBH mergers. 

In this article we use units \(G=c=1\). The binary components are labelled as \(m_1\) and \(m_2\), where \(m_1\geq m_2\). We use the following combinations of \(m_{\{1,\,2\}}\): total mass \(M=m_1+m_2\), reduced mass \(\mu = m_1\,m_2/M\),  mass-ratio \(q=m_1/m_2\), and symmetric mass-ratio \(\eta = \mu/M\). This article is organized as follows: Section~\ref{recent} contextualizes this work in light of recent efforts to model eccentric compact binary systems. Section~\ref{summary} provides a brief description of our waveform model.  In Sections~\ref{waveforms},~\ref{ecc_waveforms} we validate \texttt{ENIGMA} with a state-of-the-art, IMR quasi-circular waveform model, and a set of eccentric NR simulations, respectively. We discuss the detectability of moderately eccentric BBH mergers with aLIGO in Section~\ref{see_me}. In Section~\ref{ho_modes} we discuss the importance of including higher-order waveform multipoles for the detection of eccentric BBH mergers. We summarize our findings and outline future directions of work in Section~\ref{end}.

%%%%%%%%%%%%%%%%%%%%%%%%%%%%%%%%%%%%%%%%%%%%%
%%%%%%%%%%%%%%%%%%%%%%%%%%%%%%%%%%%%%%%%%%%%%
\section{Previous work}
\label{recent}

In this section we briefly summarize recent developments in the literature in connection to IMR waveform models that describe moderately eccentric compact binary mergers.

An IMR model describing highly eccentric compact binary mergers was introduced in~\cite{East:2013}. This model used a geodesic based description for the inspiral dynamics, and a quasi-circular merger waveform using the  phenomenological approach described in~\cite{Kelly:2011PRD}. The model was used to study the detectability of burst-like signals in LIGO data.

In Ref.~\cite{cao:2017}, the authors introduce an effective-one-body (EOB) model that combines the quasi-circular dynamics of SEOBNRv1~\cite{Tara:2012}, and eccentric post-Newtonian (PN) corrections up to 2PN order. They compare the model to three numerical relativity simulations: two of mass-ratio \(q=1\) and eccentricities \(e_0 = \{0.02, \,0.19\}\) at orbital frequencies \(\{0.0105, \,0.0147\}\), respectively; and one with \(q=5\) and \(e_0\lesssim 0.02\) at an orbital frequency \(0.0105\). 

Ref.~\cite{Hinderer:2017} presents a formalism to consistently incorporate eccentricity corrections to the EOB formalism. The radiative dynamics is restricted to 1.5PN order. No comparison to eccentric NR simulations is done. 

In Ref.~\cite{ian:2017}, the authors present an eccentric model that includes PN corrections up to 2PN in the fluxes of energy and angular momentum. They show that their model can reproduce eccentric numerical relativity simulations with mass-ratios \(q\leq3\), and eccentricities \(e_{\rm ref}\leq0.05\), measured seven cycles before merger.

In this work, we show that \texttt{ENIGMA} reproduces the most recent version of EOB models, SEOBNRv4~\cite{Bohe:2016gbl}, in the quasi-circular limit. Our model is not based on phenomenological approximations for the description of the inspiral and merger evolution. Rather, we develop a hybrid inspiral model that encodes eccentric PN corrections up to 3PN order that includes tail, tails-of-tails, and corrections due to non-linear memory that enter at 2.5PN and 3PN order, as described in~\cite{Huerta:2017a}. Furthermore, we improve the inspiral evolution by including up to 6PN quasi-circular corrections using self-force and black hole perturbation theory (BHPT) results. We model the merger phase using a machine learning algorithm that is trained with a dataset of NR simulations. This approach ensures that our merger model has the same fidelity of NR simulations. We validate our model using a set of 12 \texttt{Einstein Toolkit}~\cite{naka:1987,shiba:1995,baum:1998,baker:2006,camp:2006,Lama:2011,wardell_barry_2016_155394,ETL:2012CQGra,Ansorg:2004ds,Diener:2005tn,Schnetter:2003rb,Thornburg:2003sf} eccentric NR simulations, with mass-ratios up to \(q \leq 5.5\) and eccentricities \(e_0 \lesssim 0.2\) ten orbits before merger.

Inspiral only models have steadily increased their accuracy~\cite{Levin:2011C,Yunes:2009,Huerta:2014,Huerta:2013a,Hinder:2010,lou:2016arXiv,lou:2014PhRvD,Osburn:2016,lou:2017CQG}. However, as extensively discussed in~\cite{Huerta:2014,Huerta:2017a} and in this article, PN-based models, even including up to 3PN eccentric corrections, are not accurate enough to describe the dynamical evolution of eccentric compact binaries through merger, since the PN prescription breaks down at that point.

A key result in this article is the validation of \texttt{ENIGMA} with eccentric NR simulations. Recent NR studies on the physics of eccentric compact binary mergers include~\cite{ihh:2008PhRvD,Hinder:2010,east:2012a,east:2012,Gold:2012PG,Gold:2013,East:2015PRDa,East:2016PhRvD,Radice:2016MNRAS,2016arXiv161107531T,lewis:2017CQG,Huerta:ncsacatalog}. In this article we show that \texttt{ENIGMA} reproduces the true features of eccentric NR simulations throughout late inspiral, merger and ringdown, without requiring the use of eccentric NR simulations to calibrate it. In different words, a consistent combination of analytical relativity formalisms, boosted with a machine-learning based merger waveform provides a powerful framework to describe both quasi-circular and eccentric BBH mergers.

%%%%%%%%%%%%%%%%%%%%%%%%%%%%%%%%%%%%%%%%%%%%%
%%%%%%%%%%%%%%%%%%%%%%%%%%%%%%%%%%%%%%%%%%%%%
\section{Waveform model construction}
\label{summary}

Our eccentric waveform model has two main components. The first component is an inspiral evolution scheme that combines results from PN theory~\cite{Arun:2009PRD}, the self-force formalism~\cite{2017PhRvD..96d4005C,barus,BiniDa:2014PRD,Huerta:2011a,Huerta:2011b,Huerta:2014a,smallbody,2017arXiv171201098H} and BHPT~\cite{Fujita:2012,Huerta:2012,Huerta:2010,2015CQGra..32w2002C,Huerta:2009}. The second component is a merger waveform, which is constructed by interpolating a set of NR-based surrogate waveforms~\cite{blackman:2015} using GPR~\cite{2014PhRvL.113y1101M,gpr:2016PhRvD,gpebook,doctor:70605408D}. The training dataset of NR-based surrogate waveforms describe BBHs with mass-ratios \(1\leq q \leq 10\).

In the following sections we present a succinct description of the improved inspiral evolution scheme, and a detailed description of the stand-alone GPR-based waveform model. Thereafter, we describe how to combine the inspiral and merger models to render a unified description of the dynamical evolution of moderately eccentric compact binary systems.

\texttt{ENIGMA} is tailored to carry out searches of compact binary systems that enter the aLIGO frequency band with moderate values of eccentricity. As discussed in~\cite{Huerta:2017a}, this approximation covers an astrophysically motivated population of compact binary sources that are expected to enter the aLIGO frequency band with eccentricities \(e_0\leq 0.3\) at 10Hz, and circularize just prior to merger~\cite{Anto:2015arXiv,Samsing:2014,Anton:2014}. Furthermore, using a set of eccentric NR simulations, in Section~\ref{ecc_waveforms} we show that \texttt{ENIGMA} accurately describes BBH mergers that retain significant residual eccentricity prior to merger. This feature may prove useful anticipating astrophysically unconstrained formation mechanisms for eccentric BBH mergers.

\subsection{Inspiral evolution}
\label{inspiral}

We model the inspiral evolution within the adiabatic approximation, i.e., we assume that the radiation time scale is much longer than the orbital time scale, and therefore we use an averaged description of the radiation reaction over an orbital period~\cite{Blanchet:2006}. Furthermore, in order to combine a variety of recent results from analytical relativity in a consistent way, we express the equations of motion in a gauge-invariant manner using as an expansion parameter the gauge-invariant quantity \(x=\left(M\omega\right)^{2/3}\), where \(\omega\) represents the \textit{mean} orbital frequency.

A first attempt to accurately describe the inspiral evolution of eccentric compact binaries with mass-ratios \(1\leq q \leq 10\) was introduced in Ref.~\cite{Huerta:2017a}. Below we succinctly describe the main ingredients of that model. Thereafter, we describe the new physics that we incorporate into \texttt{ENIGMA}.

The description of compact binary dynamics involves the relative orbital separation of the binary \(r\), which is related to the eccentric anomaly \(u\) by 

\beq
\frac{r}{M} =\frac{1-e\cos u}{x}+ \sum^{i=3}_{i=1}a_{i\rm{PN}}x^{i-1}\,.
\label{r_eq}
\eeq

\noindent On the other hand, the mean anomaly \(\ell\), which is related to the mean motion \(n\) through the relation \(M\dot{\ell}= M n\), is customarily described in terms of the eccentric anomaly \(u\) as follows~\cite{Blanchet:2006,Arun:2009PRD}:
\beq
\ell = u-e \sin u + \sum^{i=3}_{i=2}b_{i\,\rm{PN}}x^{i}\,.
\label{ele}
\eeq 

The conservative dynamics, which are obtained from a 3PN order Hamiltonian for eccentric compact binary systems~\cite{Blanchet:2006,Arun:2009PRD}, determines the time evolution of the instantaneous angular velocity \(\dot{\phi}\) and the mean anomaly \(\ell\):

\bea
\label{conservative}
M \dot{\phi} &=&  x^{3/2} \sum^{i=3}_{i=0} c_{i\,\rm{PN}}x^i + {\cal{O}}(x^{11/2}),\\
\label{l_dot}
M \dot{\ell} &=& x^{3/2}\left(1 + \sum^{i=3}_{i=1} d_{i\,\rm{PN}} x^i\right)+{\cal{O}}(x^{11/2}) \,.
\eea

\noindent  The instantaneous angular velocity \(\dot{\phi}\) is related to the mean orbital frequency \(\omega\) through \(\omega = \langle \dot{\phi}\rangle = K n\), where \(\langle \cdot \rangle\) indicates average over an orbital period. \(K,\, k\) represent the periastron precession and the relativistic precession, respectively, and are related via \(K=1+k\). The radiative evolution of the binary, which is driven by the energy and angular momentum carried out by GWs from the binary system, can be specified through the time evolution of eccentricity \(e\) and the gauge-invariant expansion parameter \(x\)

\bea
\label{radiative_x}
M\dot{x} &=&  x^5 \sum^{i=3}_{i=0} y_{i\,\rm{PN}}x^i  + \dot{x}_{\rm HT} \,,\\
\label{radiative_e}
M\dot{e} &=&  x^4 \sum^{i=3}_{i=0} z_{i\,\rm{PN}}x^i + \dot{e}_{\rm HT}  \,.
\eea

\noindent The hereditary terms (HT) in Eqs.~\eqref{radiative_x}-\eqref{radiative_e} include tail, tails-of-tails and non-linear memory corrections at 1.5PN, 2.5PN and 3PN order. The 2.5PN and 3PN pieces of Eq.~\eqref{radiative_x} were derived in Ref.~\cite{Huerta:2017a}. As discussed above, we used the adiabatic approximation to model eccentric BBH mergers, therefore the functions \((y_{i\,\rm{PN}},\, z_{i\,\rm{PN}})\) only depend on \(e\), and Eqs.~\eqref{radiative_x}-\eqref{radiative_e} describe a closed system that can be solved given initial conditions for \(x(t=0)\) and \(e(t=0)\). Once this is done, \((x(t),\,e(t))\) are used to numerically integrate Eq.~\eqref{l_dot}. Thereafter,  \((x(t),\,e(t),\, \ell(t))\) are substituted in Eq.~\eqref{ele} to determine \(u\) by root-finding. Finally, all these pieces are used to determine Eqs.~\eqref{r_eq} and~\eqref{conservative}. 

Having this workflow in mind, we realize that to improve phase accuracy of any PN-based model, it is necessary to include higher-order PN corrections to Eqs.~\eqref{radiative_x} and~\eqref{radiative_e}. This becomes apparent when we write the above expressions in the quasi-circular limit. In the expressions below, we have augmented the conservative and radiative dynamics with self-force and BHPT results, i.e.,

\bea
\label{mphidot}
M\dot{\phi}\big|_{e\rightarrow 0} &=& x^{3/2}\,,\\
M\dot{x}\big|_{e\rightarrow 0} &=&   \frac{64}{5} \eta\,x^5\, \Bigg\{
1 +
\left( -\frac{743}{336} - \frac{11}{4} \eta \right) x
+ 4 \pi x^{3/2}
\nonumber\\&+& \left( \frac{34\,103}{18\,144} +\frac{13\,661}{2016} \eta + \frac{59}{18} \eta^2  \right) x^2
\nonumber \\
&+& \left( -\frac{4159 \pi}{672} -\frac{189 \pi}{8} \eta  \right) x^{5/2}
\nonumber \\
&+& \Bigg[ \frac{16\,447\,322\,263}{139\,708\,800} - \frac{1712 \gamma}{105} + \frac{16 \pi^2}{3} \nonumber\\&-& \frac{856}{105} \log (16 x) 
+ \left(-\frac{56\,198\,689}{217\,728} + \frac{451 \pi ^2}{48} \right) \eta \nonumber\\&+& \frac{541}{896} \eta^2 - \frac{5605}{2592} \eta^3  \Bigg] x^3 +\bigg[-\frac{4415}{4032}\nonumber\\&+&\frac{358675}{6048}\eta +\frac{91945}{1512}\eta^2\bigg]x^{7/2}    + \hat{a}_4 x^4 + \hat{a}_{9/2} x^{9/2} \nonumber\\&+& \hat{a}_5 x^5 + \hat{a}_{11/2} x^{11/2} + \hat{a}_6 x^6 \Bigg\}\,,
\eea

\noindent where \(\gamma\) is Euler's constant, and the coefficients \(\hat{a}_4,\,\hat{a}_{9/2},\, \hat{a}_5,\, \hat{a}_{11/2},\, \hat{a}_6\) are presented in Appendix C of Ref.~\cite{Huerta:2017a}. As is demonstrated in Section~\ref{maps}, when this hybrid inspiral formalism is combined with the GPR merger-ringdown waveform presented in Section~\ref{gpe_mer}, the GW emission from quasi-circular binaries with mass-ratios in the range \(1\leq q \leq 10\) is described with excellent accuracy. As discussed in contemporary literature, this is a basic requirement for any waveform model that aims to establish a clear cut connection between eccentricity and deviations from quasi-circular motion~\cite{Huerta:2017a,Huerta:2014,Huerta:2013a}. In other words, accurate eccentric BBH modeling can only be done with models that simultaneously incorporate an accurate description of quasi-circular motion. 

In order to obtain GW strain from inspiral trajectory evolution, we use standard leading-order, eccentric PN strain expressions for both \((h^{\rm I}_{+}, \,h^{\rm I}_{\times})\) polarizations as a baseline, ---where \(h^{\rm I}\) represent the \textbf{I}nspiral waveform--- and then include quasi-circular PN corrections up to 3PN order \(( h^{\rm I, QC}_{+},\,  h^{\rm I, QC}_{\times})\), which are given by Eqs. (320)-(323g) in Ref.~\cite{Blanchet:2006}:

\bea
\label{strain_ins}
h^{\rm I}(t) = h^{\rm I}_{+}(t) -i  h^{\rm I}_{\times}(t)\,,
\eea
\noindent with
\bea
\label{h_plus}
h^{\rm I}_{+} &=&-\frac{M\eta}{R}\Bigg\{\left(\cos^2\iota+1\right)\Bigg[\left(-\dot{r}^2+r^2\dot{\phi}^2+\frac{M}{r}\right) \cos2\Phi \nonumber\\&+& 2r\dot{r}\dot{\phi}\sin2\Phi\Bigg] + \left(-\dot{r}^2-r^2\dot{\phi}^2+\frac{M}{r}\right)\sin^2\iota\Bigg\} \nonumber\\&+&
 h^{\rm I, QC}_{+}\left(\phi, \dot{\phi},\, M,\eta,\iota, R\right)\,,
 \eea
 \bea
\label{h_cross}
h^{\rm I}_{\times} &=&-\frac{2M\eta}{R}\cos\iota\Bigg\{\left(-\dot{r}^2+r^2\dot{\phi}^2+\frac{M}{r}\right)\sin2\Phi \nonumber\\&-& 2r\dot{r}\dot{\phi} \cos2\Phi\Bigg\}+  h^{\rm I, QC}_{\times}\left(\phi, \dot{\phi},\,M, \eta,\iota, R\right)\,,
\eea

\noindent  where \(\Phi=\phi-\chi\), and \((\chi,\,\iota)\) represent the polar angles of the observer, and \(R\) is the distance to the binary. This completes the description of the inspiral portion of our \texttt{ENIGMA} model. As mentioned above, \texttt{ENIGMA} has been developed to target eccentric compact binary systems that circularize prior to merger. This is a reasonable approach, since recent studies suggest that the eccentricity distribution of aLIGO sources may be bimodal, and the population with moderate values of eccentricity may enter the aLIGO frequency band with eccentricities \(e_0\leq0.1\) at 10Hz. We show that our model is equipped to target this population, as well as compact binaries that may enter the aLIGO frequency band with eccentricities up to \(e_0\leq 0.3\) at 10Hz. Under this assumption, the following section describes the construction of a quasi-circular merger waveform using GPR. The main motivation for the use of GPR is that they may be easily extended to higher dimensional interpolation problems which will be faced in future work simulating eccentric mergers of spinning black holes.

%%%%%%%%%%%%%%%%%%%%%%%%%%%%%%%%%%%%%%%%%%%%%%%%%%%%%%
%%%%%%%%%%%%%%%%%%%%%%%%%%%%%%%%%%%%%%%%%%%%%%%%%%%%%%
\subsection{Merger and Ringdown}
\label{gpe_mer}

The main limitation of the predecessor to the \texttt{ENIGMA} model \cite{Huerta:2017a} was the treatment of the merger-ringdown signal. In this section, a new machine learning based approach to modeling the merger--ringdown signal, using NR simulations is described. A desirable feature of this approach is the ability to identify regions of parameter space where the model performs poorly, and rapidly incorporate additional NR simulations into an updated model. 
This iterative procedure of improving the model is a feature of this approach that will be illustrated throughout this section.

In this paper only non-spinning BBHs are considered (although the approach is designed to be flexible enough to allow for the extension to spinning and precessing BBH systems in future); therefore the intrinsic parameter space of the IMR model is 3--dimensional (mass ratio, $q$, eccentricity, $e$, and mean anomaly, $\ell$; the total mass, $M$, simply sets the overall dimensional scale of the problem). It is known that the mean anomaly \(\ell\) affects the amplitude and phase of the waveform strain~\cite{Brown:2010,Huerta:2013a}. For a given value of anomaly, these effects are handled by the method we use to smoothly connect the inspiral and merger waveforms, as described in the following section.

As is well known (see, e.g. \cite{Peters:1964,peters,Huerta:2017a,Clausen:2013,ian:2017}) BBHs with moderate eccentricities early in the inspiral are efficiently circularized by GW emission and are almost circular at merger. This reduces the effective dimensionality of the \emph{merger--ringdown} model to just 1 dimension (mass ratio).

An initial \emph{training set}, $\mathcal{D}^{1}$, of 19 quasi--circular NR simulations from the public SXS catalog \cite{Mroue:2013} at mass ratios $ { q\!\in\!\mathcal{Q}^{1}\!=\!\left\{ 1.0,1.5,2.0,\ldots,10.0 \right\} }$ was used to build a merger-ringdown model. Only the ${l\!=\!m\!=\!2}$ modes of the merger waveforms were used. The effect of neglecting higher-order waveform modes is quantified in Section~\ref{ho_modes}. The time series were decomposed into the amplitude, ${g_{1}(t;q)^{2} \! \equiv \! h_{+}^{\textrm{M}}(t;q)^{2}\!+\!h_{\times}^{\textrm{M}}(t;q)^{2}}$, and (unwrapped) phase, ${g_{2}(t;q) \! \equiv \! \textrm{atan} ( h_{\times}^{\textrm{M}}(t;q)/ h_{+}^{\textrm{M}}(t;q) )}$. The waveforms were rescaled (in time) such that the total mass was unity, the peak luminosity occurs at $t\!=\!0$, the amplitude was rescaled to satisfy $g_{1}(t\!=\!0,q)\!=\!1$ and the phase was shifted to satisfy $g_{2}(t\!=\!0,q)\!=\!0$. The time series were sampled at $n\!=\!2800$ points in the interval $\textrm{--}2500\!\leq\!t/M\!\leq+100$ (with a higher sampling rate around merger). The training sets for amplitude and phase are
\beq \label{eq:Dset}
\mathcal{D}_{\alpha}^{1} = \left\{ \, (q,g_{\alpha}(t;q)) \, | \, q\in \mathcal{Q}^{1} \right\}  \textrm{, where }\alpha=1,2\,.
\eeq

GPR is an interpolation (or extrapolation) technique which makes minimal assumptions about the underlying function. It is used here to interpolate the data in Eq.~\ref{eq:Dset} to obtain the waveform at any mass ratio, $q$. A \emph{Gaussian process} (GP) of a single variable $x$ is completely described by a \emph{covariance function}, $k(x,x')$ (and a mean function $\mu(x)$ which is here assumed to be zero for simplicity). 
For the first training set, $\mathcal{D}^{1}$, the parameterization $x\!=\!q$ was used.
The covariance function, and any free parameters therein, are free to be specified; however, they can also be \emph{learnt} from the training set by maximizing the probability of the training set being realized by the GP (maximizing the \emph{GP evidence}). This learning process can be computationally expensive, especially for large training sets or when comparing covariance functions with many free parameters; the techniques described in \cite{Moore160125} were used to accelerate this learning phase. 
The covariance functions considered here were the \emph{squared-exponential} and \emph{Wendland polynomial} functions used previously for waveform modeling in \cite{gpr:2016PhRvD}; these covariance functions are all \emph{stationary}, i.e.\ $k(x,x')\!=\!k(x-x')$. 
For a discussion of GPR see \cite{Rasmussen:2005:GPM:1162254}, or \cite{2014PhRvL.113y1101M,gpr:2016PhRvD} in the context of GW signal modeling.

As zero--mean GPs were used for the interpolation, the phase interpolation can be improved if a reference phase function is first subtracted from the training set phases to bring the values closer to zero; ${g_{2}(t;q)\!\rightarrow\! g_{2}(t;q)\!-\!\varphi(t)}$. After interpolation the phase can be recovered straightforwardly by adding $\varphi(t)$ to the result. The reference phase function was chosen to be the phase of the equal mass NR simulation; $\varphi(t) \!\equiv\! g_{2}(t;q\!=\!1.0)$.

\begin{figure}[H]
\begin{center}
    \includegraphics[trim=0cm 0cm 0cm 0cm, width=0.485\textwidth]{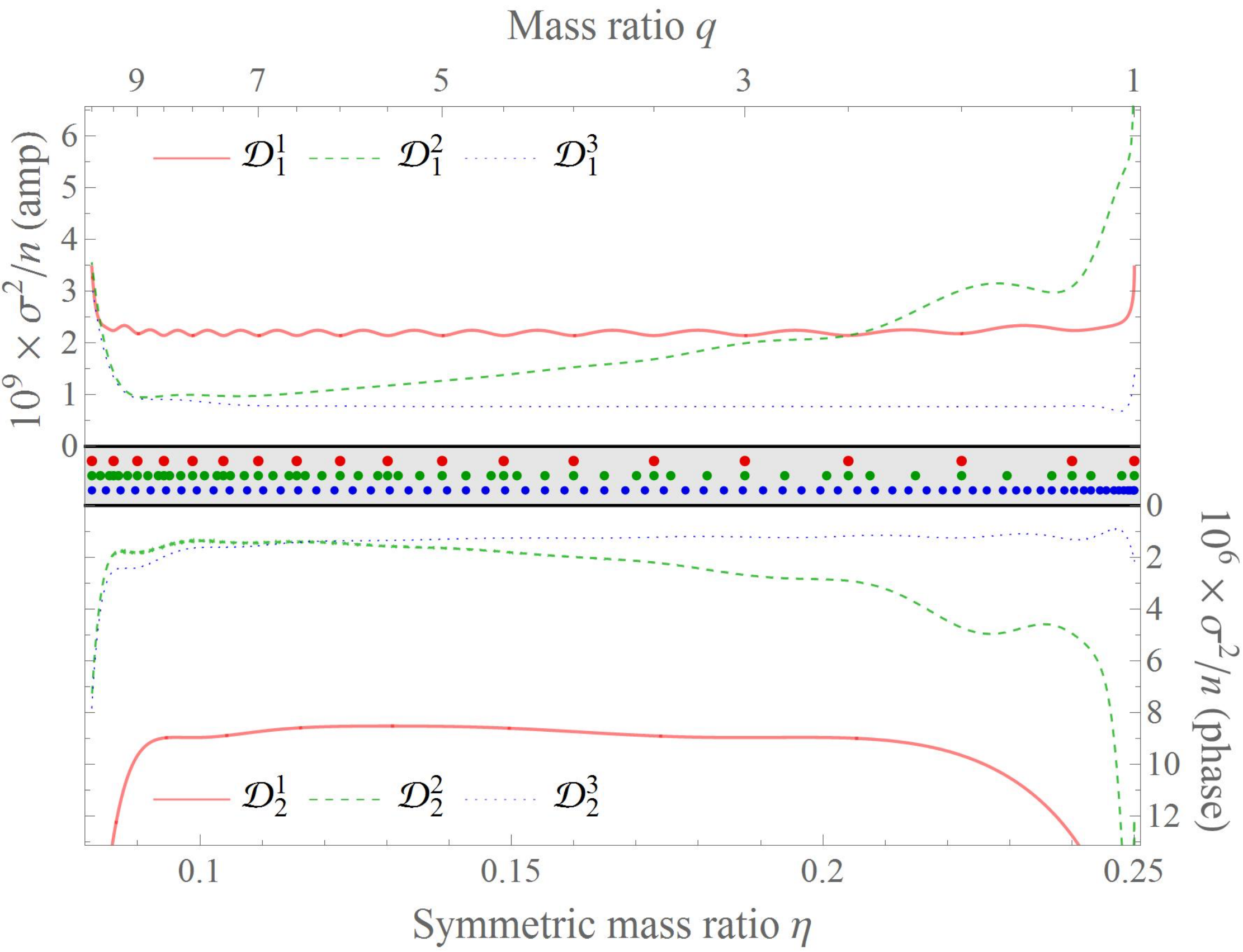}
  \caption{The GPR estimates for the interpolation error, $\sigma$, on the amplitude ($g_{1}(t;q)$, top panel) and phase ($g_{2}(t;q)$, inverted in the bottom panel) for the three training sets: $\mathcal{D}^{1}_{\alpha}$ (red), $\mathcal{D}^{2}_{\alpha}$ (green), and $\mathcal{D}^{3}_{\alpha}$ (blue). The constant $n\!=\!2800$ is the number of time samples in the waveform. The (symmetric) mass ratio values for the points in $\mathcal{Q}^{1}$, $\mathcal{Q}^{1}$ and $\mathcal{Q}^{3}$ are indicated respectively by red, green, and blue dots along the central horizontal panel. Each iteration of the training set reduces the error in the amplitude and phase interpolations. \label{fig:GPEerror}}
\end{center}
\end{figure}

It was found that the GPR interpolation using the 19 waveforms in the set $\mathcal{D}^{1}_{\alpha}$ was not sufficiently accurate; typical errors on the interpolant were $\sigma\!\sim\!0.16\,\textrm{radians}$ on the phase and $\sigma\!\sim 3\times\!10^{-3}$ on the amplitude (see Fig.~\ref{fig:GPEerror}, with $n\!=\!2800$). A second training set, $\mathcal{D}^{2}$ was created by adding more waveforms approximately uniformly in mass ratio; the new set, $\mathcal{D}^{2}_{\alpha}$, consisted of simulations at mass ratios ${ q\!\in\!\mathcal{Q}^{2}\!=\!\left\{ 1.0,1.2,1.4,\ldots,10.0 \right\}\!\cup\!\mathcal{Q}^{1}}$ (a total of 55 simulations). As sufficiently accurate, and long duration NR simulations at these mass ratios were not readily available, NR surrogate waveforms \cite{blackman:2015} were used instead. Work is ongoing to replace the surrogate waveforms used in this study with full NR simulations. On this training set GPR was tested using different parameterizations\footnote{The parameterizations tried included the mass ratio, $f(q)\!=\!q$, a simple compactification on to the interval $\textrm{--}1\!\leq\!x\!\leq\!+1$, $f(q)\!=\!\textrm{atan}\,q$, the symmetric mass ratio $f(q)\!=\!\eta\!\equiv\!q/(1\!+\!q)^{2}$, which compactifies on to the interval $0\!\leq\!\eta\!\leq\!0.25$), and $f(q)\!=\!\ln(q)$.} of the training set, $x=f(q)$; it was empirically found that $x=\ln(q)$ yielded the largest GP evidence value and the best interpolant. From Fig.~\ref{fig:GPEerror} it can be seen that $\mathcal{D}^{2}_{\alpha}$ indeed outperforms $\mathcal{D}^{1}_{\alpha}$ across most of the parameter space.

Finally, a third training set was constructed using 70 waveforms placed uniformly in the logarithm of the mass ratio; ${ q\!\in\!\mathcal{Q}^{2}\!=\!\left\{ \exp[(i-1)\ln(10)/69] \,|\, i\!=\!1,2,\ldots,70 \right\} }$. The use of a stationary covariance function $k(x-x')$ (with $x\!=\!\ln(q)$) leads to large errors near the training set boundary $x\!=\!0$ (or $q\!=\!1$). Of course, this boundary is artificial because the parameter space admits the following identifications $x\!\rightarrow\!\textrm{--}\,x$ (or $q\!\rightarrow\!1/q$). It was found that using the following \emph{non-stationary} covariance function incorporating this symmetry into the GP leads to improved performance for nearly equal mass systems;
\beq \label{eq:non-stationary}
k'(x,x')=k(x-x')+k(x+x')\,.
\eeq

The errors on the GPR interpolants for the amplitude and phase for the three training sets are shown in Fig.~\ref{fig:GPEerror}. As expected, the $\mathcal{D}^{3}$ interpolant is generally the most accurate, especially for nearly equal mass systems (the $\mathcal{D}^{3}$ interpolant is used for the remainder of the paper).
A thorough validation of the combined inspiral--merger--ringdown model is presented in Sec.~\ref{maps}. However, the finalized GPR interpolant for the merger-ringdown performs extremely well; it reproduces the NR surrogate waveforms in the mass ratio range ${1\!\leq\!q\!\leq\!10}$ with overlaps $>\!0.9998$ computed over the entire duration ${\textrm{--}2500\!\leq\! t/M\!\leq\!+100}$.

%%%%%%%%%%%%%%%%%%%%%%%%%%%%%%%%%%%%%%%%%%%%%%%%%%%%%%%%%%%%%%%
%%%%%%%%%%%%%%%%%%%%%%%%%%%%%%%%%%%%%%%%%%%%%%%%%%%%%%%%%%%%%%%
\subsection{Complete waveform model}
\label{maps}
%%%%%%%%%%%%%%%%%%%%%%%%%%%%%%%%%%%%%%%%%%%%%%%%%%%%%%%%%%%%%%%
%%%%%%%%%%%%%%%%%%%%%%%%%%%%%%%%%%%%%%%%%%%%%%%%%%%%%%%%%%%%%%%
In this Section we describe the method followed to smoothly attach the hybrid inspiral model of Section~\ref{inspiral} with the GPR-based merger waveform of Section~\ref{gpe_mer}.

The studies we have carried out to do this work indicate the regime of validity of our hybrid inspiral scheme, and furnish strong evidence that analytical and numerical relativity can be blended together to create a model that can accurately reproduce the true features of both quasi-circular and eccentric compact binaries. We now describe the construction of a map that determines the optimal frequency at which the inspiral evolution can be blended with the merger evolution for a given combination of masses \(m_{\{1,\,2\}}\). In order to do this, a number of signal processing tools are needed which will now be described.

Given two signals \(h\) and \(s\), and defining \(S_n(f)\) as aLIGO's design power spectral density (PSD)~\cite{ZDHP:2010}, and \(\tilde{h}(f)\) as the Fourier transform of \(h(t)\), the noise-weighted inner product between \(h\) and \(s\) is given by

\beq
\left( h | s\right) = 2 \int^{f_1}_{f_0} \frac{\tilde{h}^{*}(f)\tilde{s}(f) + \tilde{h}(f)\tilde{s}^{*}(f) }{S_n(f)}\mathrm{d}f\,,
\label{inn_pro}
\eeq

\noindent with \(f_0=15\,{\rm Hz}\) and \(f_1=4096\,{\rm Hz}\). The waveforms used in this study are generated with a sample rate of 8192Hz. Additionally, the normalized overlap is defined as 

\begin{align}
\label{over}
{\cal{O}}  (h,\,s)&= \underset{ t_c\, \phi_c}{\mathrm{max}}\left(\hat{h}|\hat{s}_{[t_c,\,  \phi_c]}\right)\quad{\rm with}\\
\label{n_overl}
\hat{h}&=h\,\left(h | h\right)^{-1/2}\,,
\end{align}

\noindent where \(\hat{s}_{[t_c,\,  \phi_c]}\)  indicate that the normalized waveform \(\hat{s}\)  has been time- and phase-shifted. Using these definitions, we blend our inspiral and merger models as follows:

\begin{itemize}
\item Combine the inspiral and merger codes into a single library that generates an inspiral waveform and smoothly attaches a GPR merger waveform on the fly
\item At the point of attachment, \(t_{\rm a}\), the inspiral, \(h^{\rm I}(t)\), and merger, \(h^{\rm M}(t)\), waveforms satisfy continuity and differentiability

\begin{eqnarray}
\label{cont}
h^{\rm I}(t_{\rm a}) &=& h^{\rm M}(t_{\rm a}) \,,\\
\label{differen}
\dot{h}^{\rm I}(t_{\rm a}) &=& \dot{h}^{\rm M}(t_{\rm a})\,, \quad \dot{h}=\frac{\mathrm{d}h}{\mathrm{d}t}\,.
\end{eqnarray}

\item The mass parameter space we consider to construct this map is \(m_{\{1,\,2\}}\in[5 M_{\odot},\,50M_{\odot}]\), in steps of \(1 M_{\odot}\) in the \(m_{\{1,\,2\}}\) dimensions. 
\item For each of the points of the aforementioned parameter space, we generated quasi-circular \texttt{ENIGMA} waveforms considering a wide range of frequencies to connect the inspiral and merger waveforms, namely: \(M\omega\in[0.02, \,0.1]\) in steps of \(1\times10^{-4}\)
\item Thereafter, we computed overlaps between the set of waveforms described in the previous item and their SEOBNRv4 counterparts.
\item Finally, we picked the attachment frequency value \(M\omega^*\) that maximized the overlap for each point of the BBH parameter space under consideration.
\item Our \texttt{ENIGMA} code automatically translates \(M\omega^*\) into an optimal time of attachment \(t^{*}_{\rm a}\), i.e., it determines the time in the GPR NR-based merger waveform that corresponds to \(M\omega^*\) 
\end{itemize}

Finally, the complete IMR \texttt{ENIGMA} waveform can be written as
\begin{equation}
\label{total}
h(t) = h^{\rm\bf I}(t){\cal{H}}(t^{*}_{\rm a}-t) + \textrm{e}^{\textrm{i} \Delta \Phi}h^{\rm\bf M}(t+\Delta t){\cal{H}}(t-t^{*}_{\rm a})\,,
\end{equation}

\noindent where ${\cal{H}}(t)$ is the Heaviside step function, and \((h^{\rm I}(t),\, h^{\rm M}(t))\) represent the \textbf{I}nspiral and \textbf{M}erger waveforms. \((\Delta t, \Delta \Phi)\) are time and phase shifts that need to be incorporated in the GPR merger waveforms to enforce continuity and differentiability. In the next section we explore the accuracy and robustness of this scheme in the quasi-circular limit.

%%%%%%%%%%%%%%%%%%%%%%%%%%%%%%%%%%%%%%%%%%%%%
%%%%%%%%%%%%%%%%%%%%%%%%%%%%%%%%%%%%%%%%%%%%%
\section{Validation of \texttt{ENIGMA} in the quasi-circular limit}
\label{waveforms} 
 
The algorithm we described in the previous section to connect the inspiral and merger evolution utilizes a discrete grid of 1100 points, which uniformly covers the BBH parameter space \(m_{\{1,\,2\}}\in[5\Msun,\,50\Msun]\). To ensure that this method is robust,  we now compute the overlap between our \texttt{ENIGMA} model in the quasi-circular limit, and SEOBNRv4 waveforms using a grid that covers the same region of parameter space, but now using 2500 points. For each \(m_{\{1,\,2\}}\) combination, we test $800$ possible values of attachment frequency.

To determine at which frequency we connect the inspiral and merger waveforms, we consider the following: if we were to blend in the inspiral and merger waveforms using the lowest frequency of attachment in our dataset, BBH systems with component masses similar to the first and third GW transients detected by aLIGO would be described entirely by our quasi-circular GPR merger waveforms. In that case, since we would only be using a quasi-circular waveform, we would not be able to make any meaningful statements regarding the effects of eccentricity in the detection of BBH mergers with aLIGO. However, since we want to study eccentric binary mergers, we require a model that can tolerate small eccentricity values very late in the inspiral evolution. We can only do so if we attach the merger waveform as late as possible, i.e., using the largest possible frequency of attachment.

\begin{figure}[!ht]
    \includegraphics[width=0.485\textwidth]{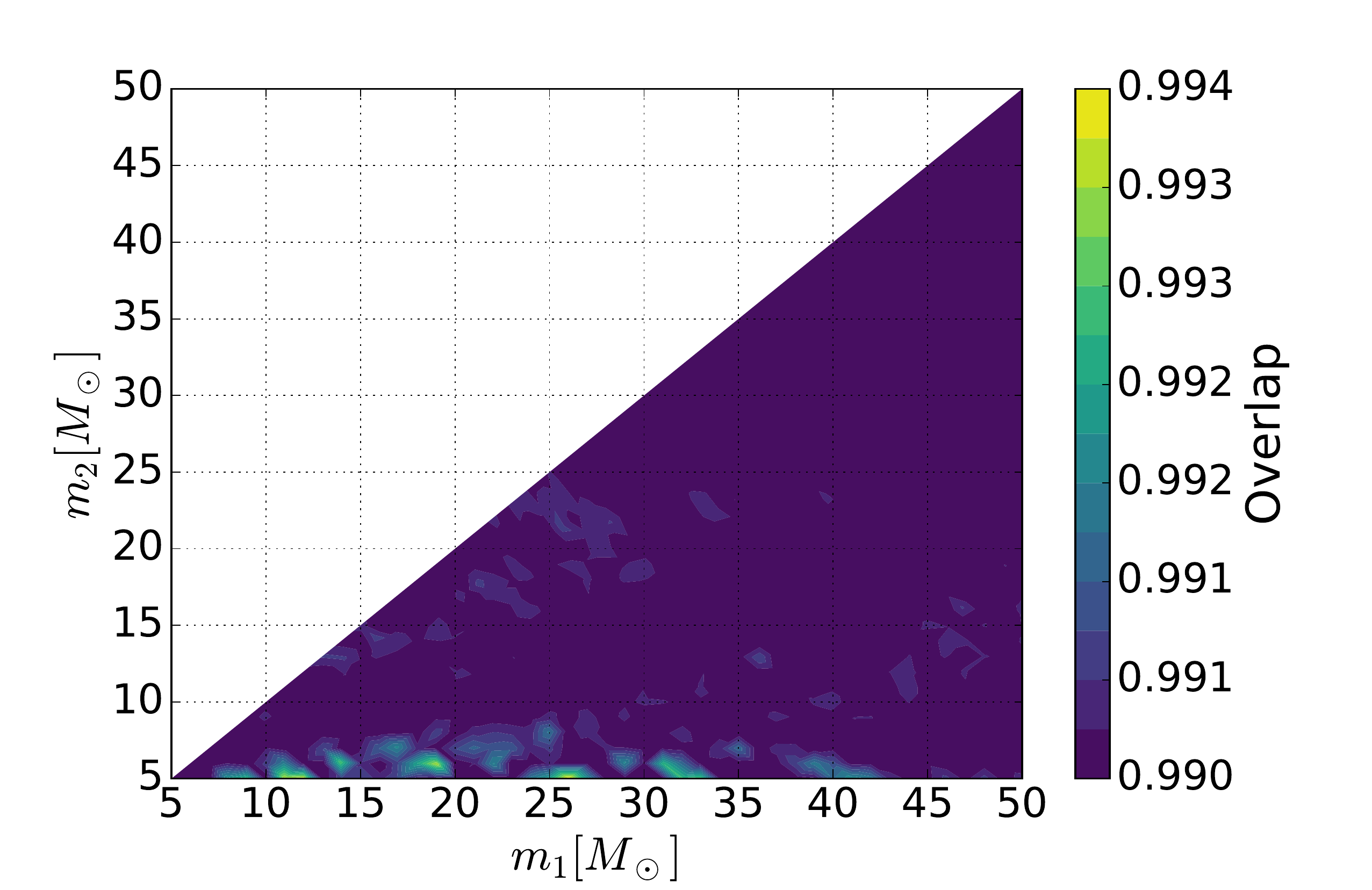}
  \caption{Overlap of quasi-circular \texttt{ENIGMA} waveforms with non-spinning, quasi-circular SEOBNRv4 waveforms. The overlaps are computed using aLIGO design sensitivity power spectral density, assuming an initial gravitational wave frequency \(f_{\rm GW}=15{\rm Hz}\).}
\label{overlap_e_0}
\end{figure}

Based on these considerations, we have constructed a map that smoothly connects the inspiral and merger evolution, using the largest frequency of attachment, and which guarantees that the overlap between quasi-circular \texttt{ENIGMA} waveforms and SEOBNRv4 waveforms is \({\cal{O}}\geq0.99\). Figure~\ref{overlap_e_0} presents these results. This is the level to which contemporary quasi-circular models agree in many regions of BBH parameter space~\cite{Bohe:2016gbl,Kumar:2016dhh}. In the following section, we will show that this map works very well when we consider binary systems with non-negligible eccentricity.

%%%%%%%%%%%%%%%%%%%%%%%%%%%%%%%%%%%%%%%%%%%%%
%%%%%%%%%%%%%%%%%%%%%%%%%%%%%%%%%%%%%%%%%%%%%
\section{Validation of \texttt{ENIGMA} with eccentric numerical relativity simulations}
\label{ecc_waveforms}

To show that \texttt{ENIGMA} reproduces the dynamics of eccentric BBHs throughout late inspiral, merger and ringdown, we use a catalog of eccentric NR simulations, generated  with the open source, community software the \texttt{Einstein Toolkit}, and post-processed with the open source software \texttt{POWER}~\cite{johnson:2017}. The \((e_0,\, \ell_0,\, x_0)\) parameters that describe these BBH NR simulations are provided in Table~\ref{results}. To determine these  parameters, we select the parameter combination that maximizes the overlap between \texttt{ENIGMA} and NR waveforms, without using any information based on the trajectory of the BHs.  Appendix~\ref{ap1} provides a brief summary of the convergence and phase error of these NR simulations, which were generated with three different grid resolutions to assess their convergence. A detailed description of this NR catalog is provided in an accompanying paper~\cite{Huerta:ncsacatalog}. To validate \texttt{ENIGMA}, we use the highest resolution run of each dataset.

\begin{table}
\caption{\label{results} Numerical relativity (NR) simulations, taken from the NCSA catalog of eccentric BBH mergers~\cite{Huerta:ncsacatalog}, used to validate \texttt{ENIGMA}. \(q\) is the mass-ratio of the BBH system. \((e_0,\, \ell_0,\, x_0)\) represent the measured values of eccentricity, mean anomaly, and dimensionless frequency parameters of the NR simulation, respectively. The overlap, \({\cal{O}}\), between \texttt{ENIGMA} waveforms and their numerical relativity counterparts is shown in the last column.}
		\footnotesize
		\begin{center}
                        \setlength{\tabcolsep}{10pt} % default is apparently 6pt
			\begin{tabular}{c c c c c c c}
				\hline 
				Simulation&$q$ & $e_0$ & $\ell_0$ & $x_0$ & ${\cal{O}}$ \\ 
				\hline
				E0001&1.0 & 0.060 & 3.50 & 0.077 & 0.998\\
				J0005&1.5 & 0.067 & 3.30 & 0.078 & 0.997\\
 				J0045&2.0 & 0.078 & 3.35 & 0.079 & 0.994\\
 				E0013&2.5 & 0.070 & 3.00 & 0.081 & 0.997\\
 				E0017&3.0 & 0.068 & 2.60 & 0.083 & 0.989\\
 				K0001&3.5 & 0.060 & 3.20 & 0.081 & 0.991\\
				J0061&4.0 & 0.065 & 2.90 & 0.086 & 0.992\\
				J0065&4.5 & 0.080 & 3.10 & 0.088 & 0.981\\
				M0004&1.0 & 0.190 & 3.20 & 0.071 & 0.993\\
				J0047&2.0 & 0.120 & 2.70 & 0.078 & 0.956\\
				K0024&4.0 & 0.200 & 2.90 & 0.084 & 0.971 \\
				L0020&5.5 & 0.210 & 3.10 &  0.087 & 0.951\\
				\hline 
			\end{tabular}
		\end{center}
		\label{sims}
	\end{table}
	\normalsize

 \begin{widetext}
 \begin{figure*}[!ht]
\centerline{
    \includegraphics[width=0.485\textwidth]{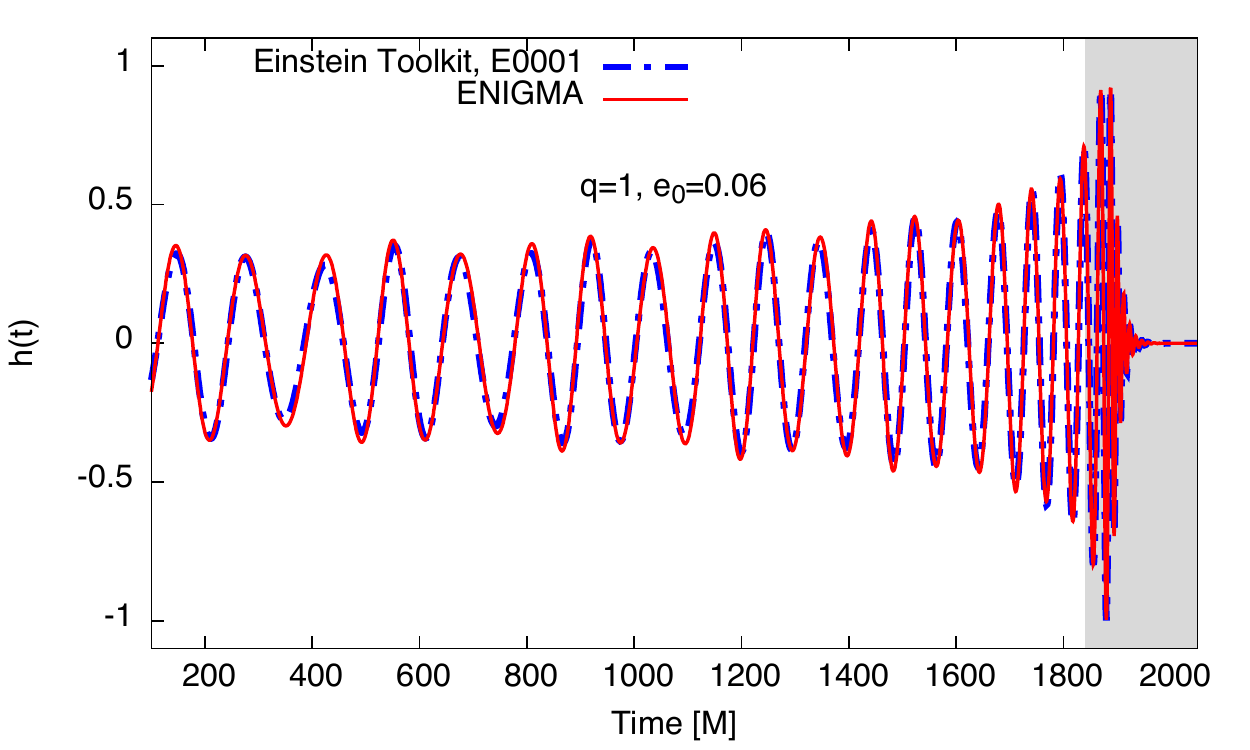}
    \includegraphics[width=0.485\textwidth]{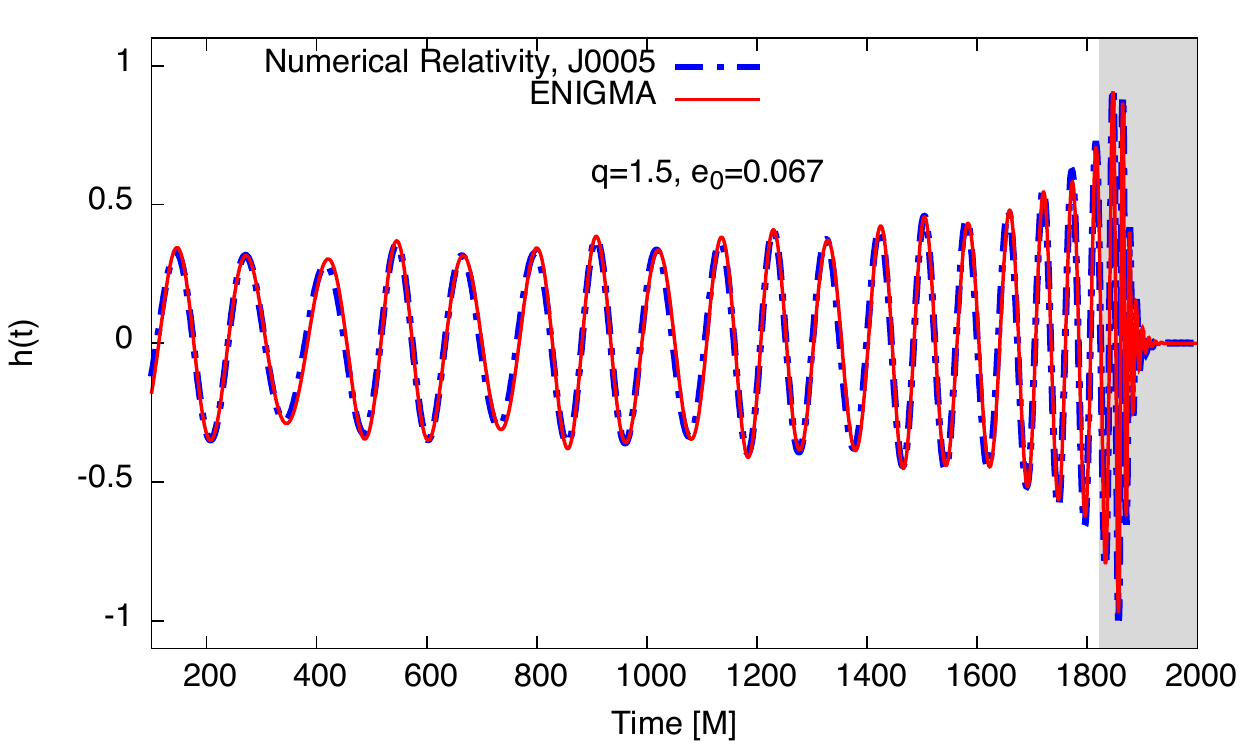}
     }
\centerline{
    \includegraphics[width=0.485\textwidth]{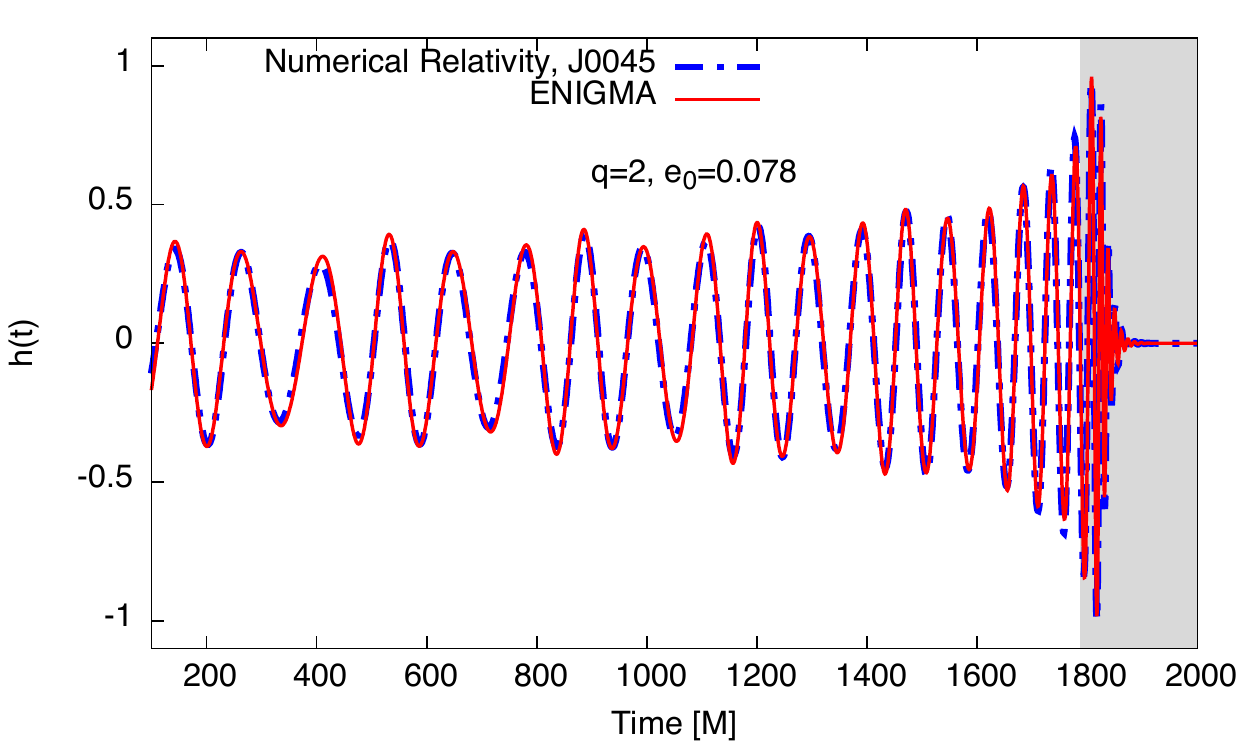}
    \includegraphics[width=0.485\textwidth]{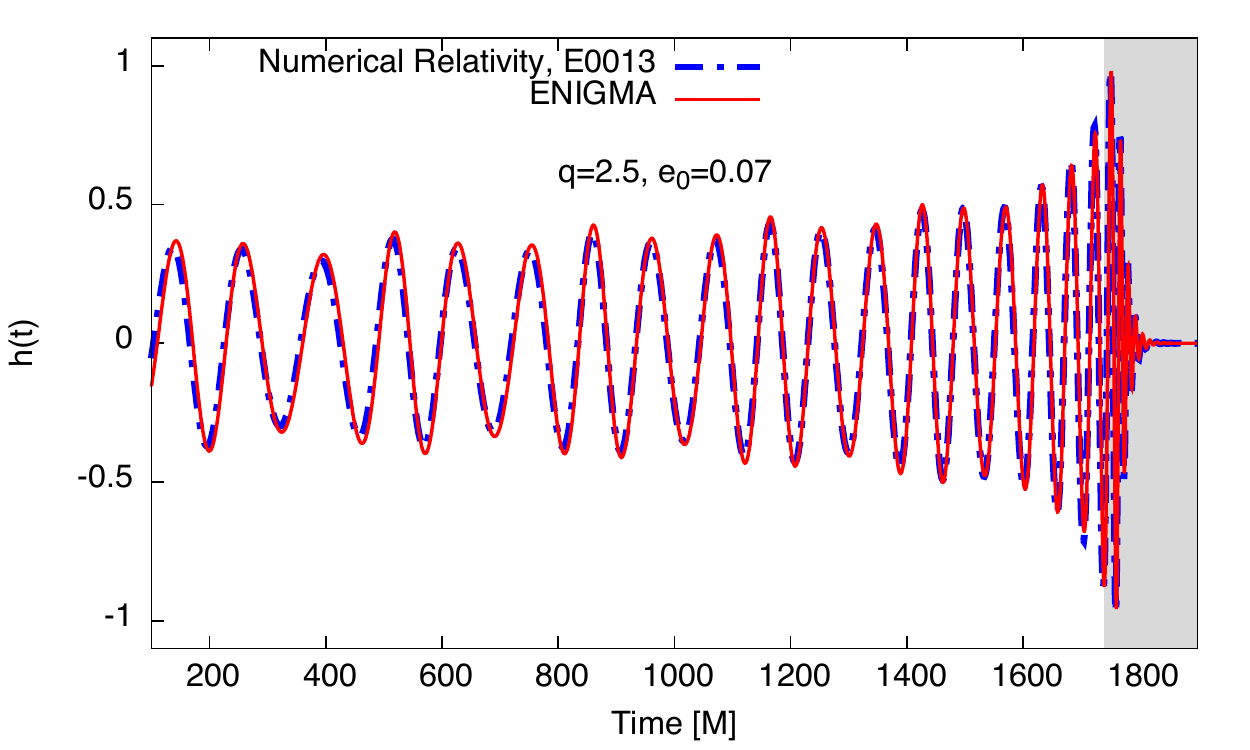}
    }  
\caption{Validation of \texttt{ENIGMA} with a set of \texttt{Einstein Toolkit} numerical relativity simulations that describe binary black hole mergers for a variety of mass-ratios, and moderate values of eccentricity---see Table~\ref{results}. The shaded region indicates the merger-ringdown part of the \texttt{ENIGMA} waveforms.}
  \label{validation_nr}
  \end{figure*}
 \end{widetext} 

\begin{widetext}
 \begin{figure*}[!ht]
\centerline{
    \includegraphics[width=0.485\textwidth]{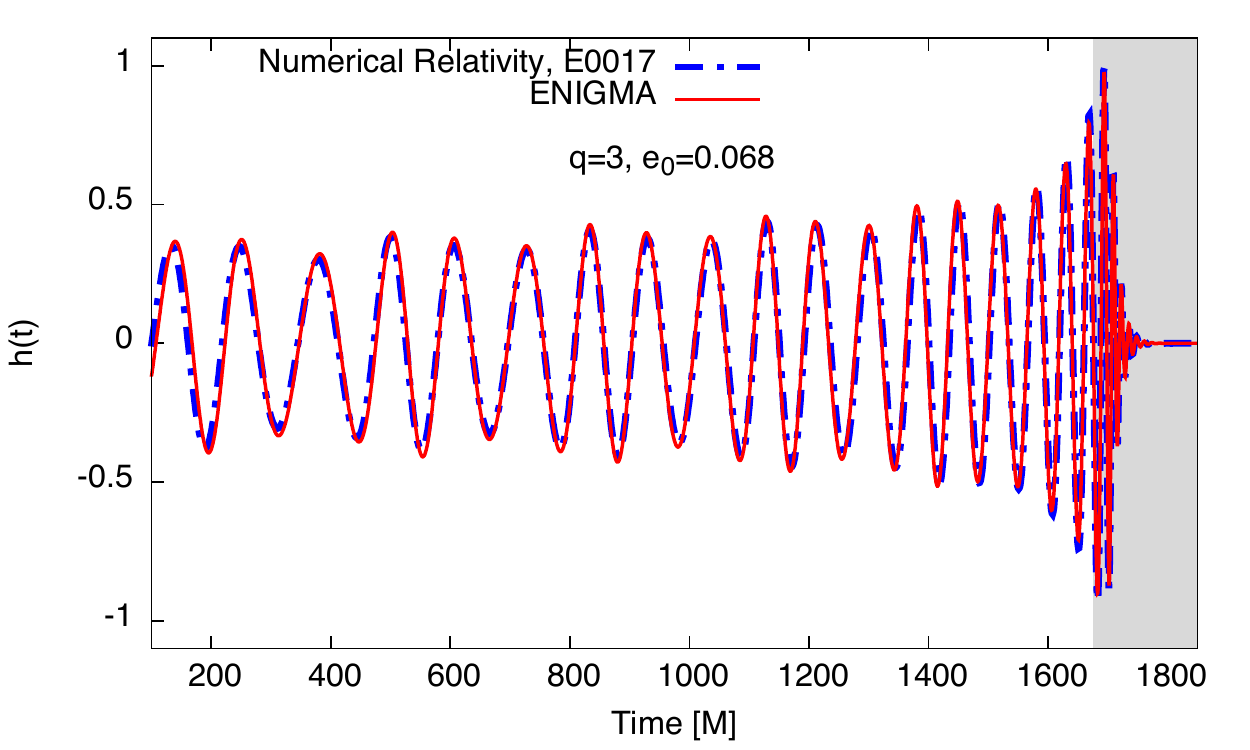}
    \includegraphics[width=0.485\textwidth]{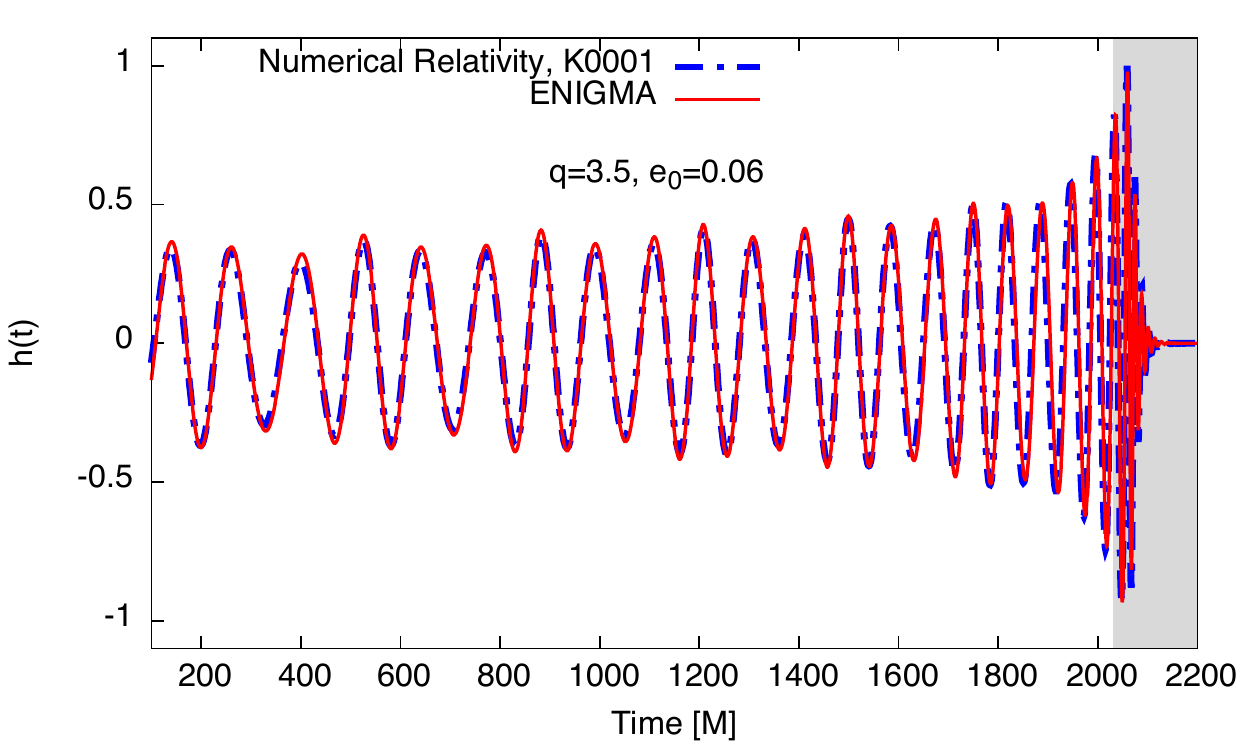}
  }
 \centerline{   
   \includegraphics[width=0.485\textwidth]{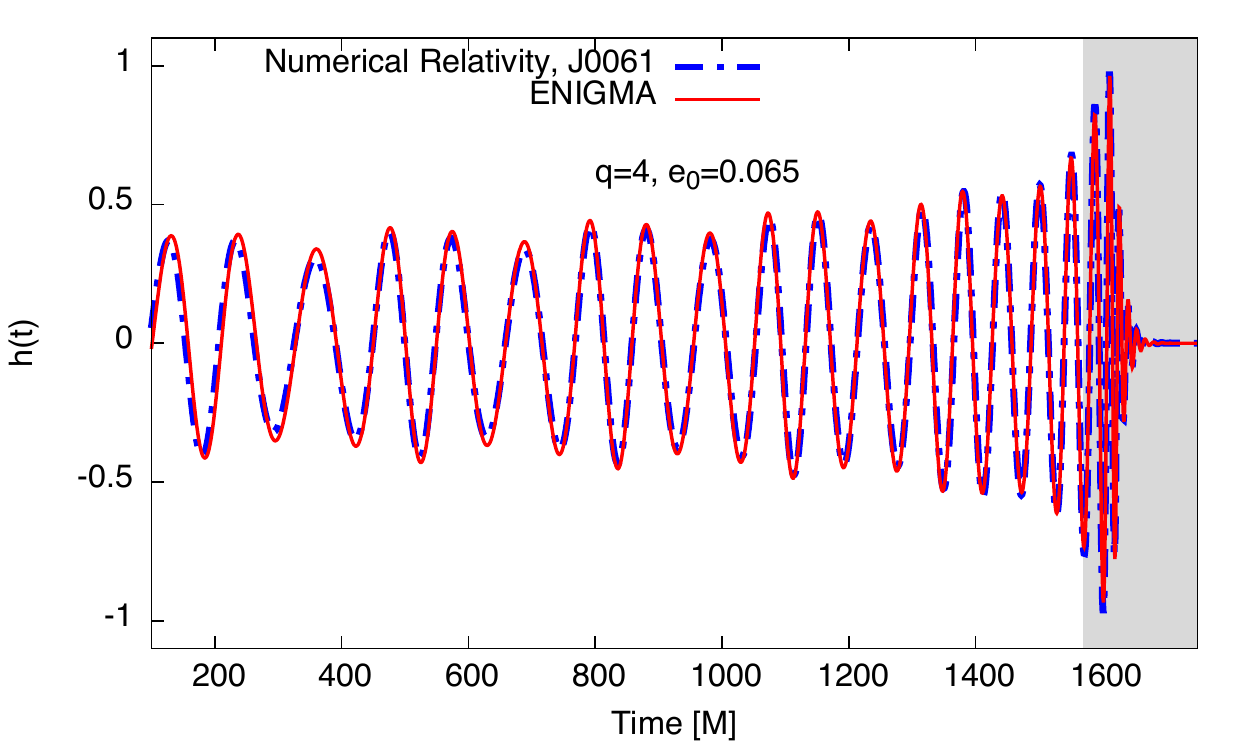}
    \includegraphics[width=0.485\textwidth]{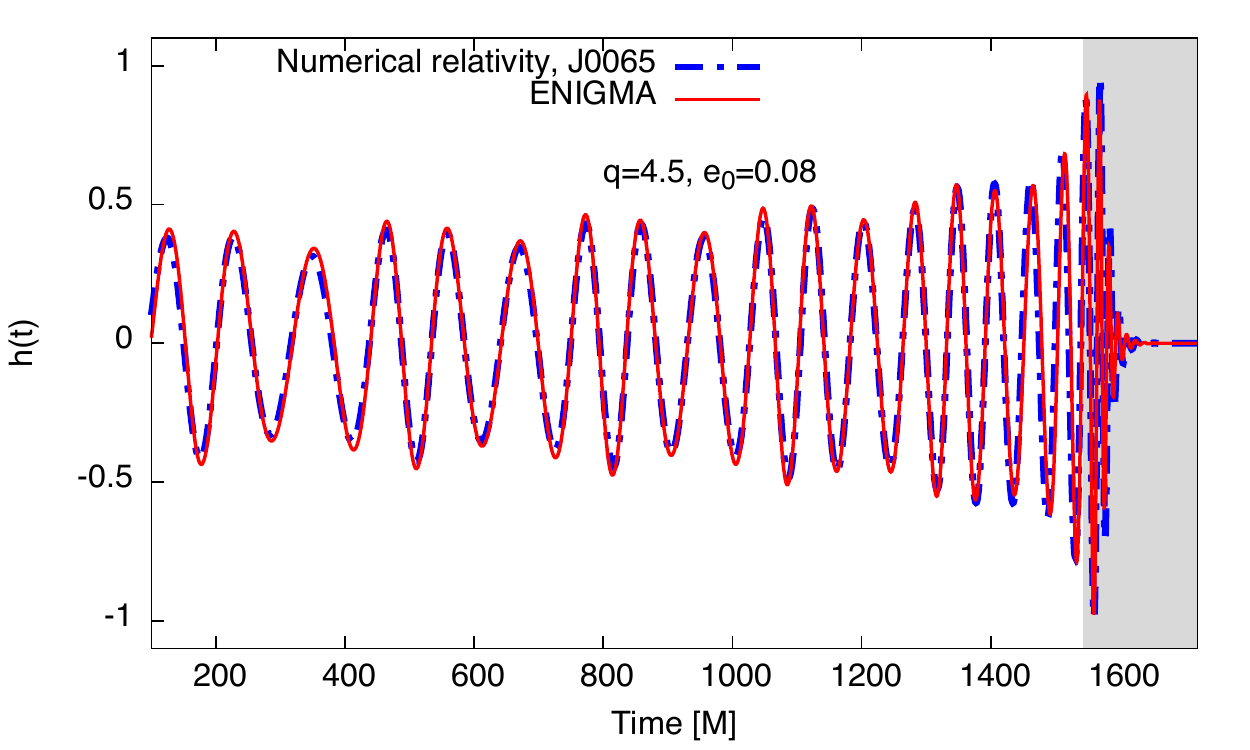}
}   
\caption{As Figure~\ref{validation_nr}, but now for BBH systems with more asymmetric mass-ratio values. }
  \label{validation_nr_two}
  \end{figure*}
 \end{widetext}

\noindent Figures~\ref{validation_nr} and~\ref{validation_nr_two} indicate that \texttt{ENIGMA} reproduces with excellent accuracy the late-time radiative evolution of NR simulations that describe BBH mergers with mass-ratios \(q \leq 4.5\) and eccentricities \(e_0\lesssim 0.08\) ten orbits before merger. The last column of Table~\ref{results} indicates that the overlap between \texttt{ENIGMA} waveforms and their NR counterparts is \({\cal{O}}\geq 0.981\).

 \begin{widetext}
 \begin{figure*}[!ht]
 \centerline{   
     \includegraphics[width=0.485\textwidth]{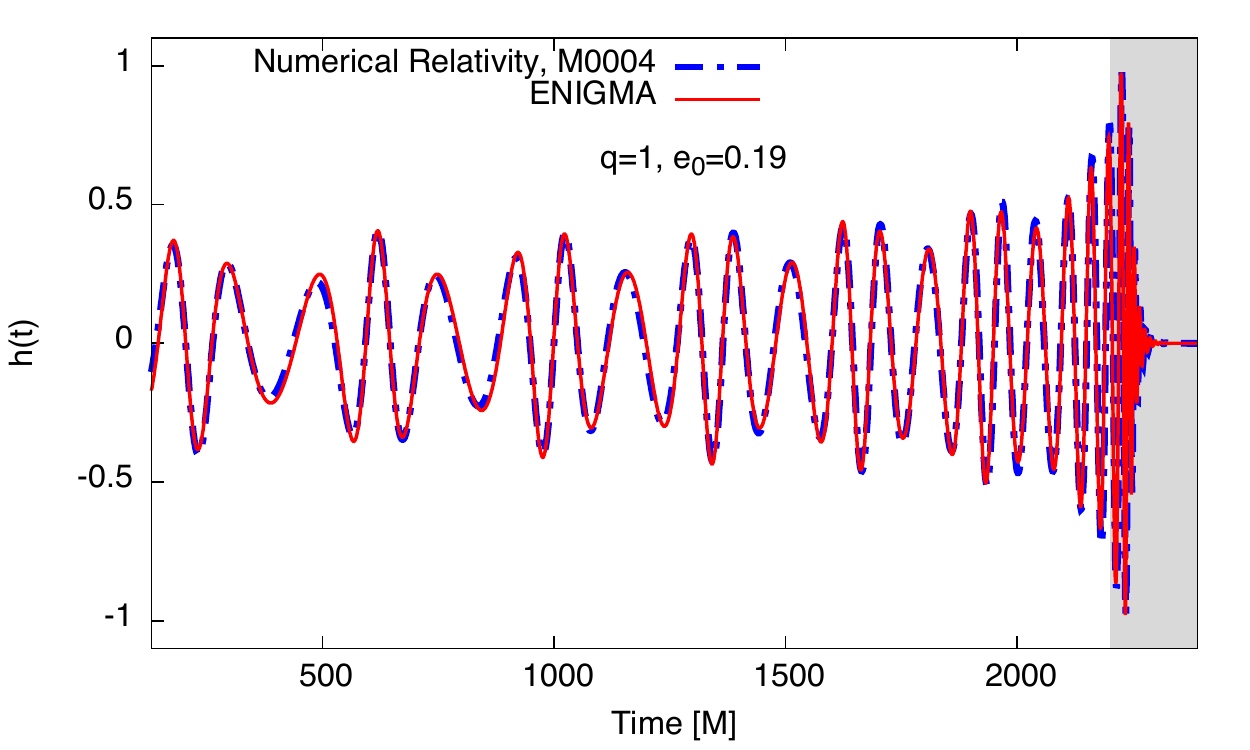}
     \includegraphics[width=0.485\textwidth]{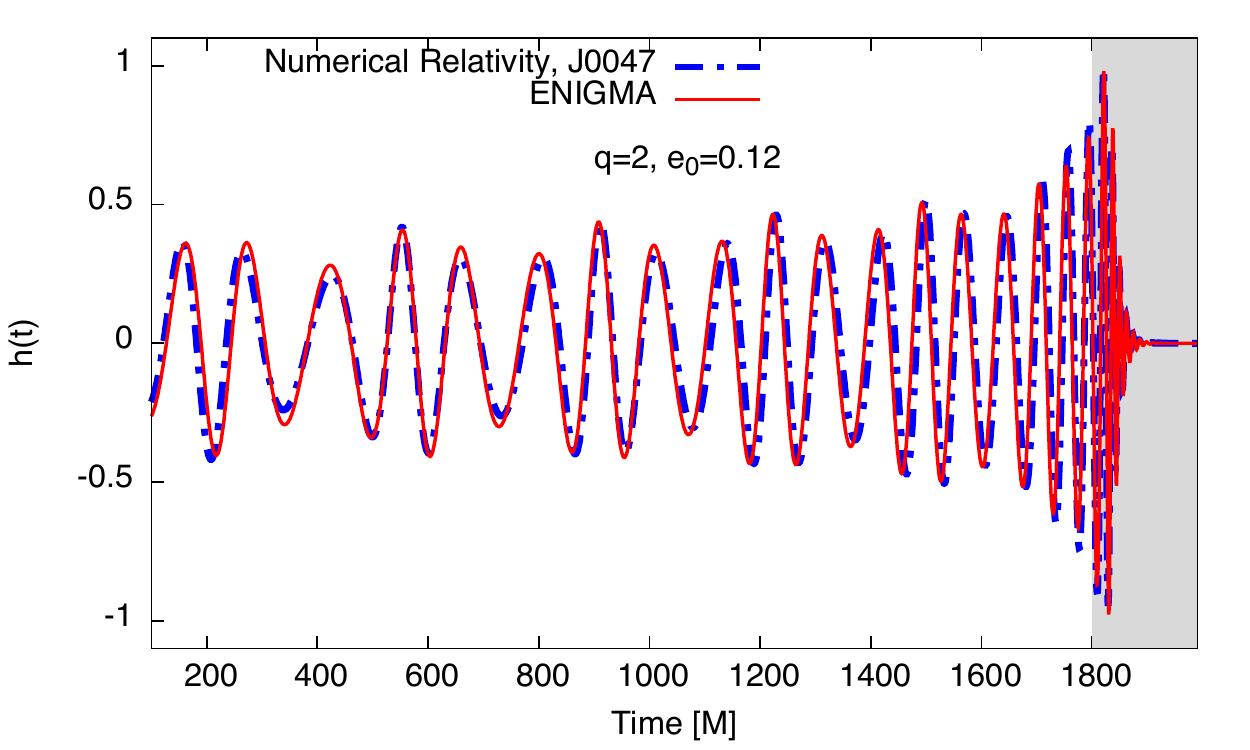}
}
\centerline{   
    \includegraphics[width=0.485\textwidth]{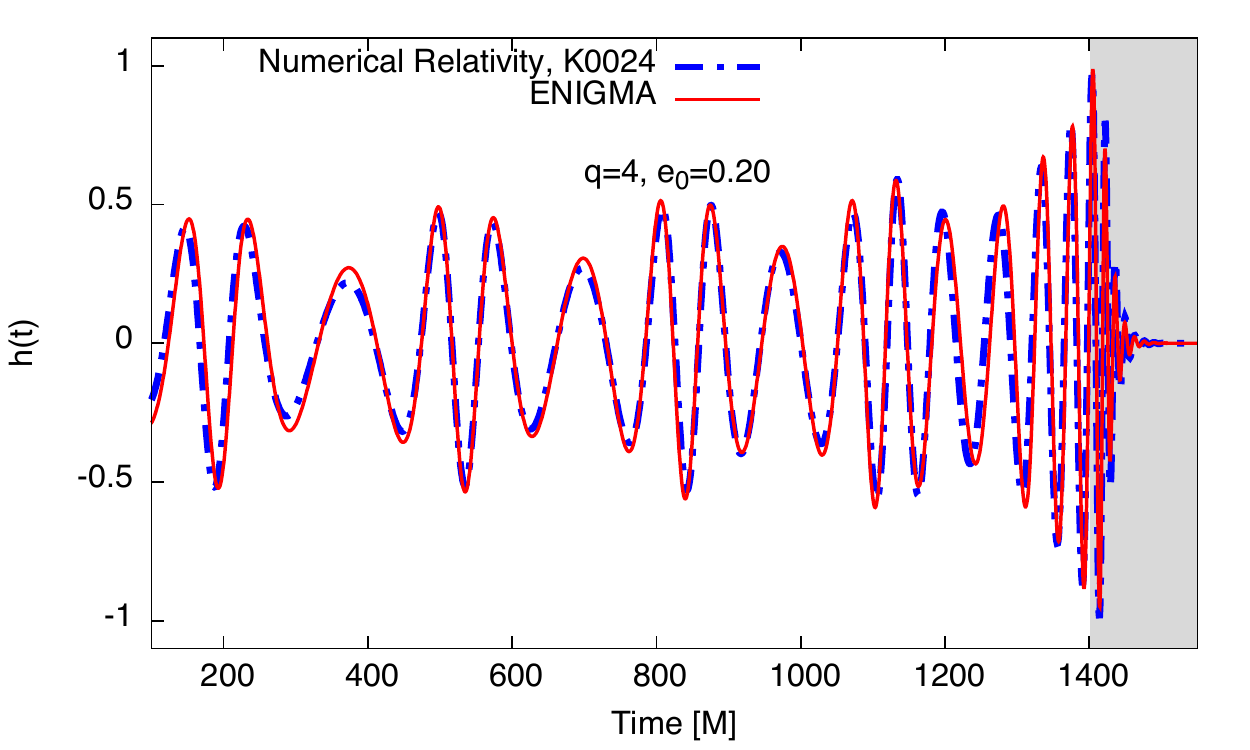}
     \includegraphics[width=0.485\textwidth]{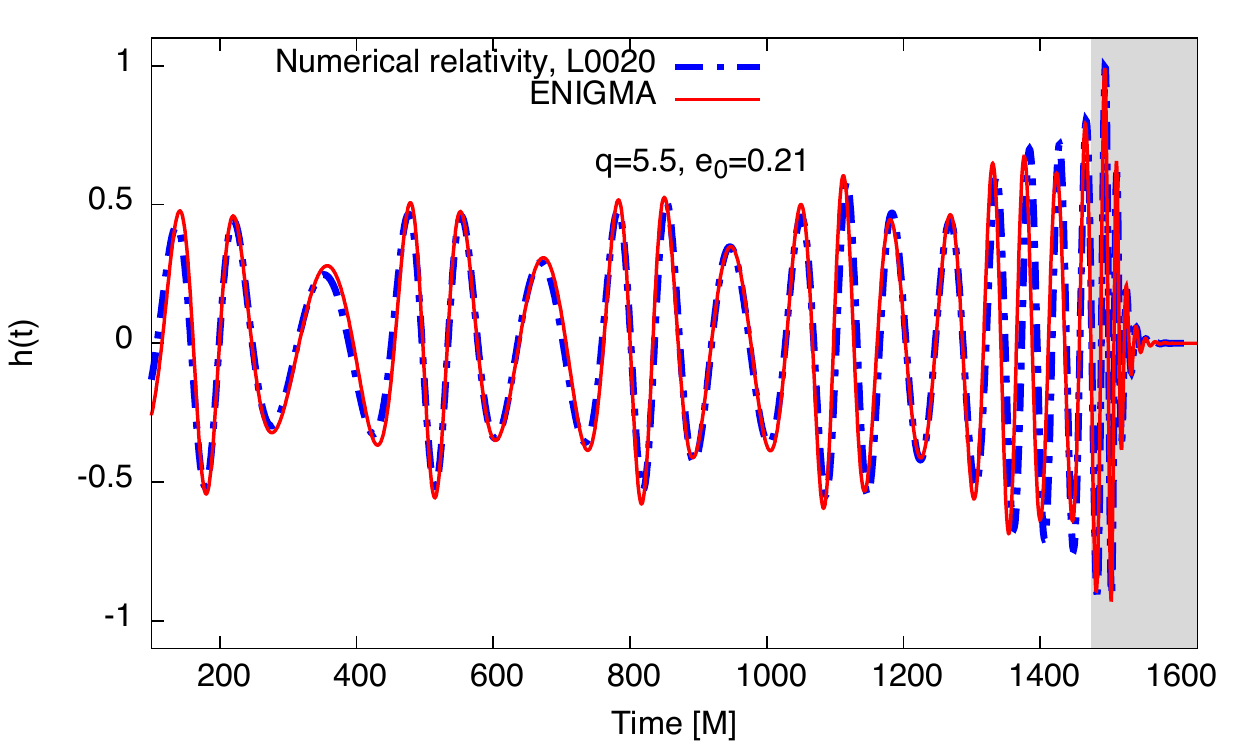}
}
\caption{As Figure~\ref{validation_nr}, but now for highly eccentric, and high mass-ratio binary black hole mergers.}
  \label{validation_nr_high_e}
  \end{figure*}
 \end{widetext}

We have also explored the performance of our new model to describe the evolution of BBH systems that retain significant eccentricity right before merger. Figure~\ref{validation_nr_high_e} and the last column of Table~\ref{results} show that \texttt{ENIGMA} can reproduce fairly well the dynamics of BBH systems with mass-ratios \(q\leq 5.5\) and eccentricities \(e_0\leq 0.21\) ten orbits before merger; with overlaps \({\cal{O}}\geq 0.95\). It is noteworthy that, even though \texttt{ENIGMA} was constructed to faithfully reproduce the dynamics of moderately eccentric BBH mergers, it can also describe BBH mergers that circularize right before merger. It is worth emphasizing that our \texttt{ENIGMA}  model can do this because our inspiral scheme can accurately reproduce the true dynamics of both quasi-circular and eccentric binaries very late in the inspiral evolution, which is clearly indicated in Figures~\ref{validation_nr}, \ref{validation_nr_two} and~\ref{validation_nr_high_e}. These results imply that \texttt{ENIGMA} can be used both for matched-filtering, burst and \texttt{Deep Filtering} searches~\cite{Tiwari:2016,dgNIPS,Sergey:2008CQG,geodf:2017a,dglitcha:2017,hshen:2017,geodf:2017b,geodf:2017c}. 
    
These results establish the accuracy of \texttt{ENIGMA} to describe eccentric BBH mergers, and indicate that the map  we have developed to smoothly connect our hybrid inspiral scheme to a quasi-circular, NR-based merger waveform is robust to describe moderately eccentric BBH mergers. 

\begin{widetext}
 \begin{figure*}[!ht]
 \centerline{
    \includegraphics[width=0.485\textwidth]{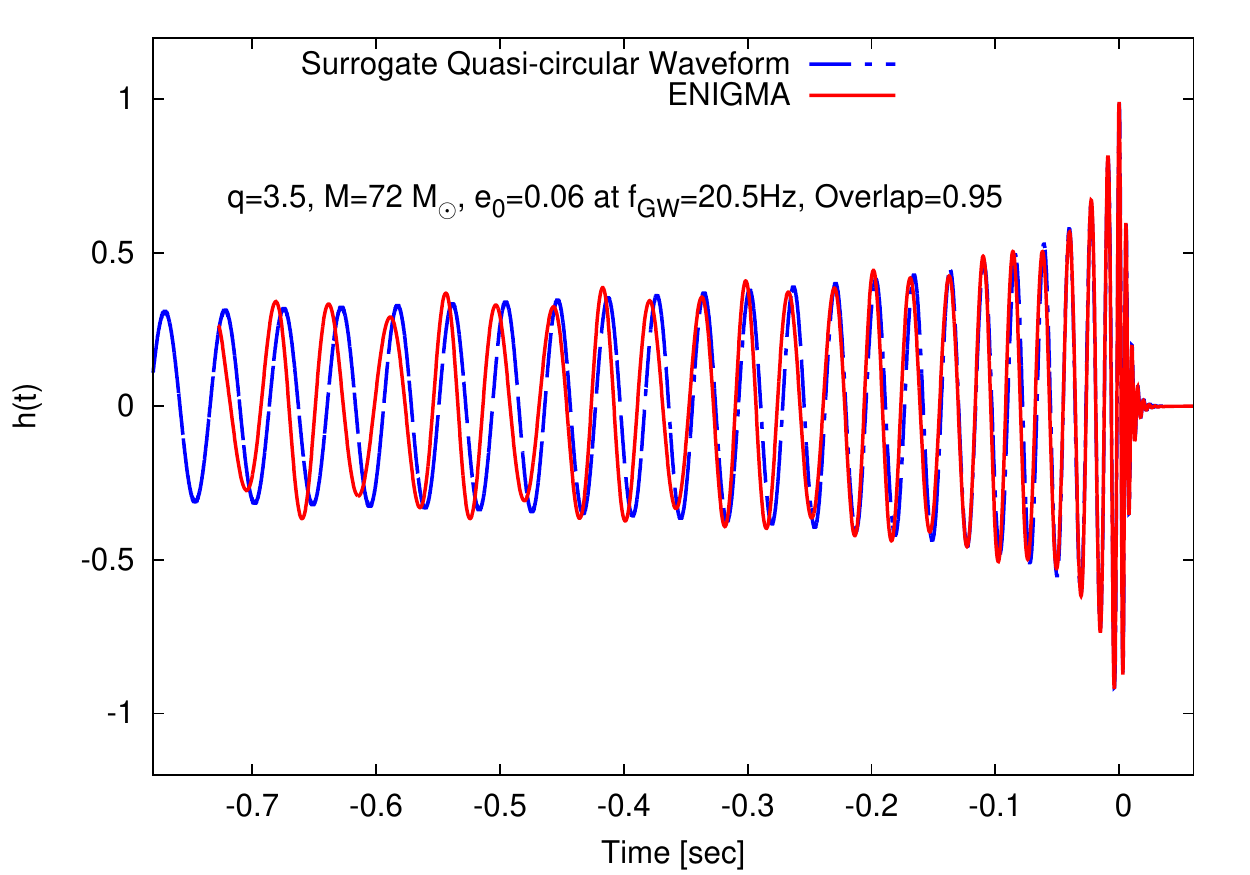}
    \includegraphics[width=0.485\textwidth]{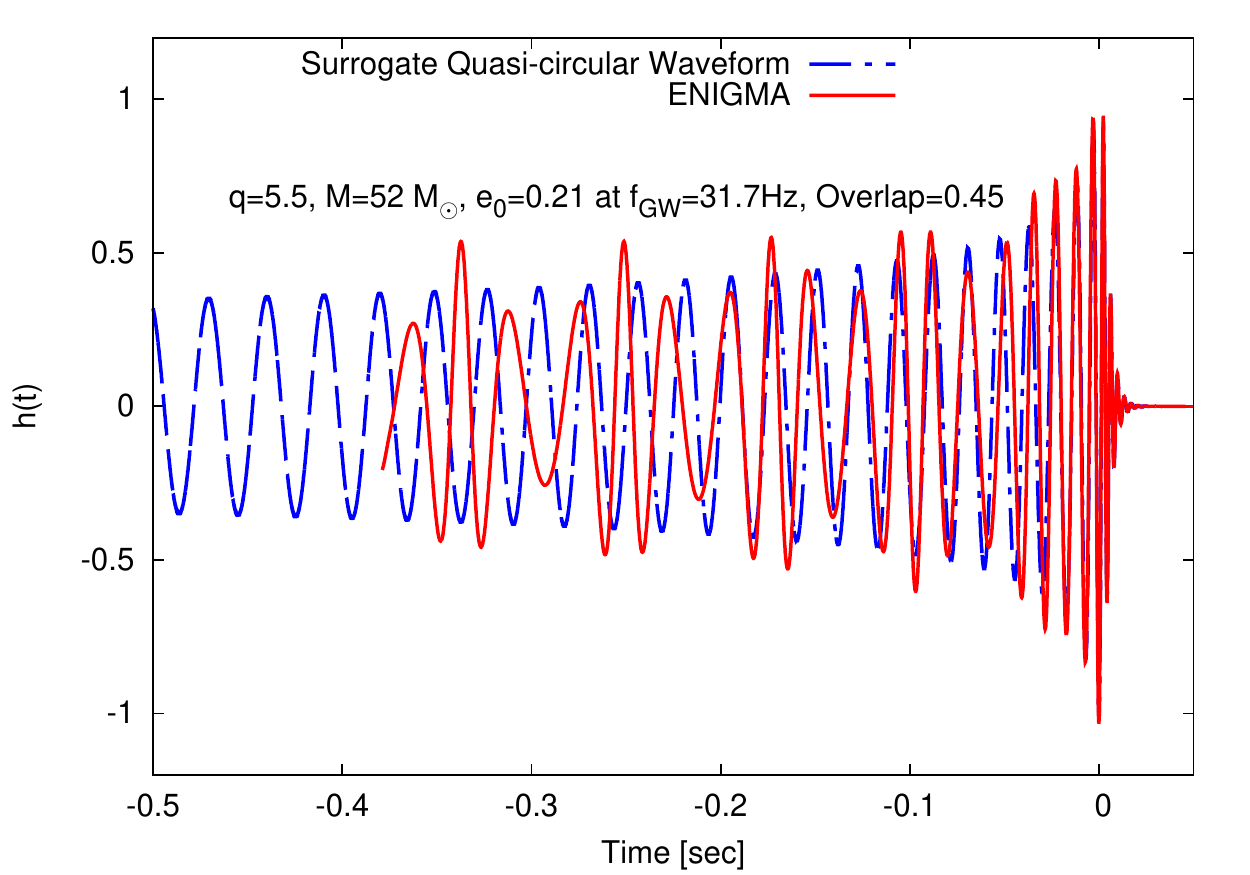}
    }
\caption{The panels show two potential astrophysical scenarios of eccentric binary black hole mergers. The left panel presents a system that enters the aLIGO frequency band on a nearly quasi-circular orbit, whereas the right panel shows an eccentric system that may form through a dynamical assembly channel in a dense stellar environment. Each panel presents the waveform signal corresponding to the quasi-circular counterpart of the eccentric system.}
  \label{astro_systems}
  \end{figure*}
 \end{widetext}

In terms of the astrophysically motivated systems that can be described with \texttt{ENIGMA}, Figure~\ref{astro_systems} presents two scenarios. Each panel includes an eccentric \texttt{ENIGMA} waveform and its quasi-circular counterpart, generated with the surrogate waveform family~\cite{blackman:2015}. The insets show the properties of the BBH systems and the overlap between the two waveforms. The left panel presents a moderately eccentric BBH with component masses \((56\Msun,\, 16\msun)\) and \(e_0=0.06\) at \(f_{\rm GW}=20.5\,{\rm Hz}\). The right panel depicts an eccentric BBH with component masses \((44\Msun,\, 8\msun)\) and \(e_0=0.21\) at \(f_{\rm GW}=31.7\,{\rm Hz}\).  We have chosen these examples to represent a variety of astrophysically motivates scenarios, i.e., BBHs that enter the aLIGO frequency band on nearly quasi-circular orbits, and moderately eccentric BBH systems that may form in dense stellar environments through a variety of dynamical mechanisms~\cite{Samsing:2014,sam:2017ApJ...840L..14S,hoang:2017arXiv170609896H}. Since the proposed formation scenarios for eccentric BBH mergers are astrophysically unconstrained, it is advantageous to use a waveform model that can cover as deep a parameter space as possible, and then let GW observations confirm or rule out potential formation channels. The analysis we carried out in this section indicates that \texttt{ENIGMA} can be used for these studies.

Figure~\ref{astro_systems} provides a clear visual description of the imprints of eccentricity. Compared to quasi-circular signals, eccentricity reduces the time-span of GW signals, and induces significant modulations in frequency and amplitude at lower frequencies. The overlap values quoted in the inset have been computed using the design sensitivity of aLIGO from the \(f_{\rm GW}\) quoted in the panels.

\noindent Having established the accuracy of our new waveform model, in the following section we use it to quantify the detectability of eccentric BBH mergers with GW observations. 

%%%%%%%%%%%%%%%%%%%%%%%%%%%%%%%%%%%%%%%%%%%%%
%%%%%%%%%%%%%%%%%%%%%%%%%%%%%%%%%%%%%%%%%%%%%
\section{Detectability of eccentric binary black hole mergers}
\label{see_me}

Having established the accuracy of \texttt{ENIGMA} in the quasi-circular limit, and its ability to describe the dynamics of eccentric systems  throughout merger with a set of eccentric NR simulations, we now want to quantify the minimum value of eccentricity at which a circular search is no longer effectual.

One can use a variety of criteria to accomplish this. In this study, we use fitting factor (\({\cal{FF}}\)) calculations to establish a connection between eccentricity and deviations from quasi-circularity. The \({\cal{FF}}\) is defined as~\cite{FittingFactorApostolatos}

\beq
\label{ff}
{\cal{FF}}=  \underset{\vec{\theta}}{\rm{max}}\;{\cal{O}}\left(h,\, h^{T}_{\vec{\theta}}\right)\,.
\eeq

\noindent The \({\cal{FF}}\)  represents the overlap between a given GW signal  \(h\), maximized \textit{continuously} over chosen intrinsic and extrinsic parameters \(\vec{\theta}\) using templates of model \(h^{T}\). If the set \(\vec{\theta}\) contains all parameters that describe the template \textit{and} model \(T\) also models the signal \(h\), our continuous \({\cal{FF}}\) would be unity by construction.  In practice, we perform this continuous maximization over \(\vec{\theta}\) using particle swarm optimization~\cite{488968}\footnote{We use an open-source Python implementation of the particle swarm optimization algorithm: \texttt{pyswarm}~\cite{pyswarm}.}. We find that using a swarm size of \(500\) the \({\cal{FF}}\) converges to one part in \(10^8\).

\begin{figure*}[h]
\begin{center}$
\begin{array}{lll}
\includegraphics[height=.23\textwidth]{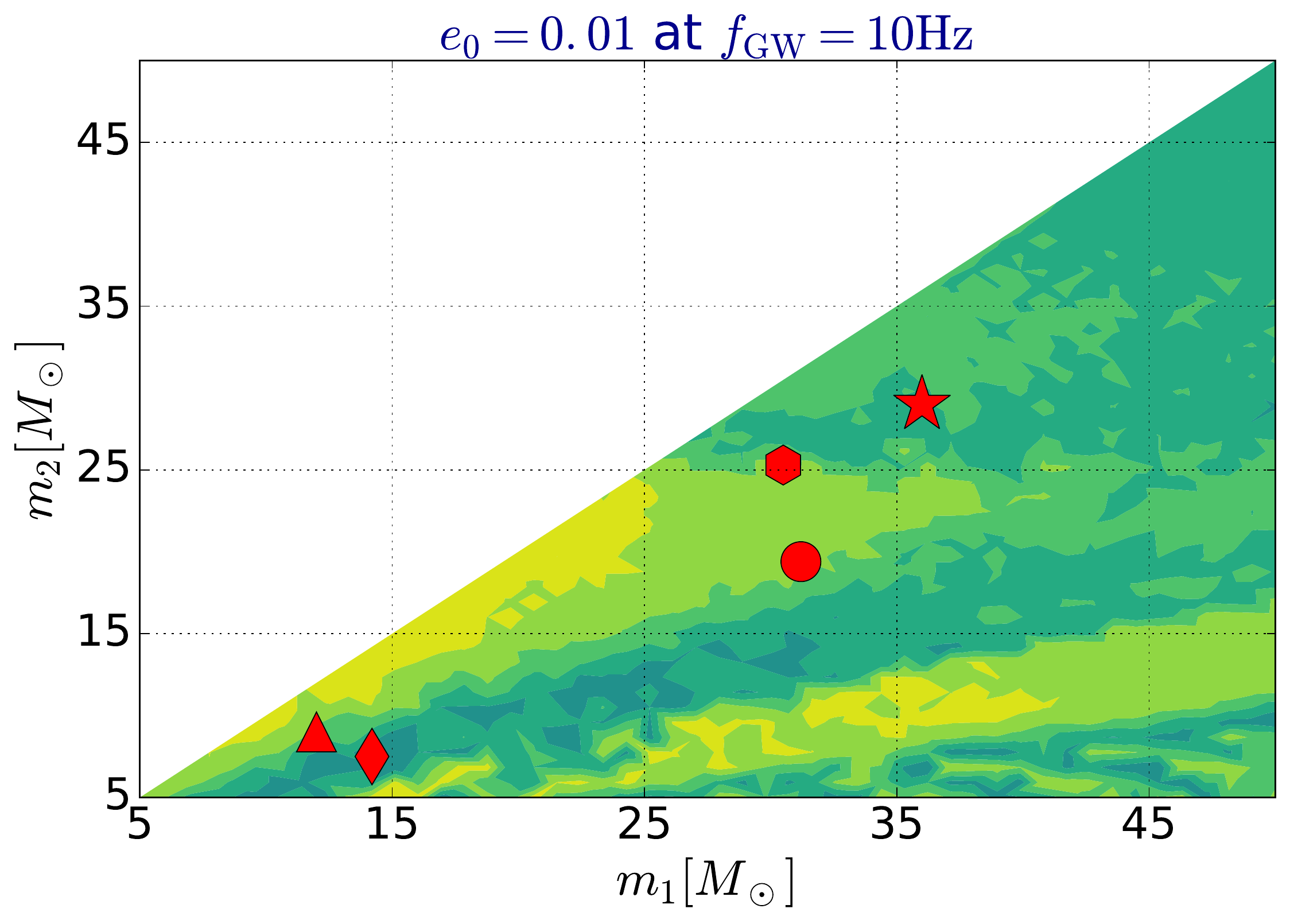}\hspace{-.2em}%
\includegraphics[height=.23\textwidth]{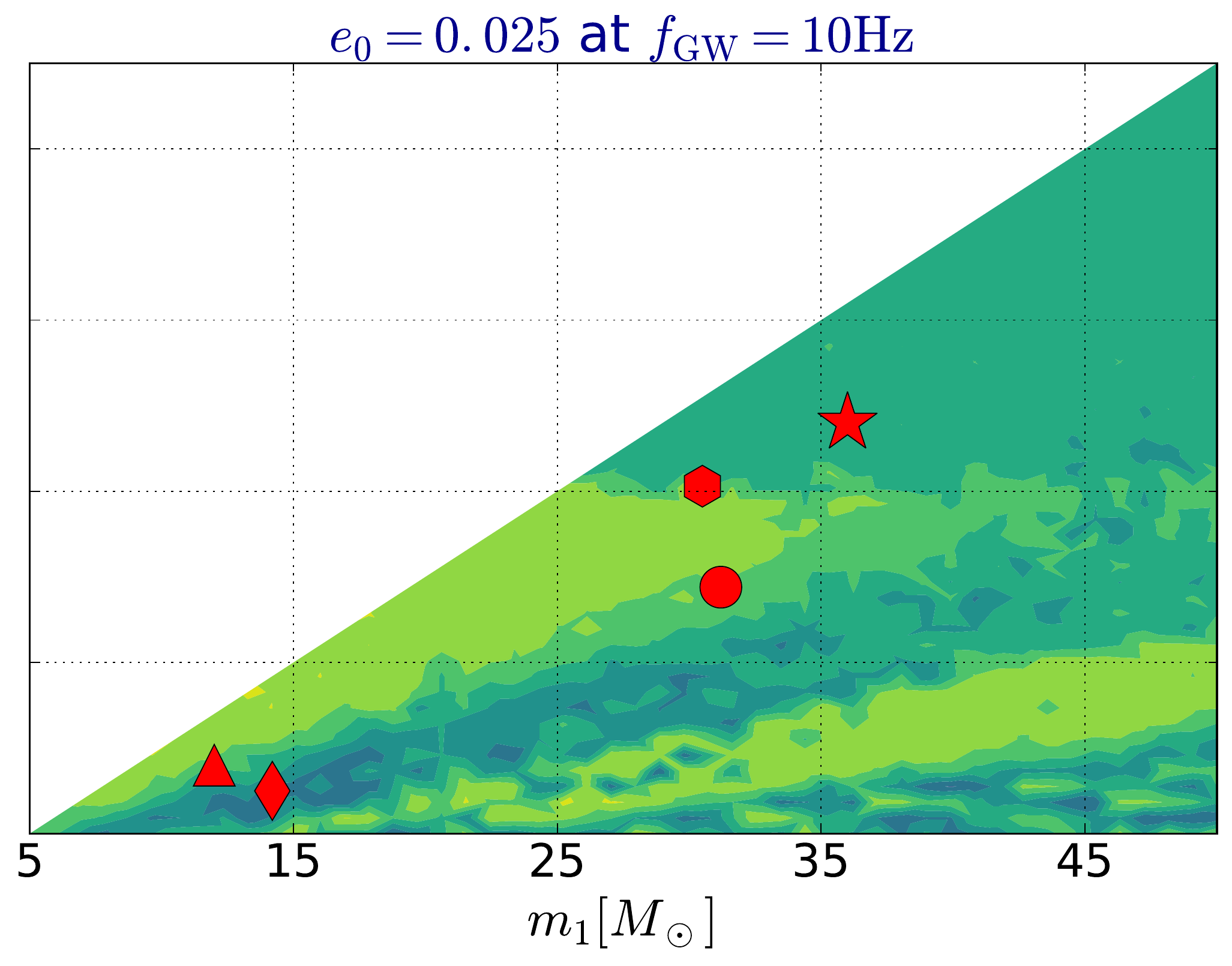}\hspace{-.2em}%
\includegraphics[height=.23\textwidth]{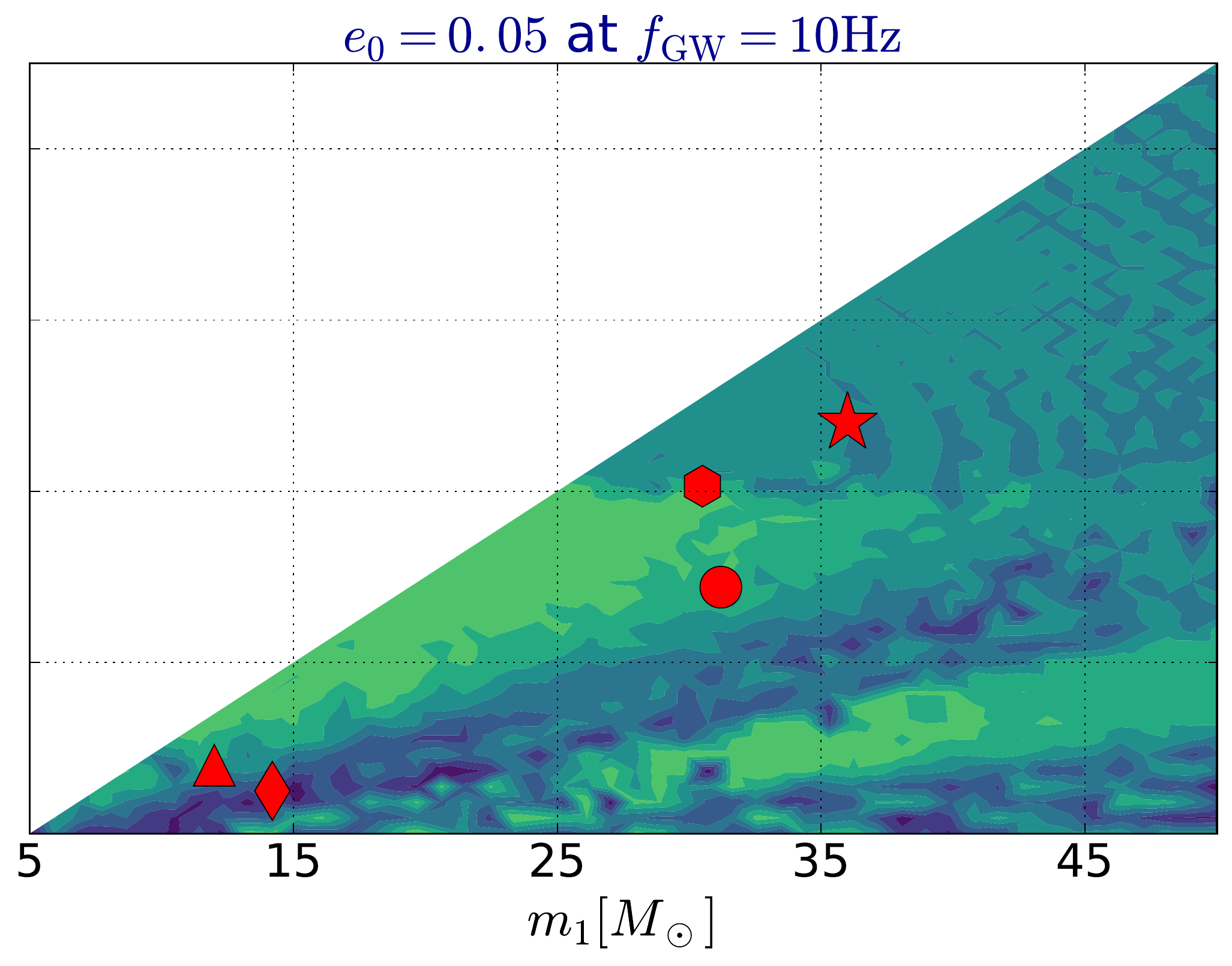}\hspace{0em}%
\includegraphics[trim= 0cm -1.3cm 0cm 0cm, height=.226\textwidth]{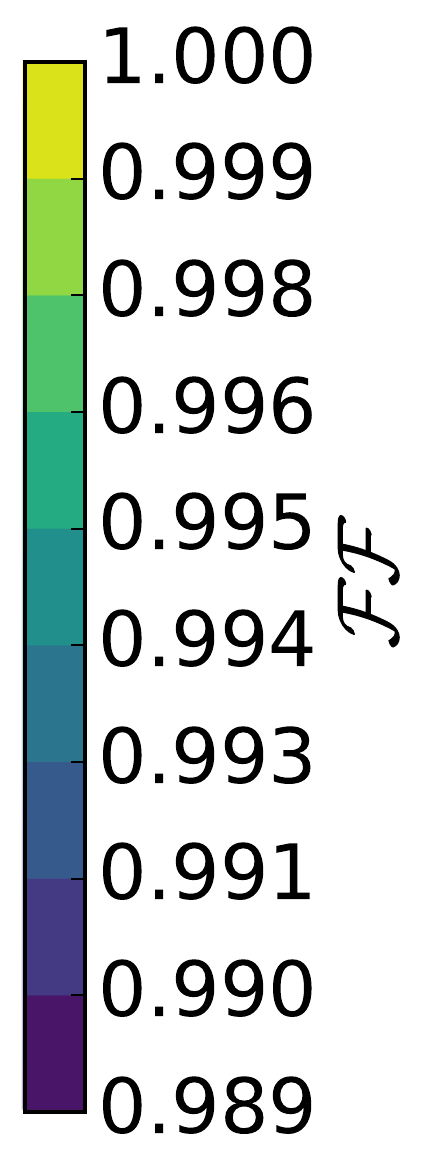}
\end{array}$
\end{center}

\begin{center}$
\begin{array}{lll}
\includegraphics[height=.23\textwidth]{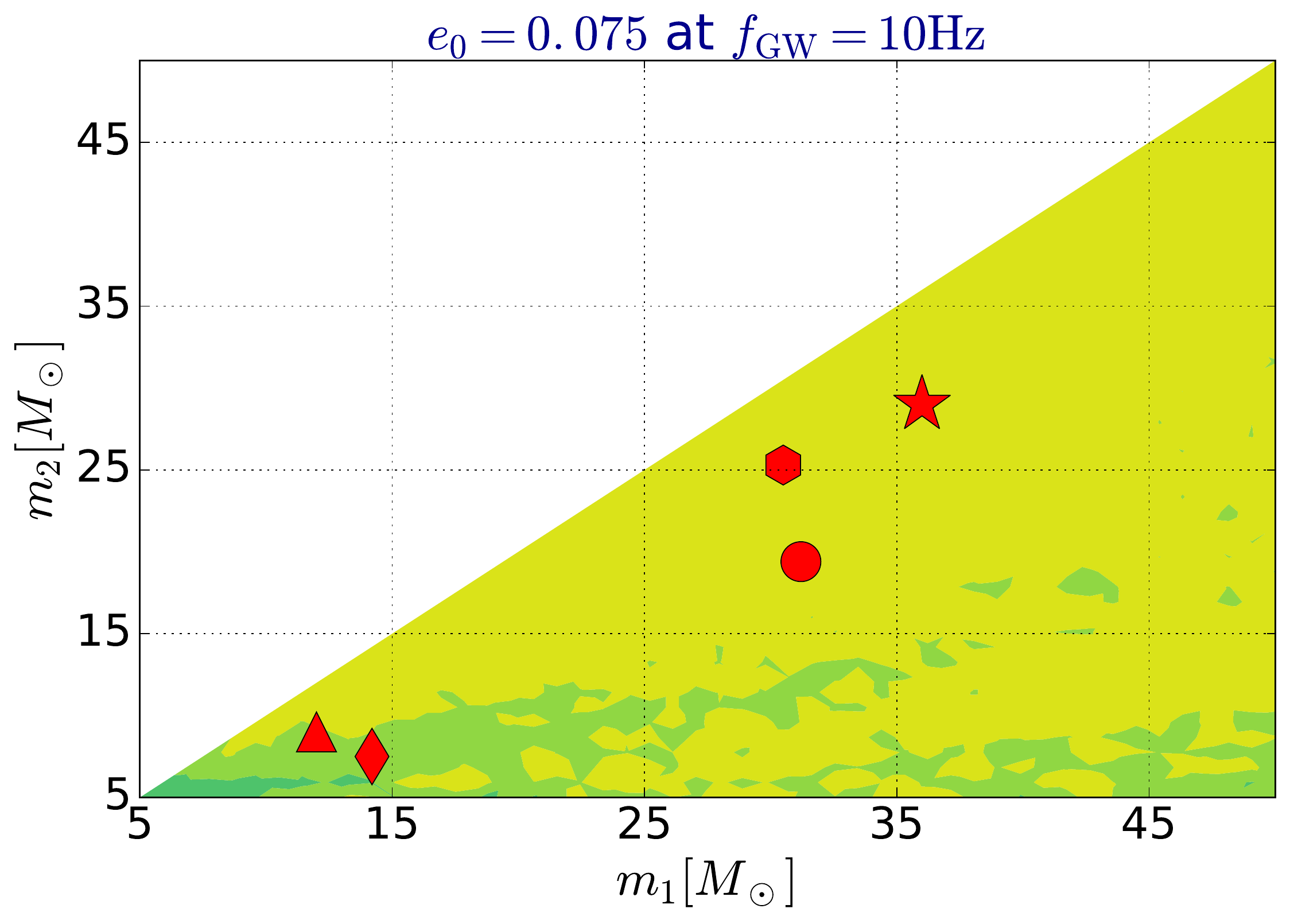}\hspace{-.2em}%
\includegraphics[height=.23\textwidth]{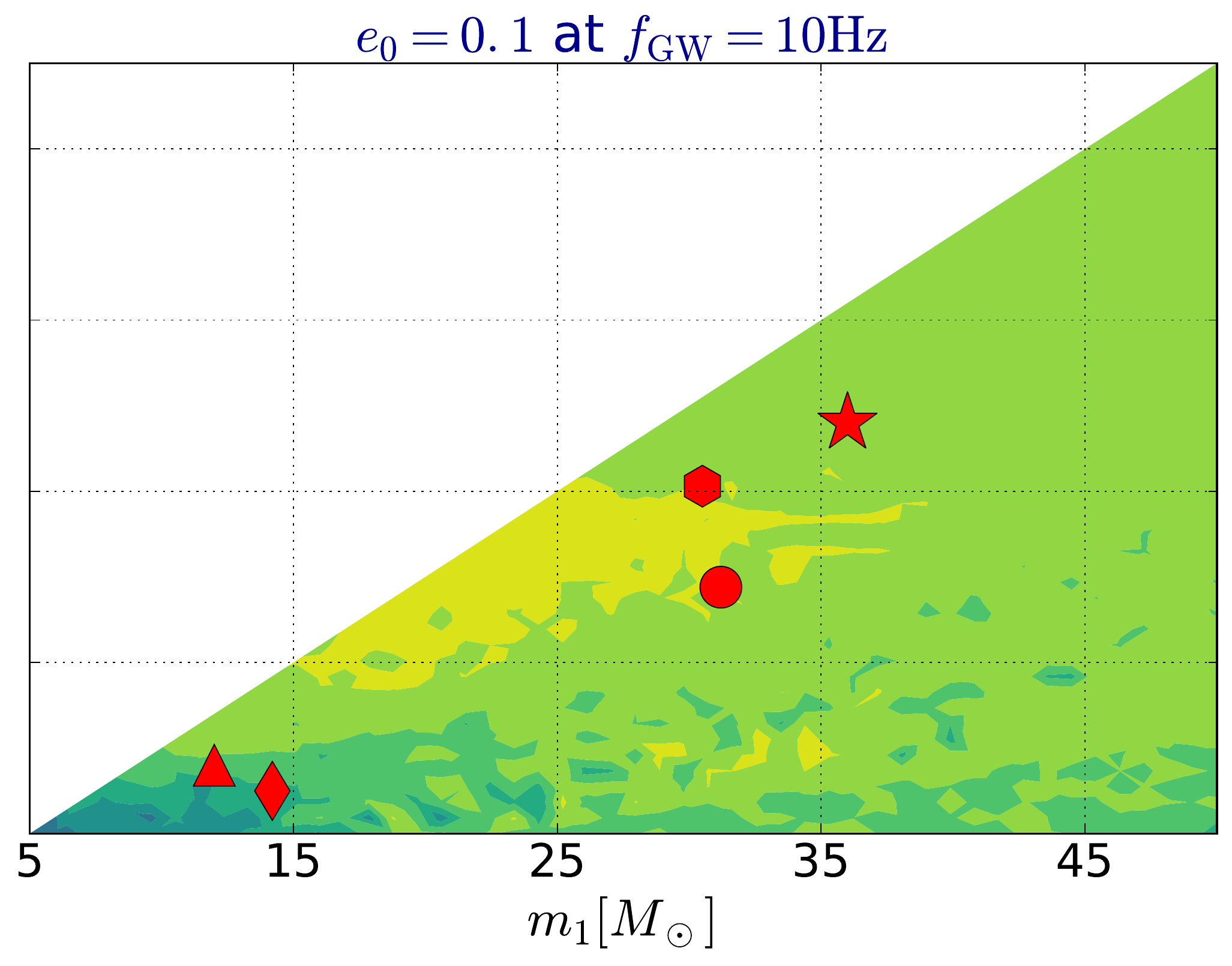}\hspace{-.2em}%
\includegraphics[height=.23\textwidth]{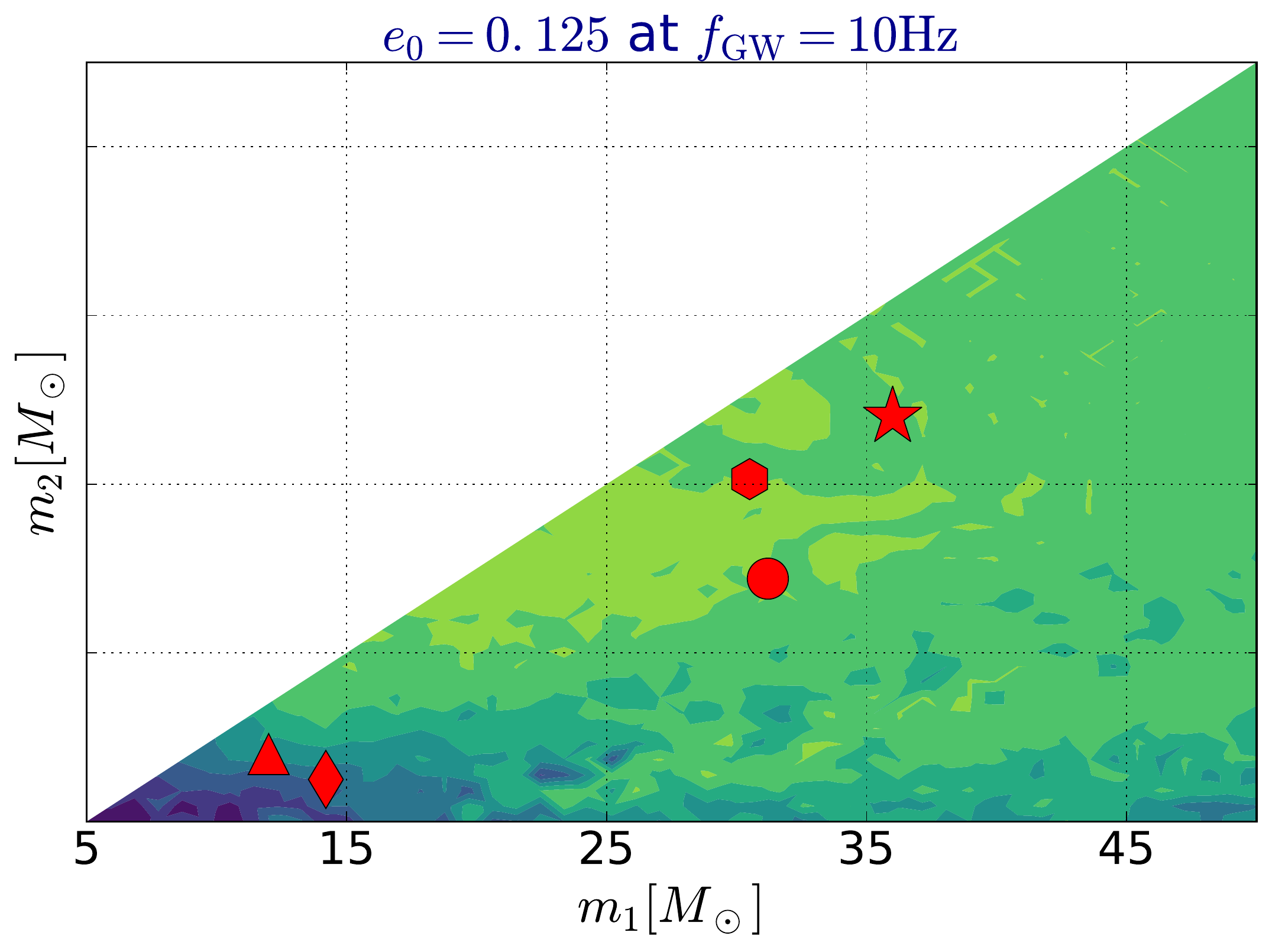}\hspace{0em}%
\includegraphics[trim= 0cm -1.3cm 0cm 0cm, height=.226\textwidth]{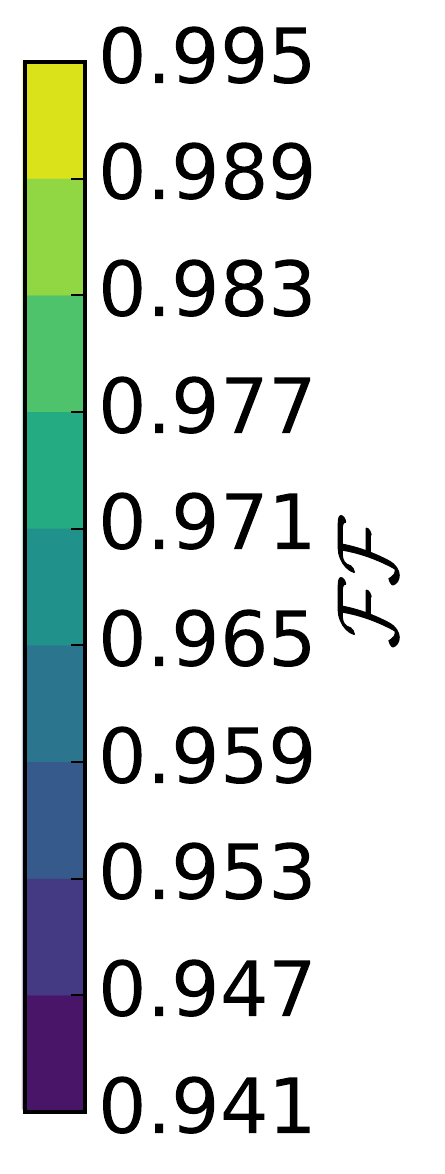}%
\end{array}$
\end{center}

\begin{center}$
\begin{array}{lll}
\includegraphics[height=.23\textwidth]{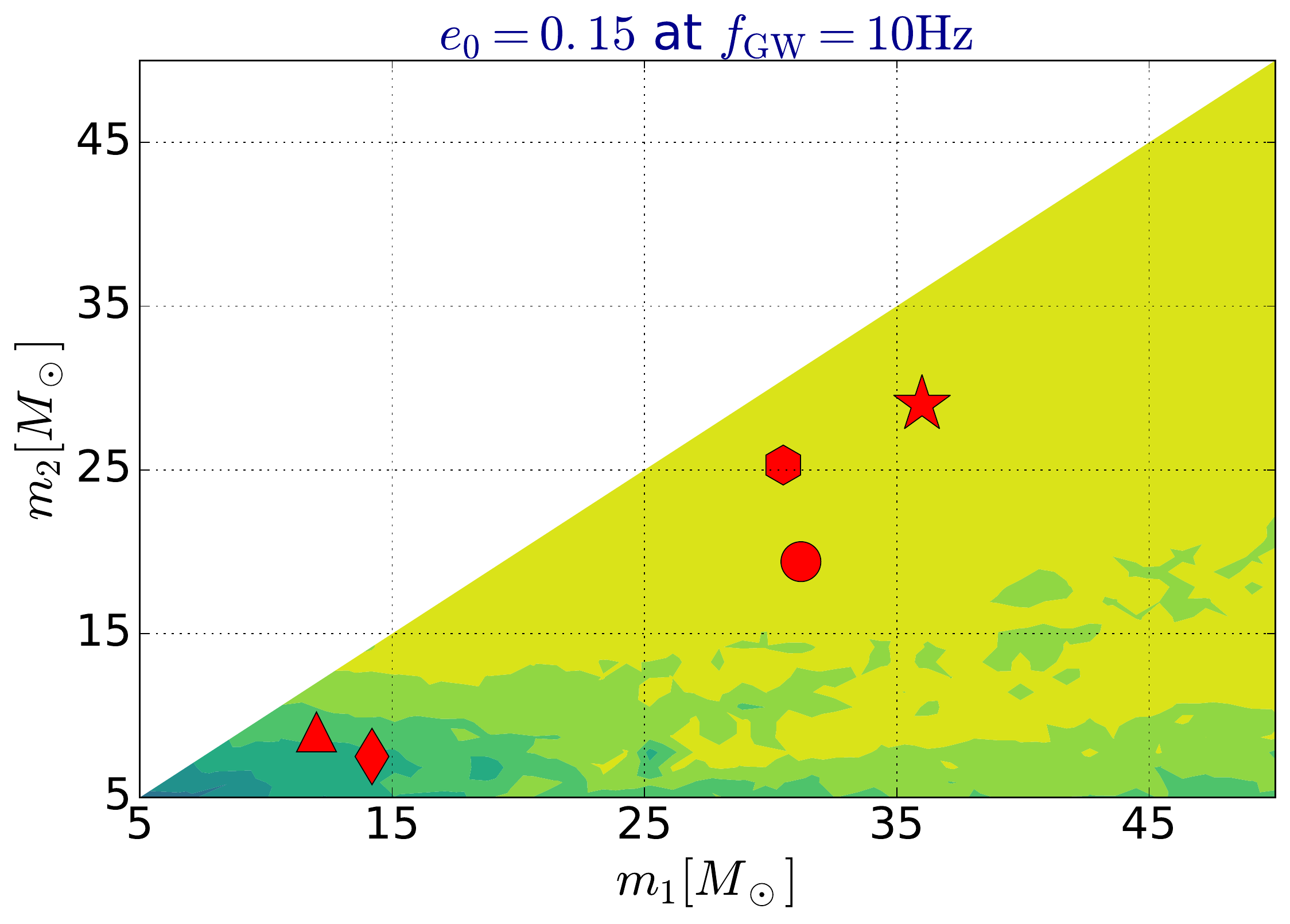}\hspace{-.2em}%
\includegraphics[height=.23\textwidth]{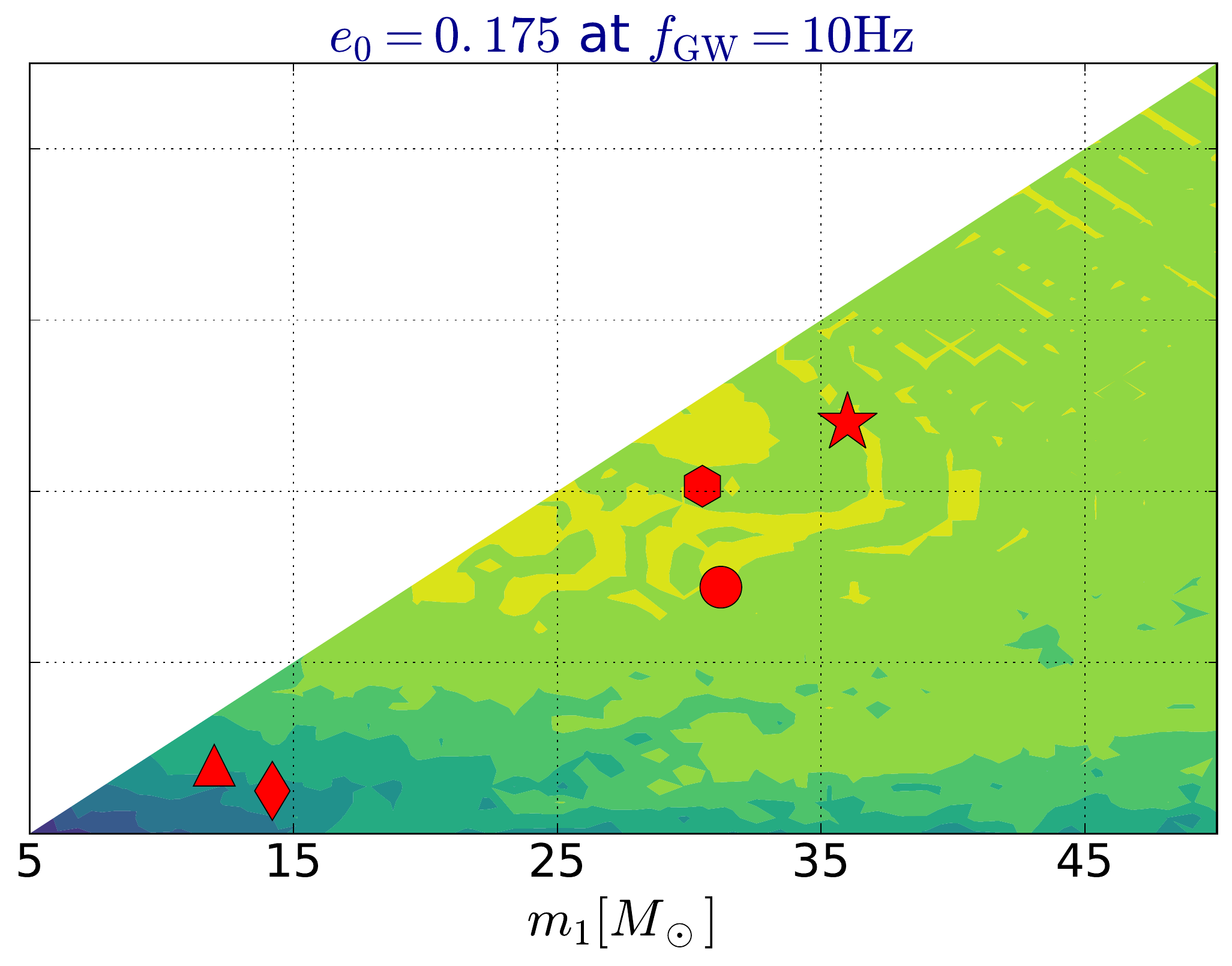}\hspace{-.2em}%
\includegraphics[height=.23\textwidth]{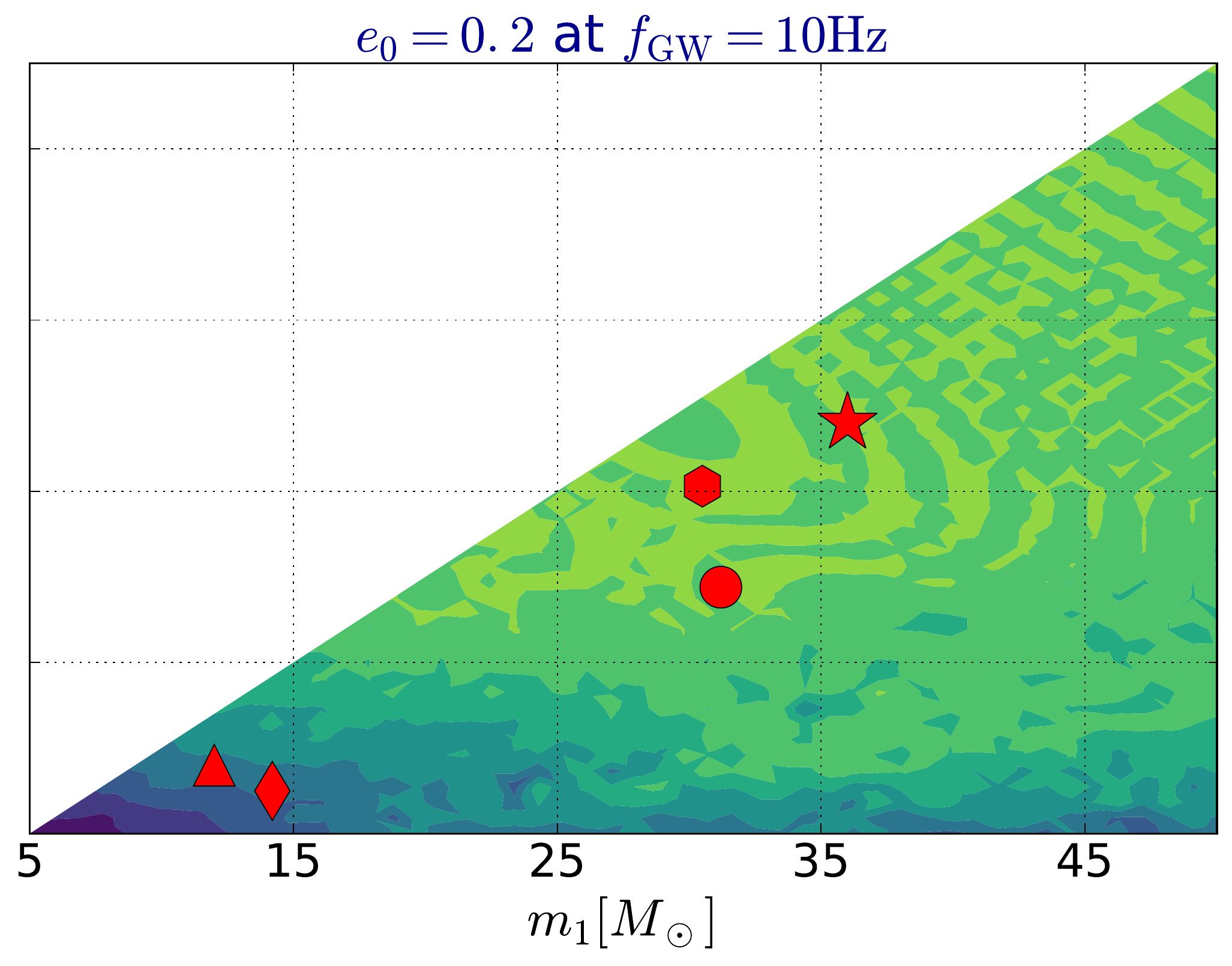}\hspace{0em}%
\includegraphics[trim= 0cm -1.3cm 0cm 0cm, height=.226\textwidth]{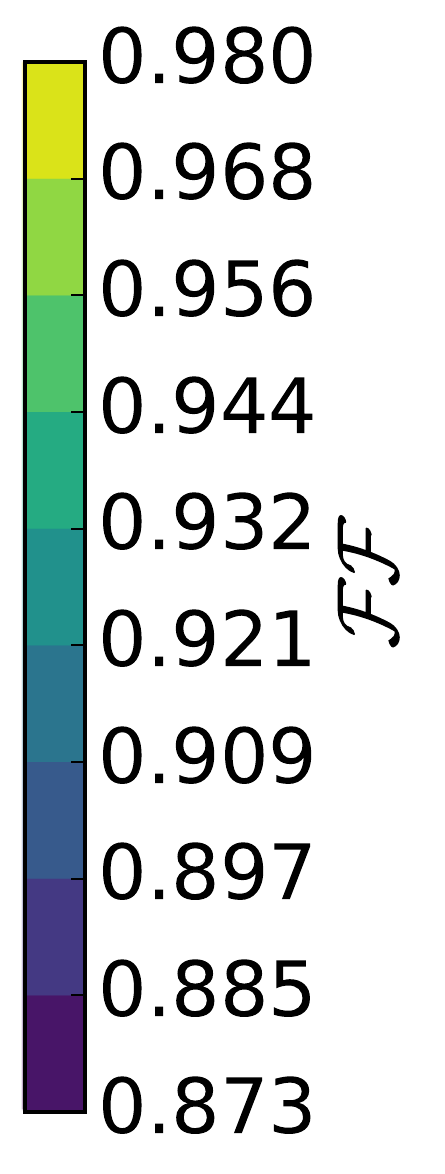}%
\end{array}$
\end{center}

\begin{center}$
\begin{array}{lll}
\includegraphics[height=.23\textwidth]{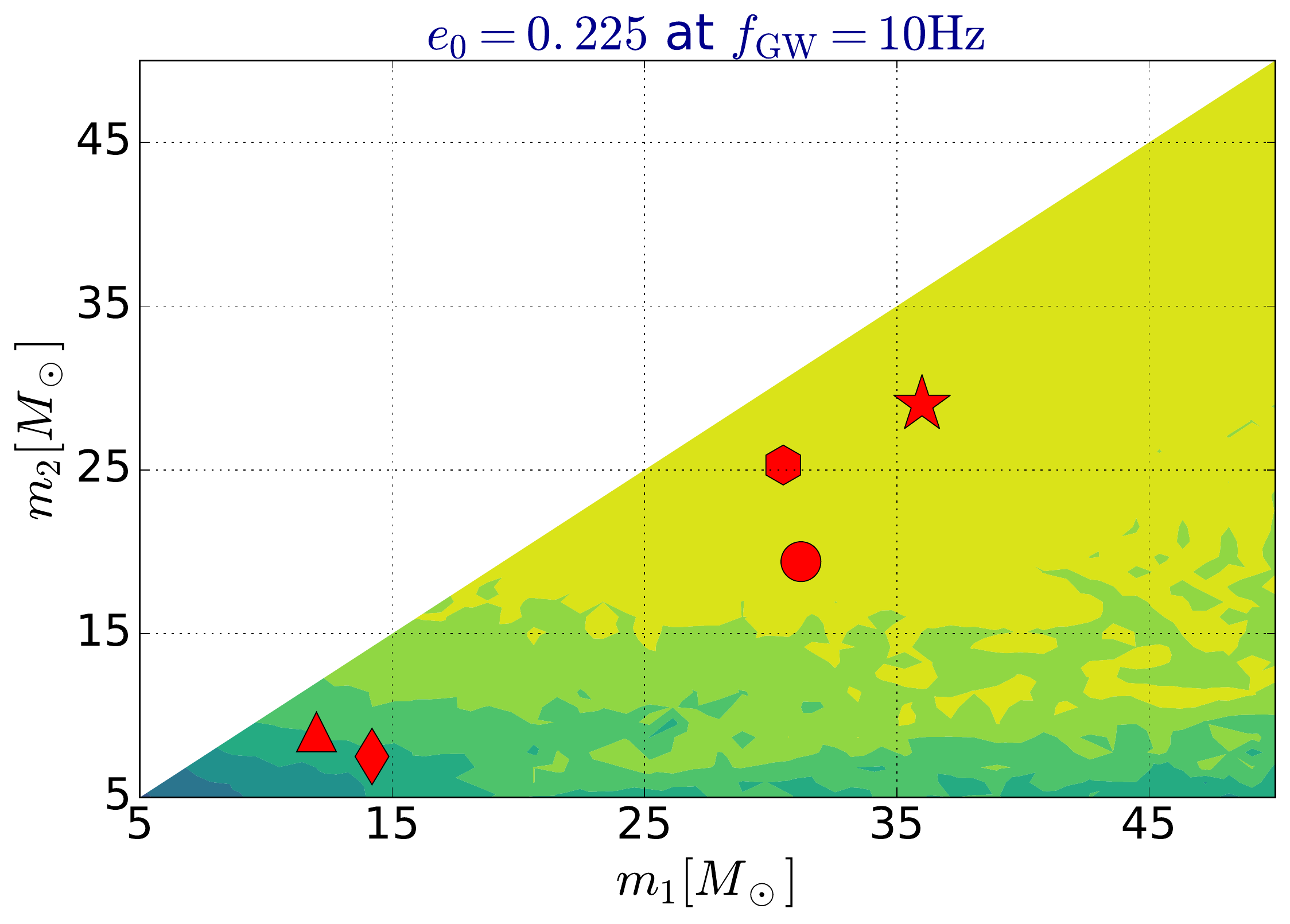}\hspace{-.2em}%
\includegraphics[height=.23\textwidth]{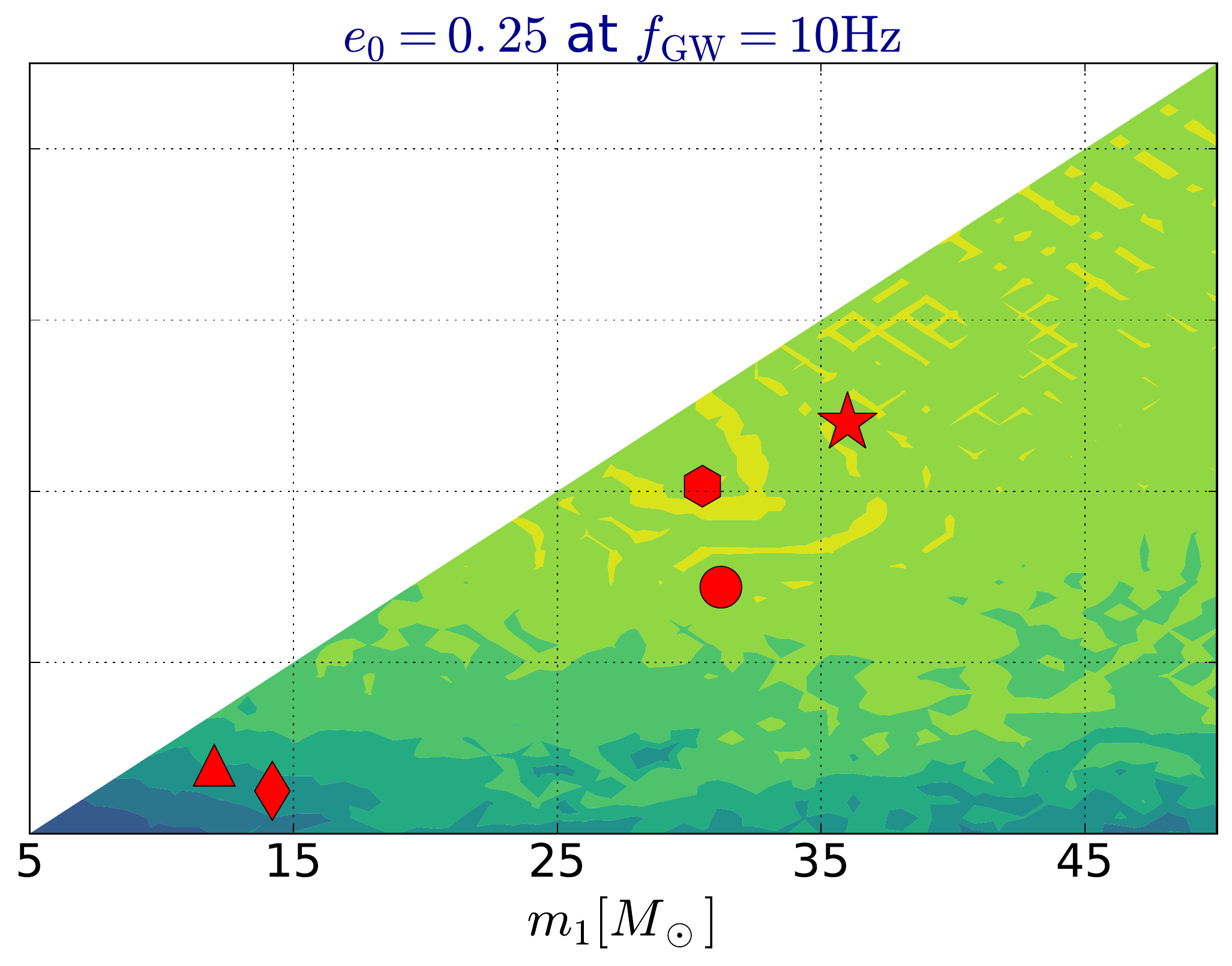}\hspace{-.2em}%
\includegraphics[height=.23\textwidth]{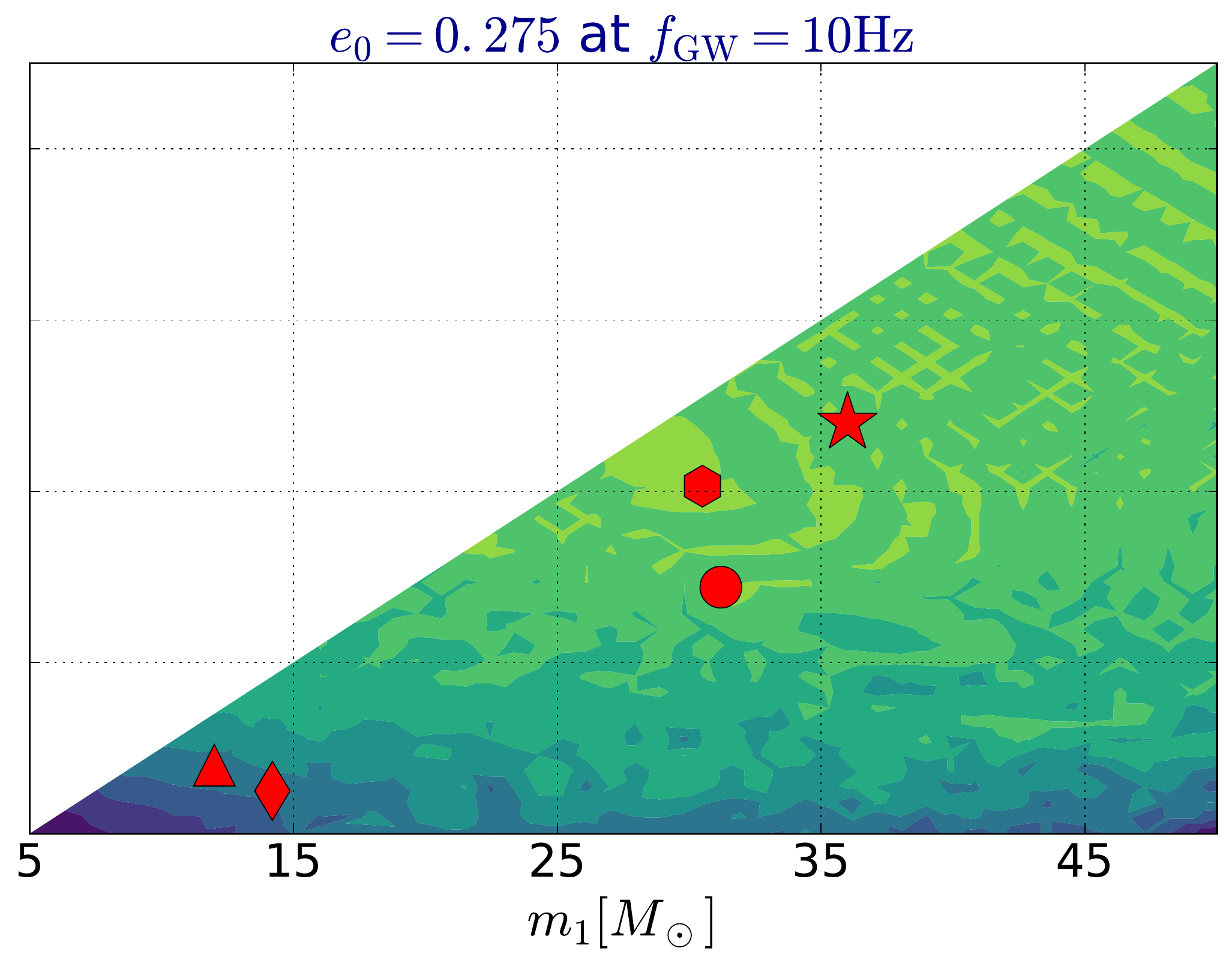}\hspace{0em}%
\includegraphics[trim= 0cm -1.3cm 0cm 0cm, height=.226\textwidth]{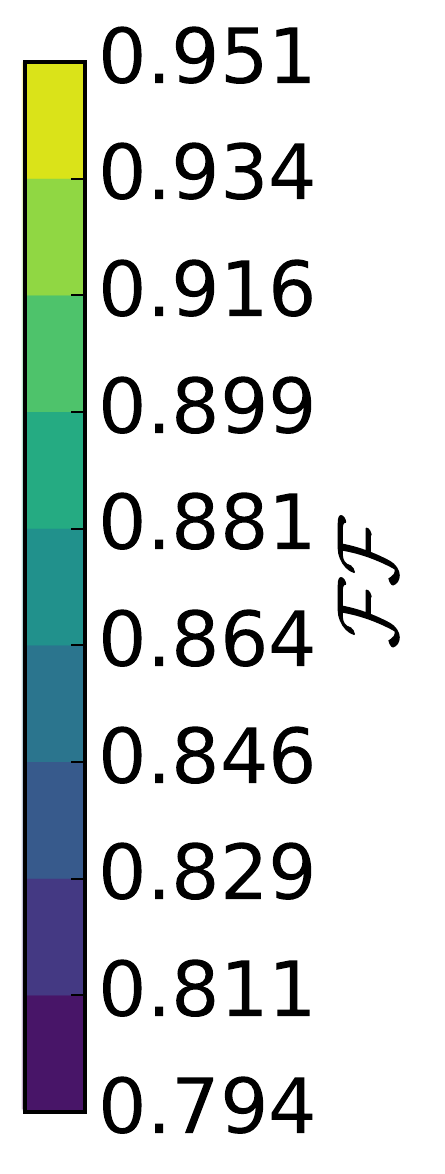}%
\end{array}$
\end{center}

\begin{center}$
\begin{array}{lll}
\includegraphics[height=.23\textwidth]{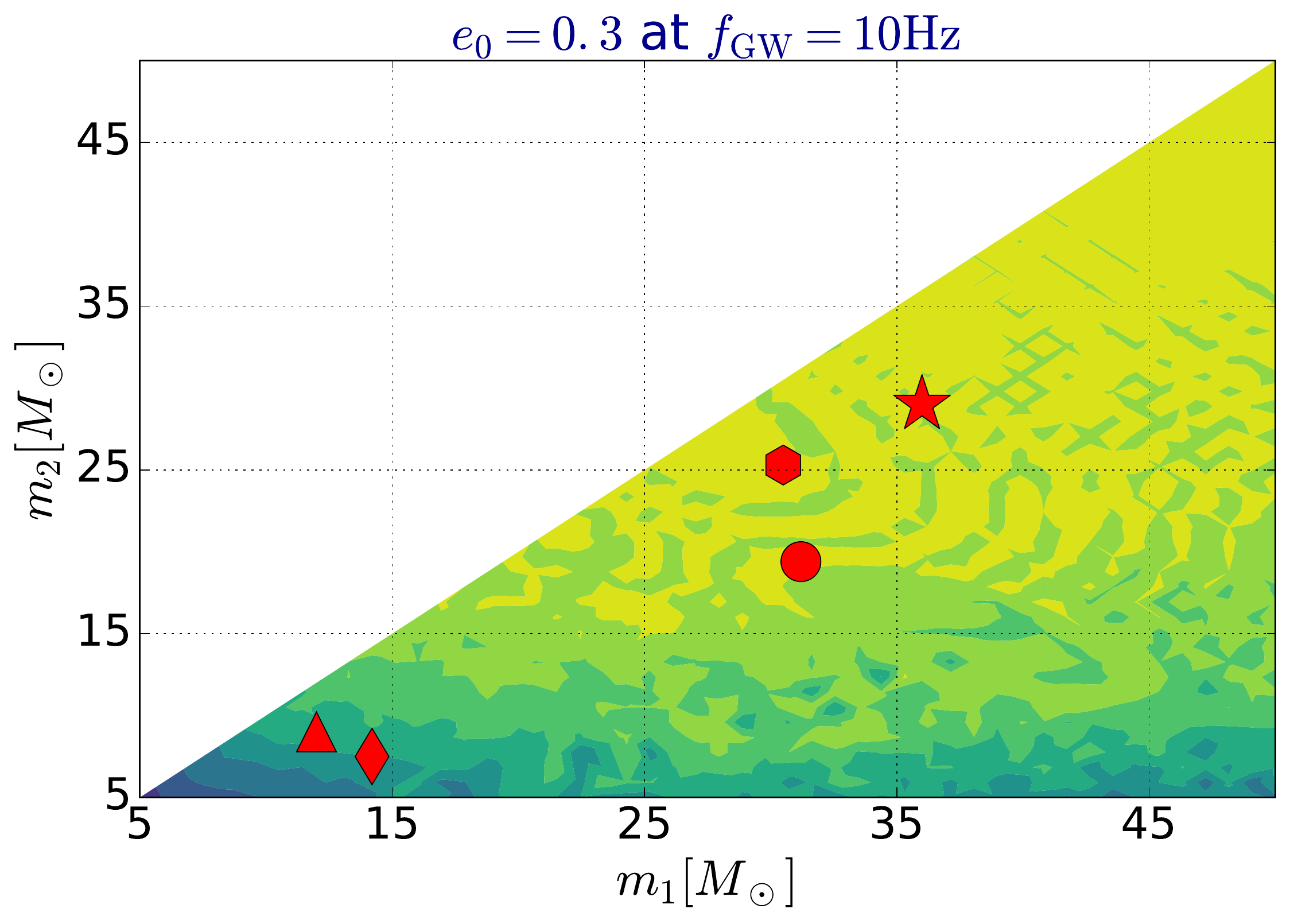}\hspace{-.2em}%
\includegraphics[height=.23\textwidth]{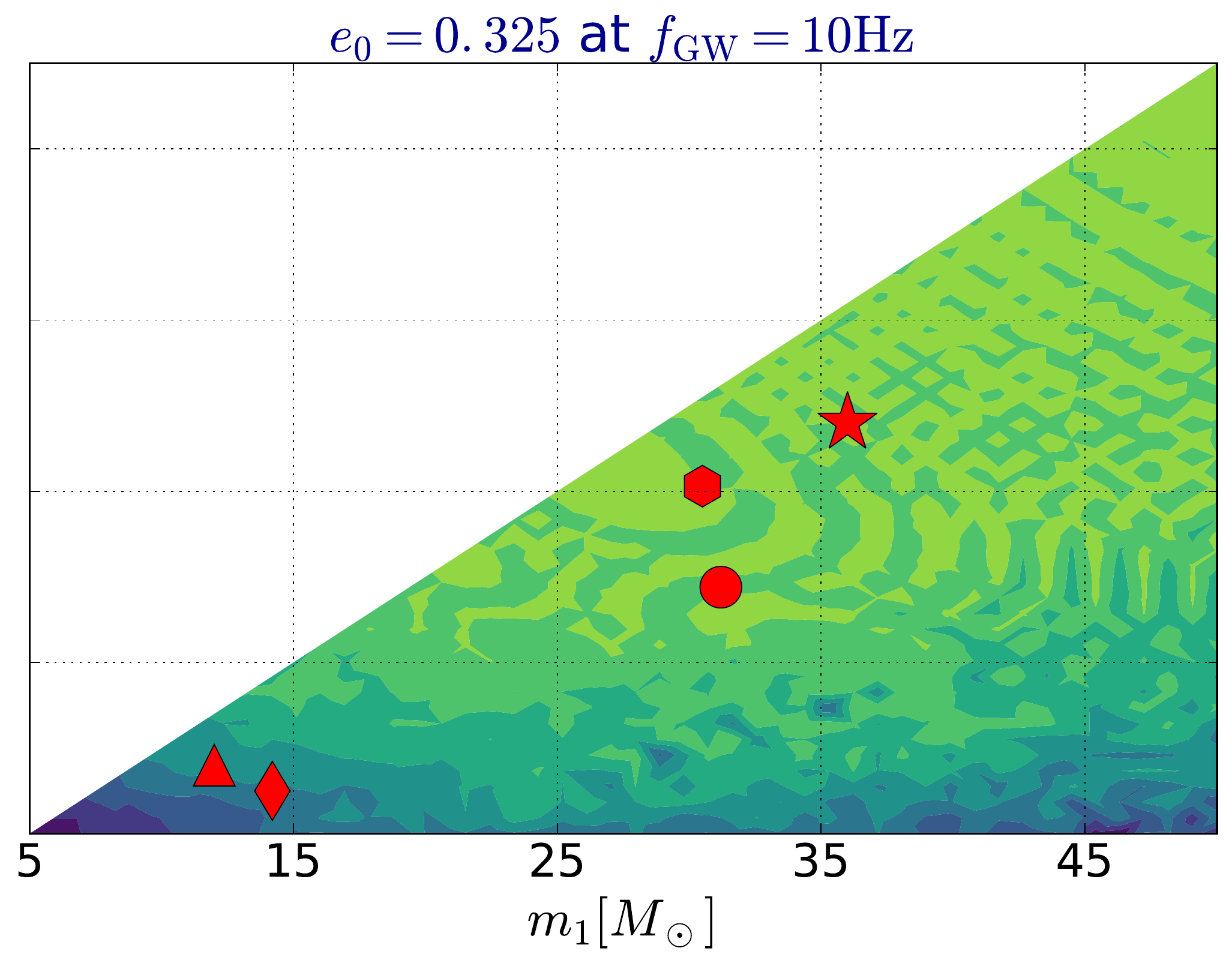}\hspace{-.2em}%
\includegraphics[trim= 0cm -1.3cm 0cm 0cm, height=.226\textwidth]{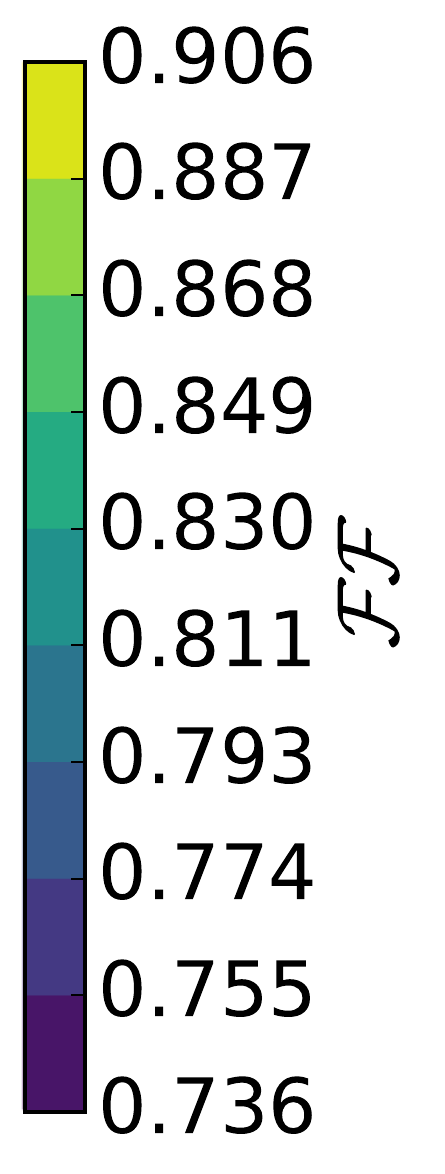}%
\end{array}$
\end{center}

\caption{Fitting Factor (\({\mathcal{FF}}\)) distribution as a function of initial eccentricity \(e_0\) at \(f_{\rm GW}=10{\rm Hz}\). The range of the color bar has been adjusted for each row to emphasize the eccentricity distribution across the binary black hole parameter space. The star, circle, triangle, diamond and hexagon represent the first five binary black hole mergers detected by aLIGO and aVirgo.}
\label{ff_all}
\end{figure*}

Figure~\ref{ff_all} presents a gallery of \({\cal{FF}}\) calculations for a variety of eccentric BBH populations. In these panels we present BBH populations with component masses \(m_{\{1,\,2\}}\in[5\msun,\,50\msun]\). The eccentricity range we explore is \(e_0\leq 0.325\) at \(f_{\rm GW}=10{\rm Hz}\). 

A word of caution is in order to interpret these results. Using a set of eccentric NR simulations, we have established that \texttt{ENIGMA} can accurately describe eccentric BBH systems throughout late inspiral, merger and ringdown. We have also shown that, in the quasi-circular limit, our model reproduces with excellent accuracy SEOBNRv4. Just as EOB models have been calibrated with quasi-circular NR simulations for the late inspiral and merger evolution, and then applied to explore the dynamics of BBH systems at lower frequencies, we expect that our model will provide an adequate description of moderately eccentric BBH systems during the early inspiral dynamics. To accomplish this, we have introduced a hybrid inspiral scheme that combines recent developments in the modeling of eccentric and quasi-circular BBHs, and which we have shown to perform very well to capture the dynamics of both eccentric and quasi-circular BBH systems very late in the inspiral evolution.

	\begin{table}[H]
		\caption{\label{results_FF} \({\cal{FF}}\) results for a variety of eccentric binary black hole populations with masses \(m_{\{1,\,2\}}=[5\msun,\,50\msun]\). The eccentricity \(e_0\) is defined at \(f_{\rm GW}=10{\rm Hz}\). See Figure~\ref{ff_all}.}
		\footnotesize
		\begin{center}
                        \setlength{\tabcolsep}{12pt} % default is apparently 6pt
			\begin{tabular}{c c c c c}
			\hline
			%\multicolumn{3}{|c|}{ \({\cal{FF}}\) distribution} \\
				\hline 
				\(e_0\) &\({\cal{FF}}^{\rm min}\)  & \({\cal{FF}}^{\rm max}\)  \\ 
				\hline
				0.010 & 0.994 & 0.999 \\
				0.025 & 0.992 & 0.999 \\
 				0.050 & 0.989 & 0.998 \\
				0.075 & 0.980 & 0.995 \\
				0.100 & 0.963 & 0.991 \\
				0.125 & 0.941 & 0.986 \\
				0.150 & 0.912 & 0.980 \\
       				$\mathbf{0.175}$ &$\mathbf{ 0.889}$ & $\mathbf{0.972}$ \\
				0.200 & 0.873 & 0.962 \\
				$\mathbf{0.225}$ & $\mathbf{0.842}$ & $\mathbf{0.951}$ \\
				0.250 & 0.829 & 0.937 \\
				0.275 & 0.794 & 0.923 \\
				$\mathbf{0.300}$ & $\mathbf{0.763}$ & $\mathbf{0.906}$ \\
				0.325 & 0.736 & 0.886 \\
				\hline 
			\end{tabular} 
		\end{center}
	\end{table}
	\normalsize

\noindent Table~\ref{results_FF} presents key results extracted from the \({\cal{FF}}\) calculations presented in Figure~\ref{ff_all}. Just as we found in~\cite{Huerta:2013a}---left panel of Figure 2 therein---the anomaly does not have a significant impact in the  \({\cal{FF}}\) results we present in this section.

If we assume that eccentricity corrections become significant when \({\cal{FF}}\lesssim0.97\), then we can highlight the following results for the BBH parameter space \(m_{\{1,\,2\}}\in[5\msun,\, 50\msun]\):

A circular search will be effectual for eccentric BBH that satisfy \(e_0\leq0.05\) at \(f_{\rm GW}=10{\rm Hz}\). We notice that eccentricity corrections become significant for \(e_0=0.175\), since at this particular threshold \({\cal{FF}}\lesssim0.97\) throughout the entire BBH parameter space. This eccentricity threshold is highlighted in Table~\ref{results_FF}. Putting this information in context with Figure~\ref{ff_all}, we notice that for this eccentricity value, some regions of the BBH parameter space have  \({\cal{FF}}\sim0.9\). These regions correspond to low mass and highly asymmetric mass-ratio BBH systems. These results are consistent with previous results reported in the literature~\cite{Huerta:2017a}, i.e., compact sources that generate long-lived GW signals in the aLIGO frequency band will be the ones that present the most significant imprints of eccentricity. Shorter signals, represented by eccentric, massive BBH systems, will be better recovered by quasi-circular templates. This is because eccentricity corrections have less time to accumulate, and thus have a lesser impact in the waveform phase. This pattern is clearly shown in the panels of Figure~\ref{ff_all}.

\noindent Table~\ref{results_FF} indicates that \({\cal{FF}}\leq0.95\) for \(e_0=0.225\). As before, whereas massive BBH systems can be recovered with \({\cal{FF}}\sim 0.95\), low mass and asymmetric mass-ratio BBH systems have \({\cal{FF}}\sim 0.84\). Finally, we notice that recovering a BBH population with \(e_0\geq 0.275\) would require the use of searches that specifically target eccentric binaries.

Figure~\ref{ff_all} presents the five GW transients, consistent with BBH mergers, currently reported by aLIGO: GW150914, GW151226, GW170104, GW170814 and GW170608. These are indicated by a star, a diamond, a circle, a hexagon and a triangle, respectively. Table~\ref{limits} indicates that these events can be recovered using spinning, quasi-circular SEOBNRv4 templates with \({\cal{FF}}\gtrsim0.96\) if \(e_0\leq \{0.175,\, 0.125,\,0.175,\,0.175,\, 0.125\}\) at \(f_{\rm GW}=10{\rm Hz}\), respectively.

	\begin{table}[!ht]
		\caption{\label{limits}  \({\cal{FF}}\) recovery of GW150914 (\(\#1\)), GW151226 (\(\#2\)), GW170104 (\(\#3\)), GW170814 (\(\#4\)) and GW170608 (\(\#5\)) with quasi-circular, spinning SEOBNRv4 waveforms.}
		\footnotesize
		\begin{center}
                        \setlength{\tabcolsep}{10pt} % default is apparently 6pt
			\begin{tabular}{c c c c c c}
			\hline
			\multicolumn{6}{|c|}{ \({\cal{FF}}\) recovery with spinning SEOBNRv4 waveforms} \\
				\hline 
				\(e_0\) & \(\#1\)  & \(\#2\) & \(\#3\) & \(\#4\) & \(\#5\) \\ 
				\hline
				0.010 & 0.996 & 0.994 & 0.998 & 0.997 & 0.995\\
				0.025 & 0.996 & 0.993 & 0.997 & 0.996 & 0.994\\
 				0.050 & 0.994 & 0.992 & 0.996 & 0.995 & 0.992\\
				0.075 & 0.992 & 0.984 & 0.993 & 0.991 & 0.986\\
				0.100 & 0.986 & 0.976 & 0.989 & 0.987 & 0.973\\
				0.125 & 0.982 & $\mathbf{0.961}$ & 0.984 & 0.982 & $\mathbf{0.956}$\\
				0.150 & 0.977 & 0.943 & 0.978& 0.974 & 0.934\\
				0.175 & $\mathbf{0.969}$ & 0.928 & $\mathbf{0.970}$& $\mathbf{0.963}$ & 0.920\\
				0.200 & 0.960 & 0.913 & 0.960& 0.960 &  0.906\\
				\hline 
			\end{tabular} 
		\end{center}
	\end{table}
	\normalsize

\begin{widetext}
 \begin{figure*}[!ht]
 \centerline{
    \includegraphics[width=0.485\textwidth]{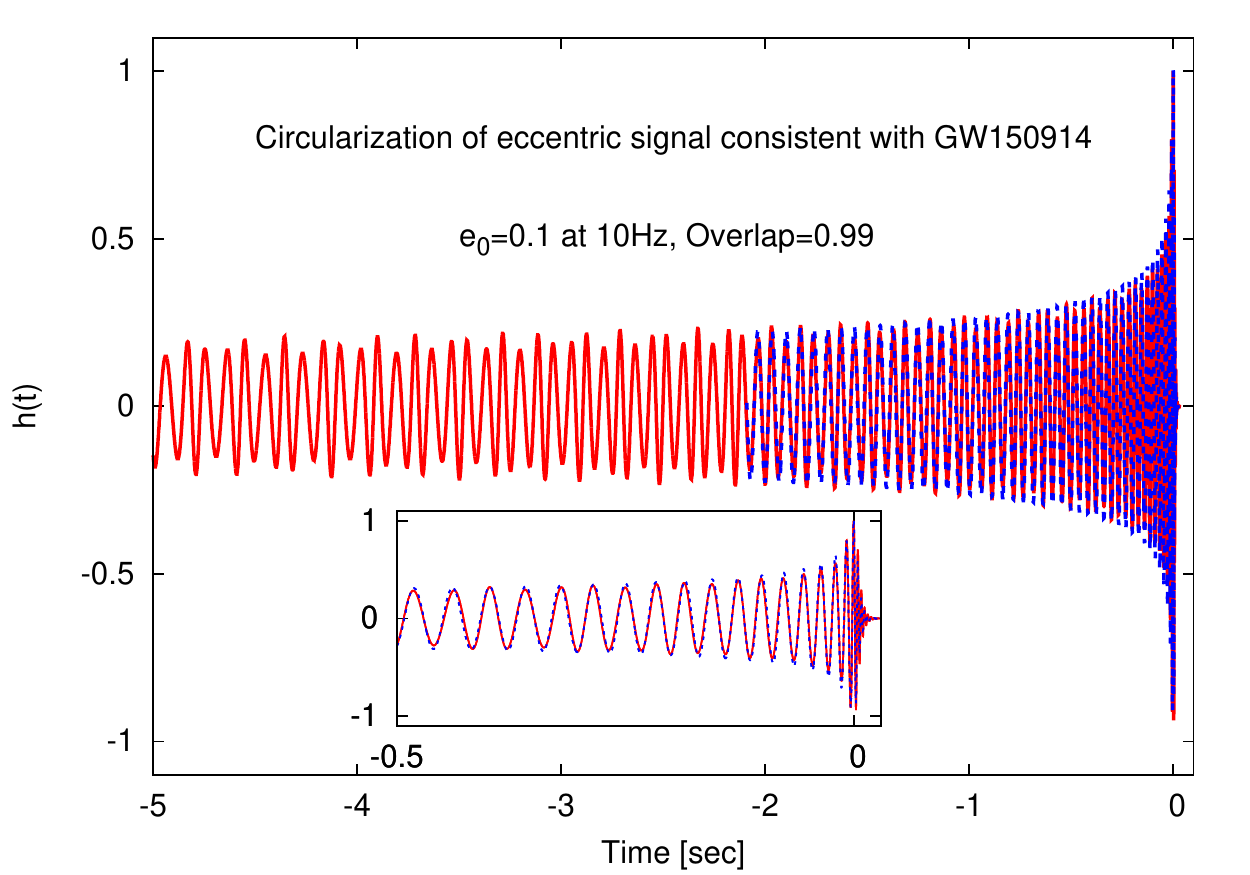}
     \includegraphics[width=0.485\textwidth]{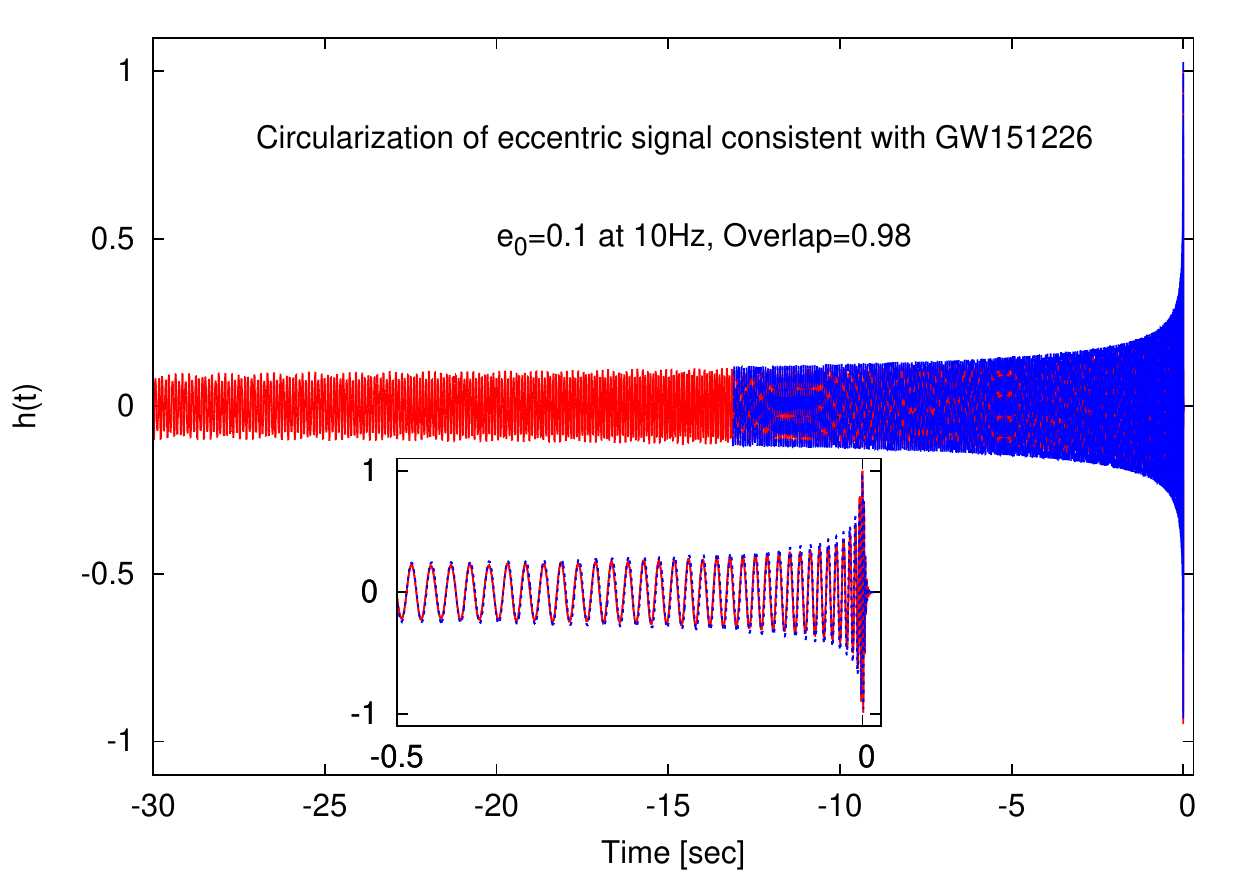}
    }
\centerline{
    \includegraphics[width=0.485\textwidth]{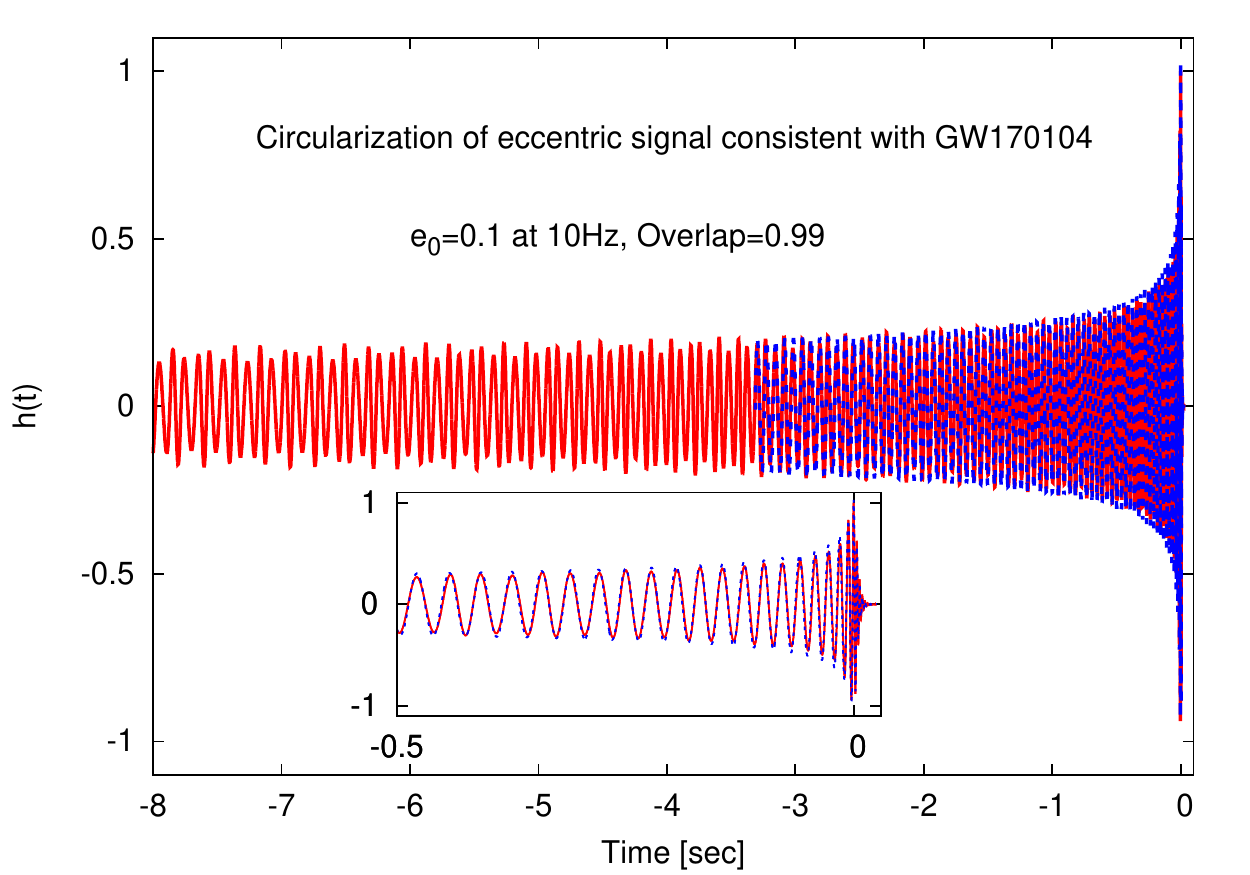}
    \includegraphics[width=0.485\textwidth]{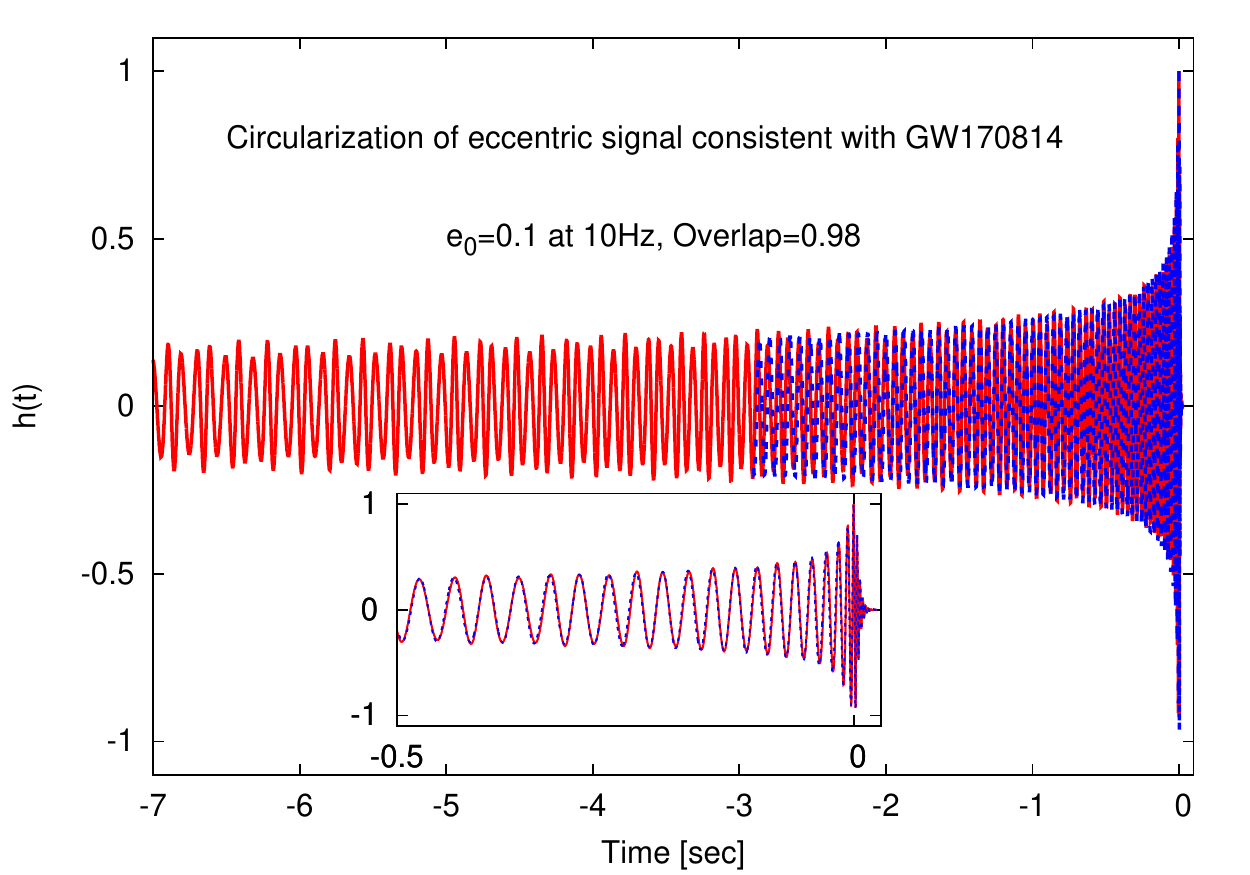}
     }  
     \centerline{
    \includegraphics[width=0.485\textwidth]{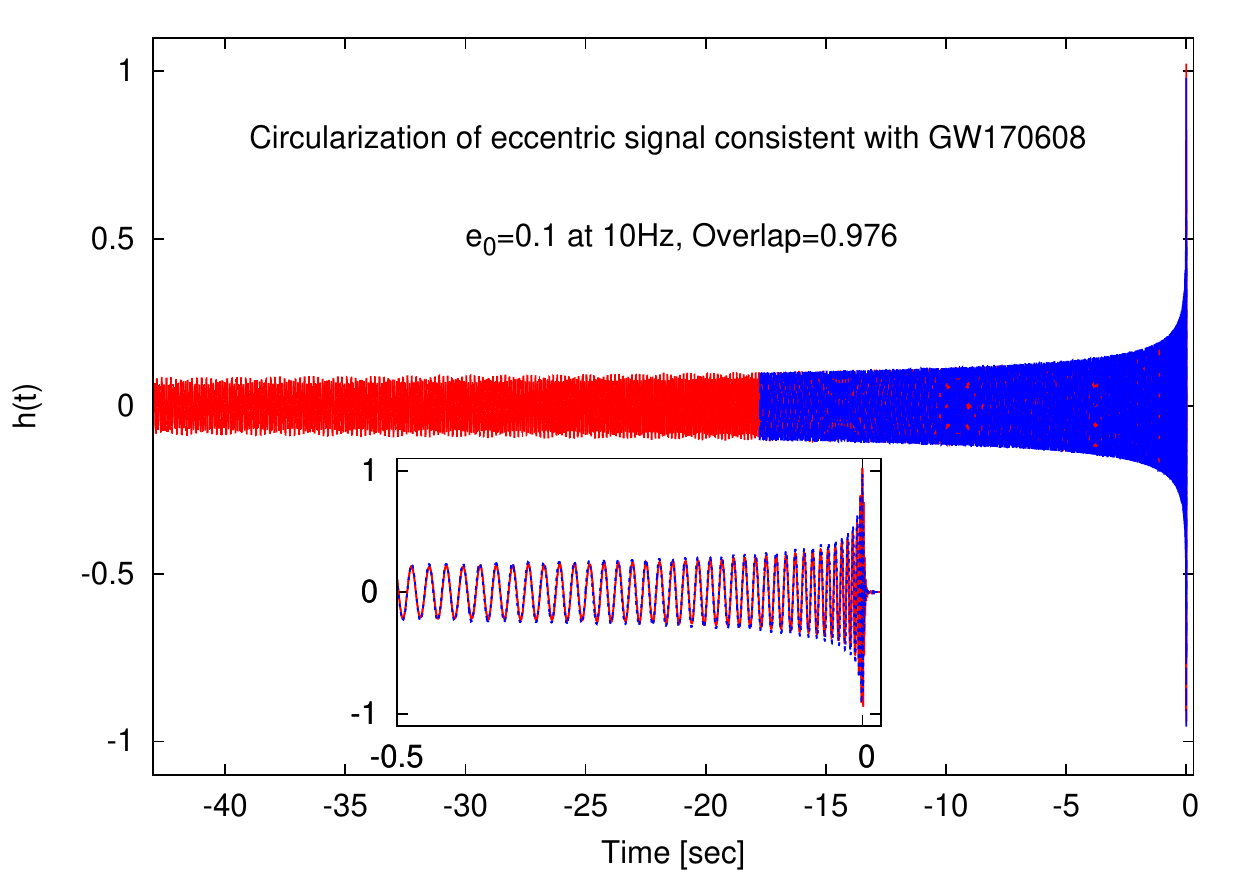}
    }
\caption{Recovery of moderately eccentric signals with spin-antialigned SEOBNRv4 waveforms. Circularization due to gravitational wave emission enables moderately eccentric signals to be misclassified as quasi-circular systems due to parameter space degeneracies between eccentricity and spin corrections.}
  \label{astro_take}
  \end{figure*}
 \end{widetext}

\noindent To clearly emphasize the content of these results, in Figure~\ref{astro_take} we present eccentric waveform signals that have \({\cal{FF}}\gtrsim0.98\) with the quasi-circular waveforms which the GW transients were assumed to be. The eccentric signals have \(e_0=0.1\) at \(f_{\rm GW}=10{\rm Hz}\). The quasi-circular SEOBNRv4 signals are generated from  \(f_{\rm GW}=14{\rm Hz}\). The overlap values quoted in each panel are computed using the design sensitivity of aLIGO from \(f_{\rm GW}=15{\rm Hz}\). These results clearly indicate that the GWs we have detected, and assumed to be quasi-circular, may have eccentricity content at lower frequencies that is not easily discerned by the time these signals become detectable by aLIGO and aVirgo. Indeed, all these signals can be recovered with \({\cal{O}}\geq0.98\). 

	\begin{table}[!ht]
		\caption{\label{recover} Recovery of GW150914 (\(\#1\)), GW151226 (\(\#2\)), GW170104 (\(\#3\)),  GW170814 (\(\#4\)) and GW170608 (\(\#5\)) with quasi-circular, spin-antialigned SEOBNRv4 waveforms.}
		\footnotesize
		\begin{center}
                        \setlength{\tabcolsep}{12pt} % default is apparently 6pt
			\begin{tabular}{c c c c c}
			\hline
			\multicolumn{5}{|c|}{ Recovery with spin-antialigned SEOBNRv4 waveforms} \\
				\hline 
				\(\#1\) & \(\delta m_1 [\%]\)  & \(\delta m_2 [\%]\) & \(s^z_1\) & \(s^z_2\)  \\ 
				\hline
				1 & 1.78 & 0.26 & 0.043 & -0.083\\
				2 & 3.07 & 3.67 & 0.449 & -0.738\\
 				3 & 0.41 & 0.96 & -0.288 & 0.454\\
				4 & 2.06 & 1.51 & 0.093 & -0.145\\
				5 & 4.75 & 5.14 & 0.051 & -0.401\\
				\hline 
			\end{tabular} 
		\end{center}
	\end{table}
	\normalsize

Furthermore, as shown in Table~\ref{recover}, the astrophysical parameters of both waveform signals, eccentric and quasi-circular, are very similar, with differences in the component masses of a few percent. The key difference is that our \texttt{ENIGMA} model currently describes non-spinning, eccentric binaries. In different words, GWs emitted by quasi-circular, spin-antialigned BBH systems can easily mimic the features of moderately eccentric, non-spinning BBH mergers. We have also computed \({\cal{FF}}\) results using a template bank of SEOBNRv4 waveforms with the spin of the binary components set to zero. Our results are consistent using both spin-aligned and non-spinning template banks, with deviations  \(\leq2\%\) level, in a similar spirit to the results reported in~\cite{Huerta:2017a}.

\noindent  In conclusion, even though recently detected GW transients can be effectively recovered with quasi-circular, spinning SEOBNRv4 templates, it is plausible that these events may have significant residual eccentricity at \(f_{\rm GW}\sim10{\rm Hz}\), and circularize by the time these systems become detectable by aLIGO, i.e., at \(f_{\rm GW}\gtrsim20{\rm Hz}\). We expect that when aLIGO~\cite{DII:2016,LSC:2015} and aVirgo~\cite{Virgo:2015} reach design sensitivity within the next few years, and we can start observing the evolution of compact binary mergers from \(f_{\rm GW}\gtrsim10{\rm Hz}\), we will be able to quantify eccentricity corrections, and clearly associate them to currently unconstrained astrophysical formation channels of compact binary populations in dense stellar environments. 

%%%%%%%%%%%%%%%%%%%%%%%%%%%%%%%%%%%%%%%%%%%%%
%%%%%%%%%%%%%%%%%%%%%%%%%%%%%%%%%%%%%%%%%%%%%
\section{Importance of higher-order waveform multipoles for the detection of eccentric binary black hole mergers}
\label{ho_modes}

There are several studies in the literature that have explored the importance of including higher-order waveform multipoles for the detection of quasi-circular binaries~\cite{Prayush:2013a,Colin:2013,london:170800404L}. There is, however, no study in the literature that has shed light on this important topic in the context of eccentric binary mergers. In this section, we provide a succinct introduction to this problem using our catalog of eccentric NR simulations. 

The NR higher-order waveform multipoles we use in this section exhibit similar convergent behavior to the \((\ell,\,m)=(2,\,2)\) we used in Section~\ref{ecc_waveforms} to validate the \texttt{ENIGMA} model. The waveform multipoles we use in this section are those that contribute more significantly to the waveform strain, namely \((\ell,\,m)=(2,\,2),\, (3,\,3),\, (4,\,4),\, (2,\,1)\) and \((3,\,2)\).

\begin{widetext}
 \begin{figure*}[!ht]
    \centerline{ 
    \includegraphics[width=0.46\textwidth]{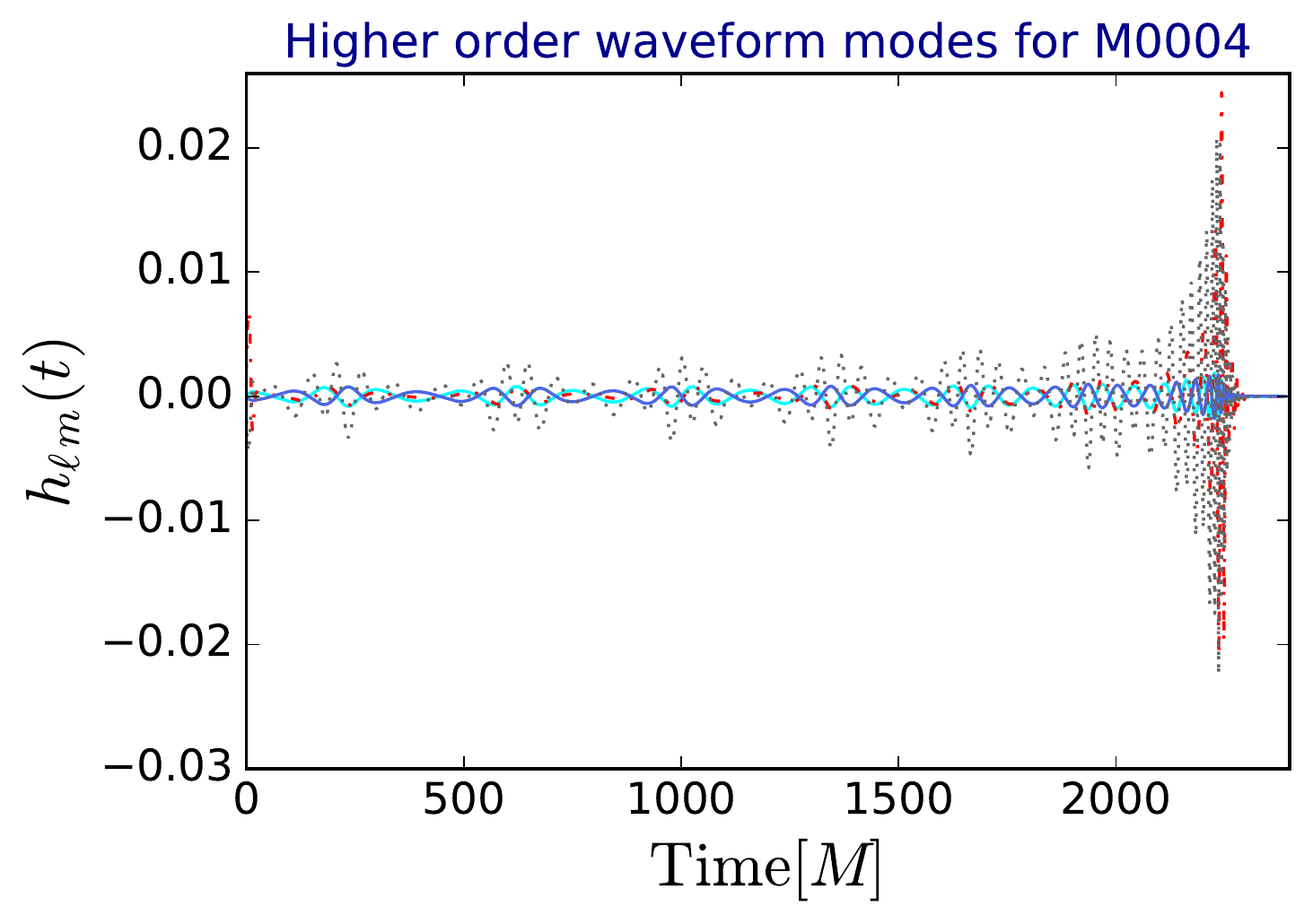}
    \raisebox{0.008\totalheight}{%
    \includegraphics[width=0.57\textwidth]{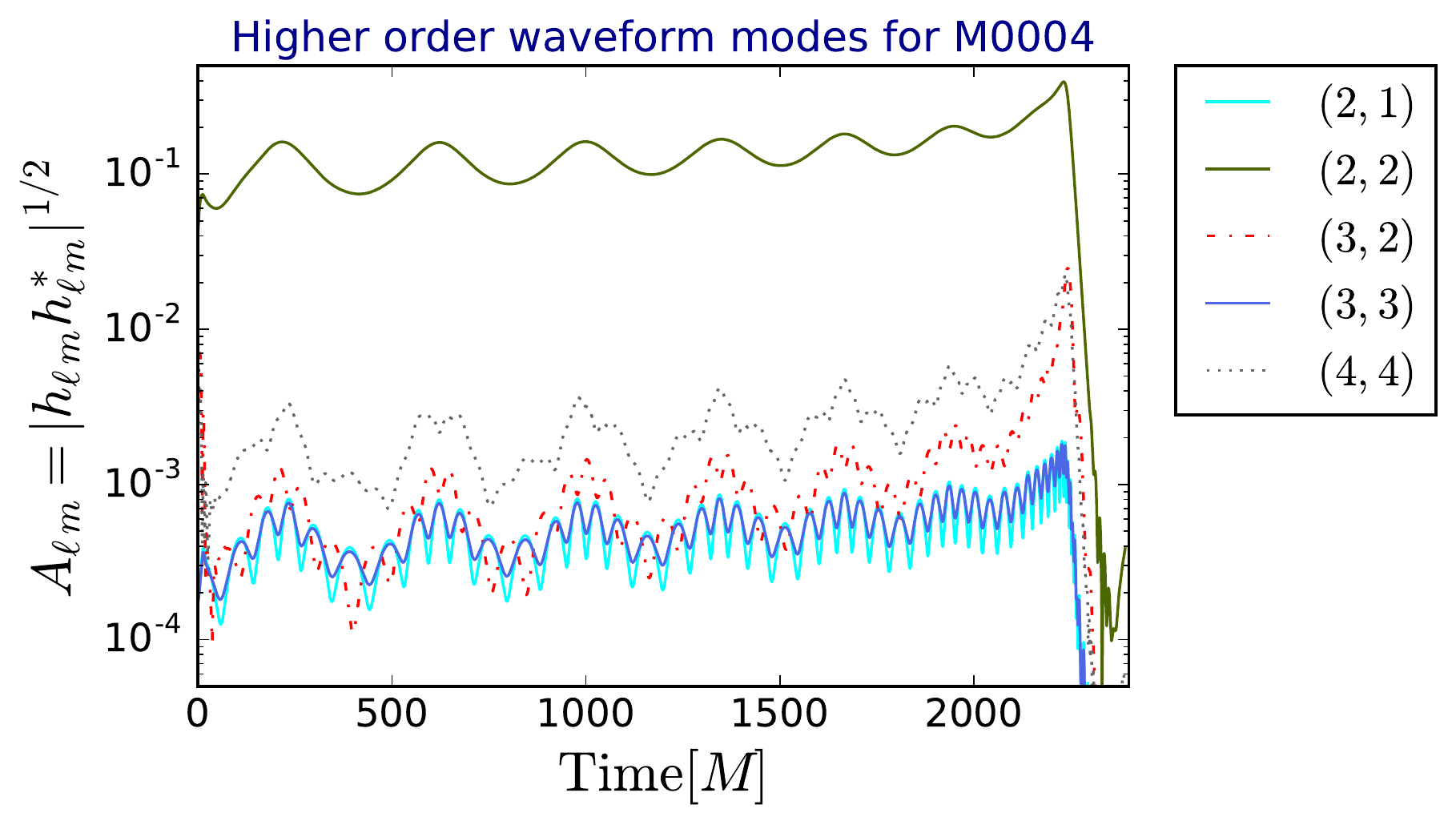}}
    }
 \centerline{
    \includegraphics[width=0.45\textwidth]{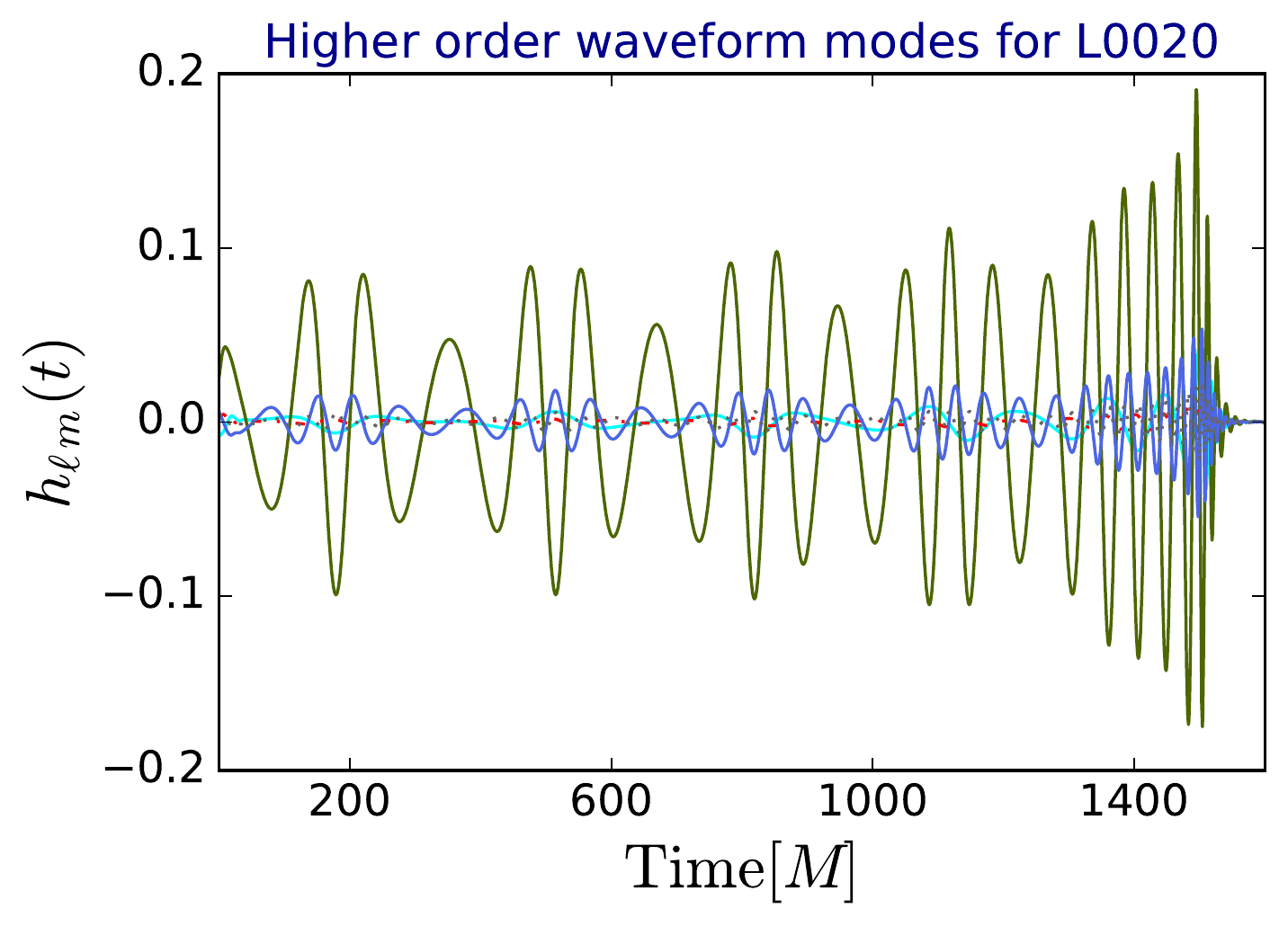}
    \raisebox{0.01\totalheight}{%
    \includegraphics[width=0.57\textwidth]{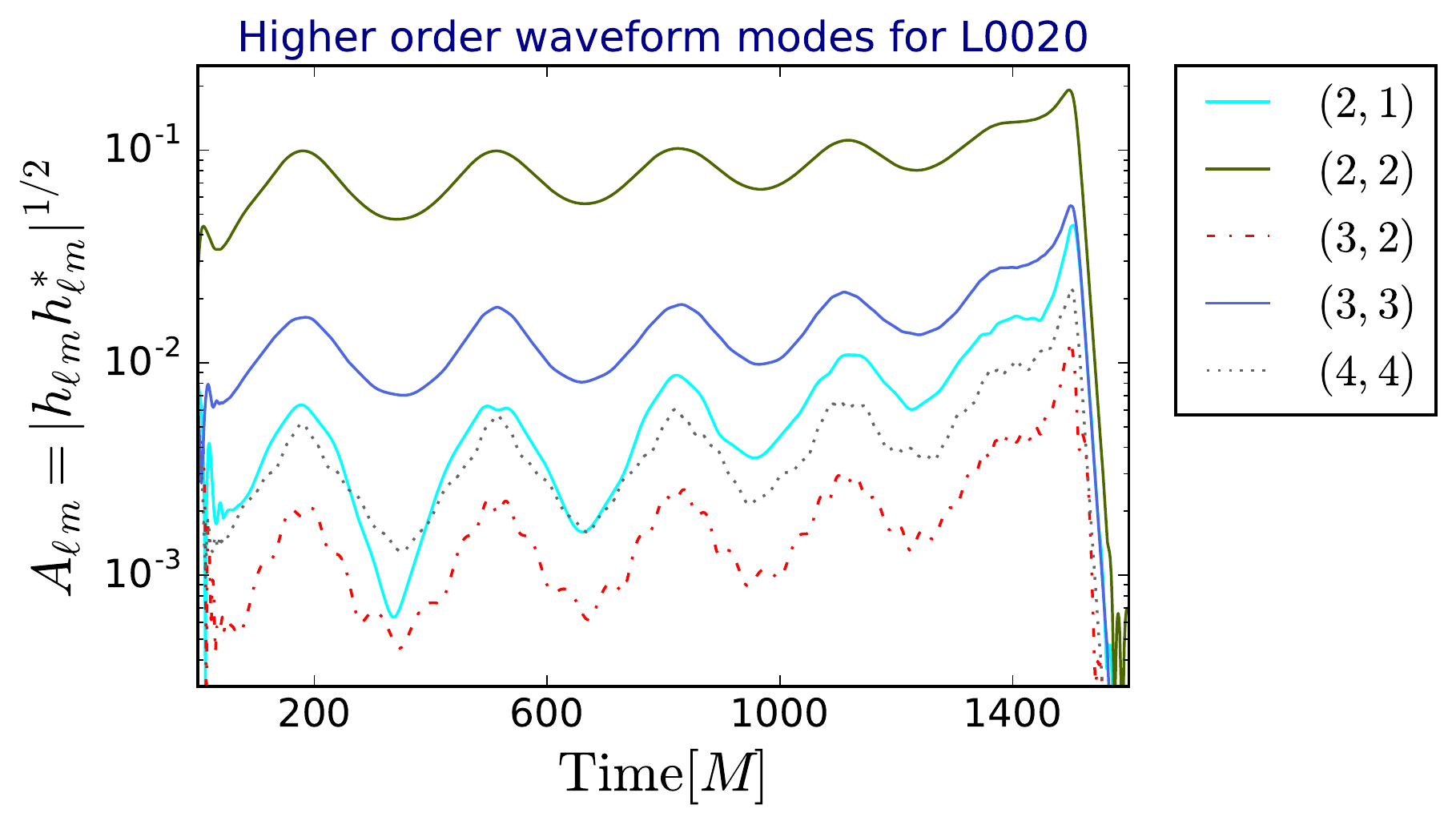}}
    }
\caption{Higher-order waveform multipoles of numerical relativity simulations that describe BBH mergers with the following properties, M0004:  \(q=1\) and \(e_0=0.19\) twelve orbits before merger; L0020: \(q=5.5\) and \(e_0=0.21\) ten orbits before merger. See Table~\ref{sims} for additional information on these numerical relativity simulations.}
  \label{L0020_M0004}
  \end{figure*}
 \end{widetext}

Figure~\ref{L0020_M0004} presents two cases. The top row presents the higher-order waveform multipoles of simulation M0004 in our catalog, as described in Table~\ref{sims}, i.e., \(q=1\) and \(e_0=0.19\) twelve orbits before merger. To clearly show the low impact of higher-order waveform multipoles for equal mass, eccentric BBH mergers, we use two different panels: the top left panel only shows the sub-dominant multipoles, whereas the right panel shows the amplitudes \(A_{\ell\,m}\) of all multipoles in comparison to the leading \((\ell,\,m)=(2,\,2)\) mode, clearly indicating that higher-order waveform multipoles contribute up to \(10\%\) of the signal power, with the greatest contribution near merger. In different words, these modes do not significantly contribute to the detectability of equal mass, eccentric BBH signals. 

The bottom panels of Figure~\ref{L0020_M0004} show the higher-order waveform multipoles of the L0020 simulation: \(q=5.5\) and \(e_0=0.21\) ten orbits before merger. These results show that the sub-dominant multipoles \((\ell,\,m)=(2,\,1),\, (3,\,3),\, (4,\,4)\) contribute up to \(20\%\) of the total signal power in the vicinity of merger. Therefore, the inclusion of higher-order waveform multipoles will be necessary for searches of eccentric BBH mergers that have asymmetric mass-ratios. This finding is consistent with results in the context of quasi-circular BBH mergers~\cite{Prayush:2013a,Colin:2013,london:170800404L}, i.e., higher-order waveform multipoles contribute significantly to the total waveform strain of asymmetric mass-ratio systems, having an important impact on their SNR, and therefore on their detectability. A detailed study of this important source modeling topic is well underway, and will be presented in an accompanying paper~\cite{Huerta:howm}. 

%%%%%%%%%%%%%%%%%%%%%%%%%%%%%%%%%%%%%%%%%%%%%
%%%%%%%%%%%%%%%%%%%%%%%%%%%%%%%%%%%%%%%%%%%%%
\section{Conclusions}
\label{end}

We have developed \texttt{ENIGMA}, a complete waveform model to search for and characterize compact binary populations 
that form in dense stellar environments, and which are expected to enter the aLIGO frequency band with
moderate values of eccentricity. 

Our model is a combination of analytical and numerical relativity results. It is constructed under the assumption
that moderately eccentric compact binary systems circularize prior to merger. In this context, we describe the inspiral
evolution using a variety of recent results from PN theory, the self-force and BHPT. On the other hand, to describe the
quasi-circular merger evolution, we use GPR to create a stand-alone, merger waveform that is trained with a
dataset of quasi-circular NR waveforms. We described a method to put together these two pieces so as to provide
a complete description of moderately eccentric compact binary systems. Our results demonstrate that \texttt{ENIGMA} 
describes with excellent accuracy the dynamics of both quasi-circular and moderately eccentric systems in a single, unified
framework. 

We have \textit{validated}  \texttt{ENIGMA} with a set of eccentric NR simulations that describe BBH mergers with mass-ratios 
\(q\leq 5.5\) and eccentricities \(e_0\lesssim 0.2\) ten orbits before merger. To the best of our knowledge, this is the only model in the literature that can reproduce the dynamics of eccentric compact binary mergers for this combination of highly eccentric, and very asymmetric mass-ratio systems. We have also validated \texttt{ENIGMA} in the quasi-circular limit using SEOBNRv4 waveforms, and have shown that both waveform families have overlaps \({\cal{O}}\geq0.99\), assuming an initial filtering frequency \(f_{\rm GW}=15{\rm Hz}\).

Having validated \texttt{ENIGMA} both in the quasi-circular limit, and with eccentric NR simulations, we used it to quantify the threshold at which the effect of eccentricity is negligible, and existing circular searches are effectual for moderately eccentric BBH mergers. Our studies show that BBH populations with \(e_0\leq 0.05\) at \(f_{\rm GW}=10{\rm Hz}\) will be recovered with spinning, quasi-circular SEOBNRv4 templates with \({\cal{FF}}\geq 0.99\). At this level, a circular search will be effectual. On the other hand, BBH populations with \(e_0\sim 0.175\) at \(f_{\rm GW}=10{\rm Hz}\) will be recovered with a maximum \({\cal{FF}}\sim 0.97\). However, within this same population, low mass, and asymmetric mass-ratio BBH populations are recovered with \({\cal{FF}}\sim 0.88\). BBH populations with eccentricities \(e_0\sim 0.225\) at \(f_{\rm GW}=10{\rm Hz}\), will be recovered with \(0.84 \lesssim {\cal{FF}}\lesssim 0.95\). Finally, BBH populations with \(e_0\geq0.275\) will require the use of dedicated eccentric searches, since these events would be poorly recovered with \(0.79 \lesssim {\cal{FF}}\lesssim 0.92\).

Our calculations indicate that GW150914, GW151226, GW170104, GW170814 and GW170608 can be recovered with
spinning, quasi-circular SEOBNRv4 templates with \({\cal{FF}}\geq0.96\) if the eccentricity of these events satisfies \(e_0\leq \{0.175,\, 0.125,\,0.175, \, 0.175,\, 0.125\}\) at \(f_{\rm GW}=10{\rm Hz}\) , respectively. We have also shown that the first five GW transients detected by aLIGO and aVirgo could be misclassified as quasi-circular systems, since spin corrections can easily mimic the properties of moderately eccentric signals when they become detectable by ground-based GW detectors. We argue that future improvements to the sensitivity of aLIGO and aVirgo will be critical to identify and quantify eccentricity content in GW signals at lower frequencies. 

We have carried out a preliminary analysis of the importance of including higher-order waveform multipoles, and have shown that these will be important for searches of eccentric BBH mergers whose components have asymmetric mass-ratios. This result is consistent with similar analysis carried out in the context of quasi-circular BBH mergers. 

Having completed the description of non-spinning, eccentric BBH mergers, ongoing work is focused on the development of an extended version of \texttt{ENIGMA} that will enable the characterization of spinning, eccentric BBH systems. This new model will be useful to explore whether the degeneracy between eccentricity and spin corrections we have found in these studies can be resolved. 

%%%%%%%%%%%%%%%%%%%%%%%%%%%%%%%%%%%%%%%%%%%%%
%%%%%%%%%%%%%%%%%%%%%%%%%%%%%%%%%%%%%%%%%%%%%
\section{Acknowledgements}
\label{ack}
This research is part of the Blue Waters sustained-petascale computing project, which is supported by the National Science Foundation (awards OCI-0725070 and ACI-1238993) and the State of Illinois. Blue Waters is a joint effort of the University of Illinois at Urbana-Champaign and its National Center for Supercomputing Applications. The eccentric numerical relativity simulations used in this article were generated with the open source, community software, the Einstein Toolkit on the Blue Waters petascale supercomputer and XSEDE (TG-PHY160053). We acknowledge support from the NCSA and the SPIN (Students Pushing Innovation) Program at NCSA. 
PK gratefully acknowledges support for this research at CITA from NSERC of Canada, the Ontario Early Researcher Awards Program, the Canada Research Chairs Program, and the Canadian Institute for Advanced Research.
CJM has received funding from the European Union's Horizon 2020 research and innovation programme under the Marie Sk\l odowska-Curie grant No 690904, and from STFC Consolidator Grant No. ST/L000636/1.
RH is supported by NSF grant 1550514. We thank the \href{http://gravity.ncsa.illinois.edu}{NCSA Gravity Group} for useful feedback and suggestions, and Ian Hinder for a painstaking review of this manuscript. 

%%%%%%%%%%%%%%%%%%%%%%%%%%%%%%%%%%%%%%%%%%%%%
%%%%%%%%%%%%%%%%%%%%%%%%%%%%%%%%%%%%%%%%%%%%%
%\clearpage
\appendix 

\section{Convergence of numerical relativity simulations}
\label{ap1}

\begin{widetext}
 \begin{figure*}[!ht]
 \centerline{   
     \includegraphics[width=0.54\textwidth]{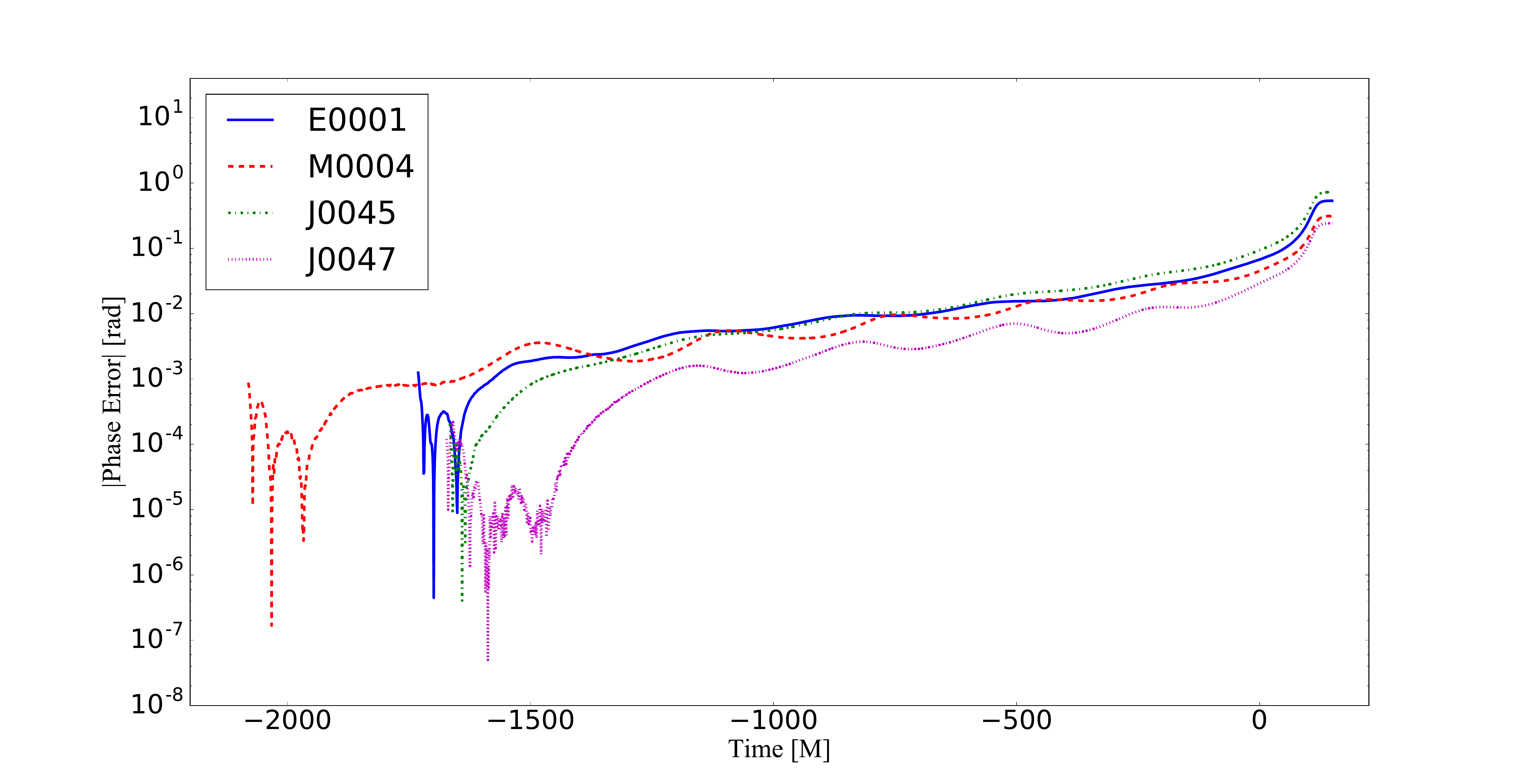}
     \includegraphics[width=0.54\textwidth]{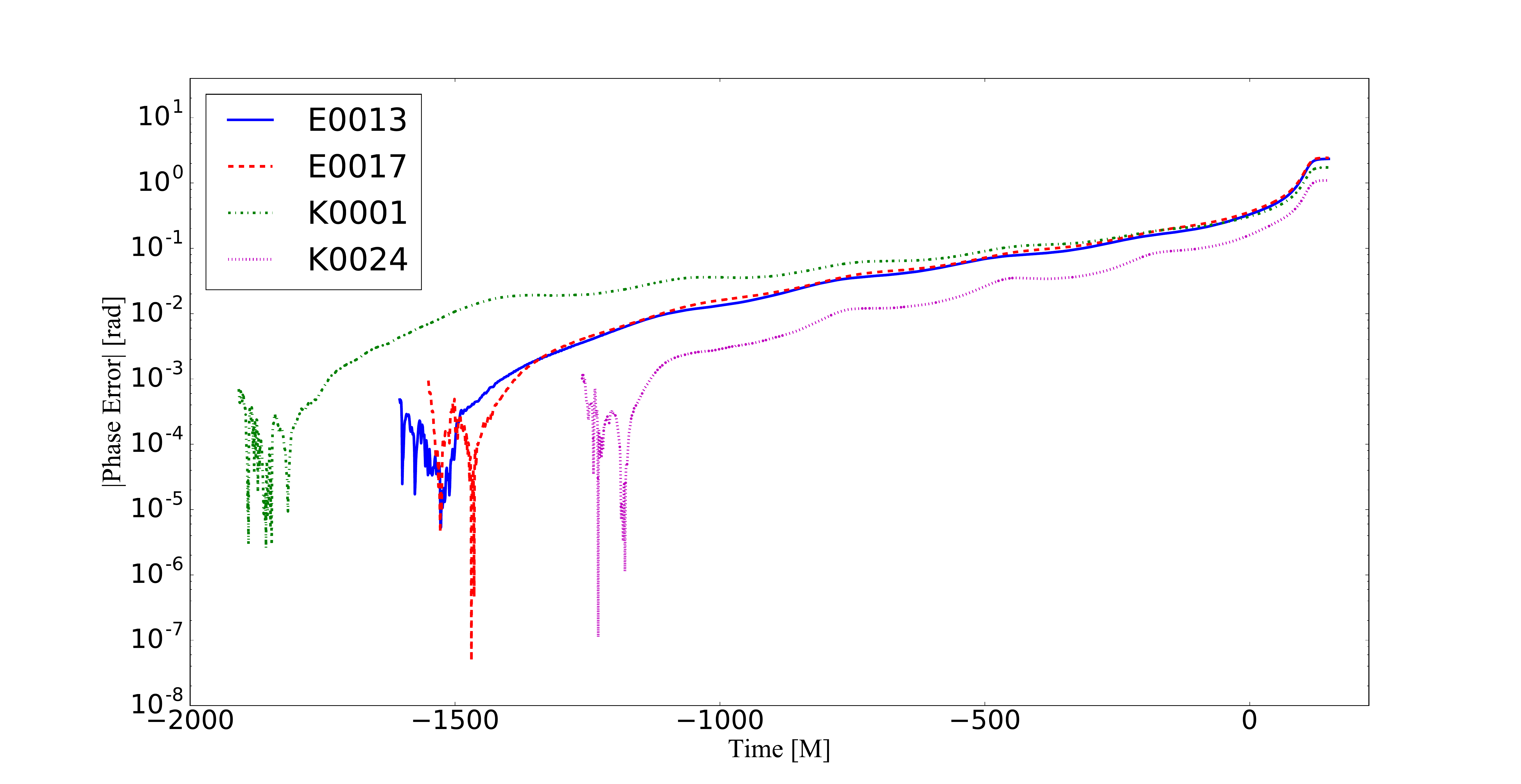}
}
\caption{Richardson extrapolation calculations to estimate the phase error of a sample of NR simulations used to validate \texttt{ENIGMA}. The merger time, given by the amplitude peak of the \((\ell,m)=(2,2)\) mode of the NR waveforms, is at \(t=0M\).}
  \label{convergence_I}
  \end{figure*}
 \end{widetext}

Using Richardson Extrapolation, Figure~\ref{convergence_I} presents the waveform phase error of a sample of eccentric NR simulations that we have generated to validate \texttt{ENIGMA}. We notice that the accumulated waveform phase error at merger, which is defined as the amplitude peak of the \((\ell,m)=(2,2)\) mode waveform, is \(\lesssim 0.2\, {\rm rads}\). A detailed analysis of these simulations is presented in an accompanying paper~\cite{Huerta:ncsacatalog}.

\bibliography{references}

\end{document}